\documentclass[11pt,a4paper]{article}

\usepackage{jcappub}
\usepackage{graphicx}
\usepackage{bm}
\usepackage{latexsym}
\usepackage{psfrag}
\usepackage{ulem}
\usepackage{textcomp}
\usepackage{color}
\usepackage{pstricks}

\makeatletter
\gdef\@fpheader{}
\makeatother

\def\benu{\begin{enumerate}}
\def\eenu{\end{enumerate}}
\def\nn{\nonumber} 
\def\pa{{\partial}}
\def\f{\frac}
\def\l{\left}
\def\r{\right}
\def\d{{\rm d}}
\def\cR{{\mathcal R}}

\def\ei{\eta_{\rm i}}
\def\es{\eta_{\rm s}}
\def\ee{\eta_{\rm e}}

\def\Ni{N_{\rm i}}
\def\Ns{N_{\rm s}}
\def\Ne{N_{\rm e}}
\def\Nf{N_{\rm f}}

\def\vk{{\bm k}}
\def\vka{{\bm k}_{1}}
\def\vkb{{\bm k}_{2}}
\def\vkc{{\bm k}_{3}}
\def\ska{{k_{1}}}
\def\skb{{k_{2}}}
\def\skc{{k_{3}}}
\def\cA{{\cal A}}
\def\cB{{\cal B}}
\def\cG{{\cal G}}
\def\cI{{\cal I}}
\def\cJ{{\cal J}}
\def\cM{{\cal M}}
\def\fnl{f_{_{\rm NL}}}
\def\hnl{h_{_{\rm NL}}}
\def\cnls{C_{_{\rm NL}}^{\mathcal R}}
\def\cnlt{C_{_{\rm NL}}^{\mathcal \gamma}}
\def\Mpl{M_{_{\rm Pl}}}
\def\Mp{M_{_{\rm Pl}}}

\newcommand{\sg}{{\cal G}}
\newcommand{\sr}{{\cal R}}
\newcommand{\g}{\gamma}                 
\newcommand{\vare}{\varepsilon}                 
\newcommand{\kT}{k_{_{\rm T}}}
\newcommand{\kt}{k_{_{\rm T}}}

\newcommand{\ko}{k_0}
\newcommand{\rr}{\rho^}
\newcommand{\ktr}{k_{_{\rm T}}^}
\newcommand{\kor}{k_0^}

\newcommand{\bea}{\begin{eqnarray}}
\newcommand{\eea}{\end{eqnarray}}
\newcommand{\gagaga}{\gamma\gamma\gamma}

\newcommand{\epij}{\epsilon_{ij}}

\begin{document}
\title{Numerical evaluation of the three-point scalar-tensor 
cross-correlations and the tensor bi-spectrum}
\author[a]{V.~Sreenath,}
\affiliation[a]{Department of Physics, Indian Institute of 
Technology Madras, Chennai~600036, India}
\emailAdd{sreenath@physics.iitm.ac.in}
\author[b]{Rakesh Tibrewala}
\affiliation[b]{School of Physics, Indian Institute of Science 
Education and Research, CET Campus,
Thiruvananthapuram~695016, India}
\emailAdd{rtibs@iisertvm.ac.in}
\author[a]{and L.~Sriramkumar}
\emailAdd{sriram@physics.iitm.ac.in}
\date{today} 
\abstract{Utilizing the Maldacena formalism and extending the earlier 
efforts to compute the scalar bi-spectrum, we construct a numerical 
procedure to evaluate the three-point scalar-tensor cross-correlations 
as well as the tensor bi-spectrum in single field inflationary models 
involving the canonical scalar field.
We illustrate the accuracy of the adopted procedure by comparing the
numerical results with the analytical results that can be obtained in 
the simpler cases of power law and slow roll inflation.
We also carry out such a comparison in the case of the Starobinsky model 
described by a linear potential with a sudden change in the slope, which
provides a non-trivial and interesting (but, nevertheless, analytically 
tractable) scenario involving a brief period of deviation from slow roll. 
We then utilize the code we have developed to evaluate the three-point 
correlation functions of interest (and the corresponding non-Gaussianity 
parameters that we introduce) for an arbitrary triangular configuration 
of the wavenumbers in three different classes of inflationary models 
which lead to features in the scalar power spectrum, as have been 
recently considered by the Planck team.
We also discuss the contributions to the three-point functions during
preheating in inflationary models with a quadratic minimum.
We conclude with a summary of the main results we have obtained.}
\maketitle


\section{Introduction}\label{sec:introduction}

The inflationary paradigm has now become a corner stone in our understanding 
of the universe at the large scales (see any of the following texts
Refs.~\cite{texts} or the reviews Refs.~\cite{reviews}). 
Inflation was originally introduced to overcome the so-called horizon problem 
of the conventional hot big bang model.
Currently though, the most attractive aspect of the inflationary scenario rests
on its ability to provide a mechanism for the origin of perturbations in the
early universe.
The perturbations generated during inflation evolve and leave their imprints 
as anisotropies in the Cosmic Microwave Background (CMB). 
With the anisotropies in the CMB being measured with constantly improving 
precision over the last decade or two, there have been emerging stronger
and stronger constraints on the inflationary models.

\par
 
Until not so long ago, constraints on inflationary models were largely 
arrived at by essentially comparing the models with the data at the 
level of the power spectrum.
A nearly scale invariant primordial power spectrum, as is generated by 
the simplest of inflationary models, such as those involving a single 
scalar field and leading to a sufficiently long period of slow roll, is 
found to be strikingly consistent with the observations of the CMB 
anisotropies by the missions such as the Wilkinson Microwave Anisotropy 
Probe (WMAP)~\cite{wmap-2009,wmap-2011,wmap-2012} and 
Planck~\cite{planck-2013-cmbps,planck-2013-ccp,planck-2013-ci} as well 
as various other cosmological data.
However, over the last decade, it was increasingly recognized that 
non-Gaussianities can play a vital role in arriving at tighter 
constraints on models of the early universe.
This expectation has been corroborated by the observations of the recent
Planck mission, which has pointed to the fact that the non-Gaussianities
are consistent with zero.
Specifically, Planck finds that the three parameters that are often used 
to characterize the scalar bi-spectrum to be: $\fnl^{\rm loc}=2.7\pm 5.8$, 
$\fnl^{\rm eq}=-42\pm 75$ and $\fnl^{\rm ortho}
=-25\pm 39$~\cite{planck-2013-cpng}.
These observations seem to imply that the data strongly favor slow roll 
inflationary models driven by a single, canonical scalar field (in this 
context, see the following rather comprehensive effort of comparing 
various models with the recent data: Ref.~\cite{martin-2013}.)

\par

Most of the efforts towards understanding non-Gaussianities generated by 
the inflationary models and arriving at constraints from the observational 
data have focused on the scalar bi-spectrum and the corresponding 
non-Gaussianity parameter $\fnl$ (for the theoretical efforts, see, for
instance, Refs.~\cite{maldacena-2003,ng-ncsf,ng-reviews}; for earlier 
work, i.e. prior to Planck, towards arriving at observational constraints, 
see, for example, Refs.~\cite{ng-da,ng-da-reviews}).
There have also been some theoretical efforts aimed at analyzing
the behavior of the tensor bi-spectrum (see, for example, 
Refs.~\cite{maldacena-2003,tensor-bs}).
But, we find that, there have been relatively limited attempts at
studying the scalar-tensor cross-correlations (see, for instance, 
Refs.~\cite{maldacena-2003,cc,cc2}). 
It will be interesting to closely examine the behavior of these quantities 
and, eventually, try to arrive at observational constraints on the 
corresponding non-Gaussianity parameters that can be constructed to 
characterize these quantities.
The fact that tensors remain to be detected at the level of the power 
spectrum could have been a dissuading factor in the limited attention 
devoted to the cross-correlations and the tensor bi-spectrum. 
However, it is important to bear in mind that, popular models, such as those
described by the quadratic and quartic potentials involving the canonical
scalar field, are being ruled out by the Planck data primarily by the upper 
limits on the tensor-to-scalar ratio~\cite{planck-2013-ci,martin-2013}. 
We believe that the three-point functions involving the scalars and the
tensors can play a similar role leading to additional constraints on the 
inflationary models.
  
\par

While, as we mentioned above, a scale independent power spectrum is rather 
consistent with the data, certain features in the power spectrum seem to 
improve the fit to the data, often, at the cost of some additional parameters.
For instance, features such as either a sharp cut-off at large 
scales~\cite{quadrupole,pi}, a burst of oscillations over an intermediate range 
of scales~\cite{l2240,hazra-2010,benetti-2011} or repeated modulations extending 
over a wide range of 
scales~\cite{pso,pahud-2009,flauger-2010,kobayashi-2011,aich-2013,peiris-2013,sm} 
were known to lead to a better fit to the WMAP data than the more conventional, 
nearly, scale invariant, power law, primordial spectrum.
Interestingly, exactly these classes of scenarios were analyzed by the Planck 
team, who find that, at the level of the power spectrum, while such features 
do improve the fit to the data, the corresponding Bayesian 
evidence~\cite{be-fim} does not exhibit any substantial 
change~\cite{planck-2013-ci}.
Moreover, it is well established that these features can be generated only 
due to deviations from slow roll~\cite{e-dfsr,starobinsky-1992,recent-fips} 
(unless one assumes that the perturbations are in an excited state above the 
Bunch-Davies vacuum~\cite{bunch-1978,ng-nvis}) which, in turn, can boost 
the extent of non-Gaussianities (in this context, see 
Refs.~\cite{ng-ne,ng-f,hazra-2013}), possibly, beyond the levels constrained 
by Planck~\cite{planck-2013-cpng}. 
Nevertheless, we believe that it is worthwhile examining the non-trivial 
scenarios further, as such exercises can aid us judge the extent of the 
constraints imposed by the Planck data.

\par

In the above backdrop, the aims of this work can be said to be three 
fold.
Firstly, extending the recent effort towards computing the scalar 
bi-spectrum~\cite{hazra-2013}, we devise a similar numerical procedure to 
evaluate the three-point scalar-tensor cross-correlations and the tensor 
bi-spectrum as well as the corresponding non-Gaussianity parameters that
we introduce.
Secondly, utilizing the developed numerical procedure, we evaluate these
quantities in the models leading to features in the scalar power spectrum, 
which do not permit analytical calculation of these quantities.
Thirdly, we consider the contribution to these quantities during the period 
of preheating, viz. the epoch which immediately follows 
inflation~\cite{prh,ep-d-prh}.

\par

The plan of this paper is as follows. 
In the next section, we shall begin by sketching a few essential points
concerning the Maldacena formalism to arrive at the action governing the 
perturbations at the cubic order in inflationary scenarios involving a 
single canonical scalar field~\cite{maldacena-2003}.
As should be evident, when the scalars as well as the tensors are taken 
into account, at the cubic order, the action can consist of either purely
three scalars or tensors or it can contain cross-terms comprising of two 
scalars and a tensor or one scalar and two tensors.
We shall describe the three-point scalar-tensor cross-correlations and 
the tensor bi-spectrum that these actions lead to and also discuss the 
corresponding non-Gaussianity parameters.  
In Sec.~\ref{sec:nm}, we shall outline the numerical procedure that we
adopt for evaluating the scalar-tensor cross-correlations and the tensor 
bi-spectrum.
We shall begin by showing that, as in the case of the scalar bi-spectrum 
(in this context, see, for instance, Refs.~\cite{hazra-2013,hazra-2012}), 
the super-Hubble contributions to the other three-point correlation 
functions too turn out to be negligible.
Further, as in the pure scalar case, one needs to introduce a suitable 
cut-off in the sub-Hubble domain in order to deal with the continued
oscillations that would otherwise arise.
Under these conditions, we shall illustrate that, it proves to be sufficient 
to evolve the scalar and the tensor modes from sufficiently inside the Hubble
radius to a suitably late time when they are well outside, and evaluate the 
integrals involved over this period.
In order to demonstrate the accuracy of the numerical procedure, we shall
compare our numerical results with the analytical results available in the 
cases of power law and slow roll inflation as well as in the case of the 
Starobinsky model~\cite{starobinsky-1992,martin-2012a,arroja-2011-2012}.
(We should clarify here that the Starobinsky model that we are interested in is
the one which involves a linear inflaton potential with a sharp change in
its slope~\cite{starobinsky-1992}.
This clarification seems necessary at this stage, since another Starobinsky model, 
involving an extended theory of gravity~\cite{starobinsky-1980}, has drawn a lot 
of attention recently in the light of the Planck data~\cite{planck-2013-ci}.)
In Sec.~\ref{sec:mwf}, we shall use the validated numerical code to study the 
three-point functions that arise in three types of models which involve 
deviations from slow roll, viz. the punctuated inflationary 
scenario~\cite{pi}, potentials with a step~\cite{l2240,hazra-2010,benetti-2011} 
and the so-called axion monodromy 
model~\cite{flauger-2010,aich-2013,peiris-2013}. 
In Sec.~\ref{sec:c-d-prh}, we consider the contributions to the cross-correlations
and the tensor bi-spectrum during preheating and show that the contributions prove
to be completely insignificant. 
We conclude the paper in Sec.~\ref{sec:c} with a quick summary of the results
we have obtained. 
We relegate some of the details pertaining to the evaluation of the 
three-point functions in the Starobinsky model to the appendix.

\par

Let us now make a few remarks concerning our notations and conventions.
We shall work with units such that $\hbar=c=1$ and assume the Planck mass 
to be $\Mp=\l(8\,\pi\, G\r)^{-1/2}$.
We shall choose the metric signature to be $(-,+,+,+)$. 
Greek indices will be used to denote the spacetime coordinates, while Latin
indices will denote the three spatial coordinates (except for the index $k$ 
which would be reserved for representing the wavenumber of the perturbations). 
The quantities $a$ and $H$ shall denote the scale factor and the Hubble
parameter of the spatially flat, Friedmann-Lema\^{\i}tre-Robertson-Walker (or, 
simply, Friedmann, hereafter) universe.
At various stages, we shall use different notions of time in the Friedmann
universe, viz. the cosmic time $t$, the conformal time $\eta$, or the number 
of e-folds $N$, as is convenient. 
An overdot and an overprime shall denote differentiation with respect to the 
cosmic and the conformal time coordinates, respectively.


\section{The cubic order actions and the three-point correlation
functions}\label{subsec:action3}

In this section, we shall briefly summarize the essential aspects of 
the Maldacena formalism to arrive at the three-point scalar-tensor 
cross-correlations and the tensor bi-spectrum.
We shall also introduce the corresponding non-Gaussianity parameters, 
which are basically dimensionless ratios of the three-point functions 
and the scalar and the tensor power spectra.


\subsection{The actions at the cubic order}

The primary aim of the Maldacena formalism is to obtain the cubic order 
action that governs the scalar and the tensor perturbations.
The action that describes the perturbations is arrived at by using the 
conventional Arnowitt-Deser-Misner (ADM) formalism~\cite{arnowitt-1960}. 
Then, based on the action, one arrives at the corresponding three-point 
functions using the standard rules of perturbative quantum field 
theory~\cite{maldacena-2003,ng-ncsf,ng-reviews}.

\par

Recall that, in the ADM formalism, the spacetime metric is expressed in 
terms of the lapse function~$N$, the shift vector $N^{i}$ and the spatial 
metric $h_{ij}$ as follows:
\begin{equation} 
ds^{2}=-N^{2}\,  \l(\d x^{0}\r)^{2}
+h_{ij}\,\l(N^{i}\, \d x^{0}+\d x^{i}\r)\, 
\l(N^{j}\, \d x^{0}+\d x^{j}\r),\label{eq:adm-m}
\end{equation}
where $x^0$ and $x^{i}$ denote the time and the spatial coordinates,
respectively. 
The system of our interest is Einsteinian gravity which is sourced by a
canonical and minimally coupled scalar field, viz. the inflaton $\phi$, 
that is described by the potential $V(\phi)$.
For such a case, the action describing the complete system can be written 
in terms of the metric variables $N$, $N^i$ and $h_{ij}$ and the scalar
field $\phi$ as follows~\cite{maldacena-2003,ng-ncsf,ng-reviews,martin-2012a}:
\bea
S[N,N^{i},h_{ij},\phi] 
&=& \int \d x^{0}\,\int \d^{3} {\bm x}\,N\,\sqrt{h}\;
\Biggl\{\frac{\Mp^{2}}{2} \l[\f{1}{N^2}\,\l(E_{ij}E^{ij} -E^2\r) 
+ ^{(3)}\!\!R\r]\nn\\
& & + \biggl[\f{1}{2\, N^2}\, \l(\pa_0\phi\r)^2 
-\f{N^i}{N^2}\; \pa_0\phi\; \pa_i\phi 
- \f{1}{2}\; h^{ij}\; \pa_i\phi\; \pa_j\phi\nn\\
& & +\, \f{N^iN^j}{2\,N^2}\, \pa_i\phi\, \pa_j\phi - V(\phi)\biggr]\biggr\},
\label{eq:adm-a}
\eea
where $\pa_0\phi=(\pa\phi/\pa x^0)$, $h \equiv {\rm det}~(h_{ij})$ and 
$^{(3)}\!R$ is the spatial curvature associated with the metric $h_{ij}$. 
The quantity $ E_{ij}$ is given by
\begin{equation}
E_{ij} = \f{1}{2}\, \l(\pa_0 h_{ij} - \nabla_iN_j - \nabla_j N_i\r),
\end{equation}
with $E = h_{ij}\,E^{ij}$.
As is well known, the variation of the action~(\ref{eq:adm-a}) with respect
to the Lagrange multipliers $N$ and $N^{i}$ leads to the so-called Hamiltonian 
and momentum constraints, respectively.
Solving these constraint equations and substituting the solutions back in 
the original action~(\ref{eq:adm-a}) permits one to arrive at the action 
governing the dynamical variables of interest.

\par

As we have mentioned, our aim is to evaluate the action describing the scalar 
and the tensor perturbations in the spatially flat, Friedmann universe.
In the absence of the perturbations, the Friedmann universe is described by 
the line-element
\begin{equation} 
\d s^{2}=-\d t^{2}+a^{2}(t)\, \d \bm{x}^{2},\label{eq:frw-le}
\end{equation}
where $\bm{x}$ represents the spatial coordinates. 
When the scalar and the tensor perturbations to the Friedmann metric are taken 
into account, it proves to be convenient to work in a specific gauge to arrive 
at the action governing the perturbations. 
As is often done in this context, we shall choose to work in the so-called 
co-moving gauge~\cite{maldacena-2003}.
In such a gauge, the perturbation in the scalar field is assumed to be absent, 
so that the quantity $\phi$ that appears in the action~(\ref{eq:adm-a}) 
actually depends only on time.
Moreover, the three metric $h_{ij}$ is written as
\begin{equation}
h_{ij}=a^{2}(t)\; {\rm e}^{2\,{\cal R}(t,\bm{x})}\,
\l[{\rm e}^{\gamma(t,\bm{x})}\r]_{ij},
\end{equation}
where $\cR$ denotes the curvature perturbation describing the scalars,
while $\gamma_{ij}$ represents the transverse and traceless 
(i.e. $\partial_{j}\gamma_{ij}=\gamma_{ii}=0$) tensor perturbations.
These assumptions for the scalar field $\phi$ and the spatial metric $h_{ij}$ 
as well as the solutions to the constraint equations then allows one to 
arrive at the action describing the perturbations, viz. the quantities $\cR$ 
and $\gamma_{ij}$, at a given order~\cite{maldacena-2003,ng-ncsf}.
The action at the quadratic order leads to the linear equations of motion 
governing the perturbations.
In Fourier space, the scalar and the tensor modes satisfy the differential
equations
\begin{subequations}\label{eq:de-p}
\begin{eqnarray}
\cR_\vk''+2\, \f{z'}{z}\, \cR_\vk' + k^2\, \cR_\vk 
&=& 0,\label{eq:de-fk}\\
\gamma_\vk''+2\, \f{a'}{a}\, \gamma_\vk' + k^2\, \gamma_\vk 
&=& 0,\label{eq:de-gk}
\end{eqnarray}
\end{subequations}
respectively, where $z=\sqrt{2\,\epsilon_1}\,a\,\Mp$, with $\epsilon_1 =
-{\dot H}/H^2$ being the first slow roll parameter.
We should add that, when no confusion can arise, here and hereafter, we 
shall suppress the indices of the tensor perturbations for convenience.

\par 

In the co-moving gauge, at the cubic order in the perturbations, evidently,
the action will consist of terms of the form ${\cal RRR}$, ${\cal RR}\gamma$, 
${\cal R}\gamma\gamma$ and $\gagaga$.
The term involving ${\cal RRR}$ leads to the scalar bi-spectrum.  
Since we are interested only in the scalar-tensor cross-correlations and 
the tensor bi-spectrum, we shall focus on the actions that lead to these 
quantities. 
The actions that lead to correlations involving two scalars and one tensor, 
one scalar and two tensors and three tensors are given by (see, for example, 
Refs.~\cite{maldacena-2003,tensor-bs,cc})
\begin{subequations}
\bea
S^3_{{\cal RR}\gamma}[\cR,\gamma_{ij}]
&=& \Mp^2\,\int\d \eta\, \int \d^3 \bm{x}\;
\biggl[a^2\, \epsilon_1\,\gamma_{ij}\,\pa_i{\cal R}\,\pa_j{\cal R}
+ \f{1}{4}\,\pa^2\gamma_{ij}\,\partial_i\chi\,\partial_j\chi\nn\\
& & +\, \f{a\,\epsilon_1}{2}\,\gamma_{ij}'\,\partial_i{\cal R}\,\partial_j\chi 
+ {\mathcal F}^2_{ij}({\cal R})\,
\f{\delta{\cal L}^2_{\g\g}}{\delta\gamma_{ij}}
+ {\mathcal F}^3(\cR,\gamma_{ij})\,
\f{\delta{\cal L}^2_{\cR\cR}}{\delta{\cal R}} \biggr],\label{eq:a-sst}\\
S^3_{{\cal R}\gamma\gamma}[\cR,\gamma_{ij}]
&=& \f{\Mp^2}{4}\,\int\d \eta\,\int \d^3 \bm{x}\;
\biggl[\f{a^2\,\epsilon_1}{2}\, {\cal R}\,\gamma_{ij}'\,\gamma_{ij}' 
+ \f{a^2\, \epsilon_1}{2}\,{\cal R}\,\pa_l\gamma_{ij}\,\pa_l\gamma_{ij}\nn\\
& &-\,a\,\gamma_{ij}'\,\pa_l\gamma_{ij}\,\pa_l\chi 
+ {\mathcal F}^4_{ij}(\cR,\gamma_{mn})\,
\f{\delta {\cal L}^2_{\g\g}}{\delta\gamma_{ij}}\biggr],\label{eq:a-stt}\\
S^3_{\gagaga}[\gamma_{ij}] 
&=& \frac{\Mp^{2}}{2}\,\int\d\eta\,\int \d^{3}\bm{x}\; 
\biggl[\frac{a^2}{2}\,\gamma_{lj}\,\gamma_{im}\,\pa_l\pa_m\gamma_{ij}
-\,\frac{a^2}{4}\,\gamma_{ij}\,\gamma_{lm}\,
\pa_l\pa_m\gamma_{ij}\biggr].\label{eq:a-ttt}
\eea
\end{subequations}
In these actions, the quantity $\chi$ is determined by the relation
$\partial^{2}\chi=a\,\epsilon_1\,{\cal R}'$, and the quantities
$\mathcal{L}^2_{\sr\sr}$ and $\mathcal{L}^2_{\g\g}$ are the 
second order Lagrangian densities comprising of two scalars and tensors 
which lead to the equations of motion~(\ref{eq:de-p}).
One can show that the terms proportional to $(\delta{\cal L}^2_{\sr\sr}
/\delta{\cal R})$ and $(\delta {\cal L}^2_{\g\g}/\delta\gamma_{ij})$ can 
be removed by suitable field redefinitions (for further details, including 
the explicit forms of the functions ${\mathcal F}^2_{ij}(\cR)$, 
${\mathcal F}^3(\cR,\gamma_{ij})$ and  ${\mathcal F}^4_{ij}(\cR,\gamma_{mn})$, 
see Refs.~\cite{maldacena-2003,ng-ncsf,ng-reviews,tensor-bs,cc}).

\par

In order to calculate the three-point correlation functions using the 
methods of quantum field theory, one requires the interaction Hamiltonian
corresponding to the above actions. 
We note that, at the cubic order, the interaction Hamiltonian $H_{\rm int}$ 
is related to the interaction Lagrangian $L_{\rm int}$ through the relation: 
$H_{\rm int}=-L_{\rm int}$~\cite{maldacena-2003,ng-ncsf,ng-reviews,tensor-bs,cc}. 
In what follows, we shall refer to $H_{\rm int}$ corresponding to the various 
actions as $H_{\sf ABC}$, where each of $({\sf A},{\sf B},{\sf C})$ can be 
either a $\cR$ or a~$\gamma$. 
In the next sub-section, we shall make use of the above actions (actually, the
corresponding interaction Hamiltonians) to compute the three-point functions
of our interest.


\subsection{The three-point functions of interest and the different 
contributions}\label{subsec:dbs-dc}

Given the interaction Hamiltonian, the corresponding three-point function 
can be evaluated using the standard rules of perturbative quantum field 
theory.
The three-point functions can be expressed in terms of products of operators 
which satisfy the linear equations of motion.
In the cases of our interest, the quantum operators associated with the
classical variables, viz. the curvature perturbation $\cR$ and the tensor 
perturbation $\gamma_{ij}$, can be decomposed in terms of the corresponding
Fourier modes as follows~\cite{texts,reviews}:
\begin{subequations}\label{eq:st-m-dc}
\bea 
\hat{\cR}(\eta, {\bf x}) 
&=& \int \frac{\d^{3}{\bm k}}{\l(2\,\pi\r)^{3/2}}\,
\hat{\cR}_{\bm k}(\eta)\, {\rm e}^{i\,{\bm k}\cdot{\bm x}}\nn\\
&=& \int \frac{\d^{3}{\bm k}}{\l(2\,\pi\r)^{3/2}}\,
\l(\hat{a}_{\bm k}\,f_{k}(\eta)\, 
{\rm e}^{i\,{\bm k}\cdot{\bm x}}
+\hat{a}^{\dagger}_{\bf k}\,f^{*}_{k}(\eta)\,
{\rm e}^{-i\,{\bm k}\cdot{\bm x}}\r),\label{eq:s-m-dc}\\
\hat{\gamma}_{ij}(\eta, {\bf x}) 
&=& \int \frac{\d^{3}{\bm k}}{\l(2\,\pi\r)^{3/2}}\,
\hat{\gamma}_{ij}^{\bm k}(\eta)\, {\rm e}^{i\,{\bm k}\cdot{\bm x}}\nn\\
&=& \sum_{s}\int \frac{\d^{3}{\bm k}}{(2\,\pi)^{3/2}}\,
\l(\hat{b}^{s}_{\bm k}\, \varepsilon^{s}_{ij}({\bm k})\,
g_{k}(\eta)\, {\rm e}^{i\,{\bm k}\cdot{\bm x}}
+\hat{b}^{s\dagger}_{\bf k}\,\varepsilon^{s*}_{ij}({\bm k})\, g^{*}_{k}(\eta)\,
{\rm e}^{-i\,{\bm k}\cdot{\bm x}}\r),\label{eq:t-m-dc}
\eea
\end{subequations}
where the scalar and tensor modes $f_k$ and $g_k$ satisfy the equations of 
motion~(\ref{eq:de-p}).
The quantity $\varepsilon^{s}_{ij}({\bm k})$ represents the polarization tensor
of the gravitational waves, with the index~$s$ denoting the helicity of the 
graviton.
In the gauge we are working in, the polarization tensor obeys the relations 
$\varepsilon^{s}_{ii}({\bm k})=k_{i}\,\varepsilon_{ij}^s({\bm k})=0$, and we 
choose to work with the normalization: $\varepsilon_{ij}^{r}({\bm k})\,
\varepsilon_{ij}^{s}({\bm k})=2\,\delta^{rs}$~\cite{maldacena-2003}.
In the above equations, the pairs of operators $(\hat{a}_{\bm k},
\hat{a}^{\dagger}_{\bm k})$ and $(\hat{b}_{\bm k}^{s},
\hat{b}^{s\dagger}_{\bm k})$ denote the annihilation and creation operators 
corresponding to the scalar and the tensor modes associated with the 
wavevector ${\bm k}$.
They obey the following non-trivial commutation relations: $[\hat{a}_{\bm k},
\hat{a}^{\dagger}_{\bm k'}]=\delta^{(3)}({\bm k}-{\bm k}')$ and
$[\hat{b}^{r}_{\bm k},\hat{b}_{\bm k'}^{s\dagger}]
=\delta^{(3)}({\bm k}-{\bm k}')\, \delta^{rs}$. 

\par

It should be mentioned here that the scalar and the tensor power spectra, 
viz. ${\mathcal P}_{_{\rm S}}(k)$ and ${\mathcal P}_{_{\rm T}}(k)$, are 
defined as follows:
\begin{subequations}
\begin{eqnarray}
\langle 0 \vert {\hat \cR}_{\bm k}(\eta)\,
{\hat \cR}_{\bm k'}(\eta)\vert 0\rangle
&=&\f{(2\,\pi)^2}{2\, k^3}\, {\mathcal P}_{_{\rm S}}(k)\;
\delta^{(3)}({\bm k}+{\bm k'}),\label{eq:sps-d}\\
\langle 0 \vert {\hat \gamma}_{ij}^{\bm k}(\eta)\,
{\hat \gamma}_{ij}^{\bm k'}(\eta)\vert 0\rangle
&=&\f{(2\,\pi)^2}{2\, k^3}\, {\mathcal P}_{_{\rm T}}(k)\;
\delta^{(3)}({\bm k}+{\bm k'}),\label{eq:tps-d}
\end{eqnarray}
\end{subequations}
where the vacuum state $\vert 0\rangle$ is defined as ${\hat a}_{\bm k}
\vert 0\rangle=0$ and ${\hat b}_{\bm k}^{s}\vert 0\rangle=0$ for all 
${\bm k}$ and $s$.
Upon using the decompositions~(\ref{eq:st-m-dc}), the power spectra can 
be expressed in terms of the modes $f_k$ and $g_k$ as 
\begin{subequations}
\begin{eqnarray}
{\mathcal P}_{_{\rm S}}(k)
&=&\f{k^3}{2\, \pi^2}\, \vert f_k\vert^2,\label{eq:sps}\\
{\mathcal P}_{_{\rm T}}(k)
&=&4\,\f{k^3}{2\, \pi^2}\, \vert g_k\vert^2,\label{eq:tps}
\end{eqnarray}
\end{subequations}
with the right hand sides evaluated at late times when the amplitude of the
modes have frozen when they are well outside the Hubble radius during the 
inflationary epoch.


\subsubsection{The scalar-tensor cross-correlations}

The two scalar-tensor cross-correlations in Fourier space, viz. 
$\cB_{\cR\cR\gamma}^{m_3n_3}(\vka,\vkb,\vkc)$ which involves two scalars and 
a tensor and $\cB_{\cR\gamma\gamma}^{m_2n_2m_3 n_3}(\vka,\vkb,\vkc)$ 
which involves one scalar and two tensors, evaluated, say, towards the 
end of inflation, at the conformal time $\ee$, are defined as
\begin{eqnarray}
\langle {\hat \cR}_{\vka}(\eta _{\rm e})\, {\hat \cR}_{\vkb}(\eta _{\rm e})\,
{\hat \gamma}_{m_3n_3}^{\vkc}(\eta _{\rm e})\, \rangle 
&=&\l(2\,\pi\r)^3 \cB_{\sr\sr\g}^{m_3n_3}(\vka,\vkb,\vkc)\;
\delta^{(3)}\l(\vka+\vkb+\vkc\r),\label{eq:sst-cc}\\
\langle {\hat \cR}_{\vka}(\eta _{\rm e})\, 
{\hat \gamma}_{m_2n_2}^{\vkb}(\eta _{\rm e})\, 
{\hat \gamma}_{m_3n_3}^{\vkc}(\eta _{\rm e})\rangle 
&=&\l(2\,\pi\r)^3 \cB_{\sr\g\g}^{m_2 n_2 m_3 n_3}(\vka,\vkb,\vkc)\;
\delta^{(3)}\l(\vka+\vkb+\vkc\r).\quad\label{eq:tts-cc}
\end{eqnarray}
(As we have mentioned earlier, occasionally, for ease of notation, 
we may drop the tensor indices such as $mn$ when they do not lead to 
ambiguities.)
At the leading order in perturbation theory, these quantities can be
expressed in terms of the corresponding interaction Hamiltonians 
${\hat H}_{\cR\cR\g}$ and ${\hat H}_{\cR\g\g}$ [obtained from the
actions~(\ref{eq:a-sst}) and~(\ref{eq:a-stt})] as
follows~\cite{maldacena-2003}:
\bea 
\langle {\hat \cR}_{\vka}(\ee)\, {\hat \cR}_{\vkb}(\ee)\,
{\hat \gamma}_{m_3n_3}^{\vkc}(\ee)\rangle 
&=& -i\,\int_{\ei}^{\ee}\d\eta\; 
\langle[{\hat \cR}_{\vka}(\ee)\, {\hat \cR}_{\vkb}(\ee)\,
{\hat \gamma}_{m_3n_3}^{\vkc}(\ee),\hat{H}_{\sr\sr\g}(\eta)]\rangle,\qquad\\
\langle {\hat \cR}_{\vka}(\ee)\, {\hat \g}_{m_2n_2}^{\vkb}(\ee)\,
{\hat \gamma}_{m_3n_3}^{\vkc}(\ee)\rangle 
&=& -i\,\int_{\ei}^{\ee}\d\eta\; 
\langle[ {\hat \cR}_{\vka}(\ee)\, {\hat \g}_{m_2n_2}^{\vkb}(\ee)\,
{\hat \gamma}_{m_3n_3}^{\vkc}(\ee),\hat{H}_{\sr\g\g}(\eta)]\rangle,\quad\nn\\
\eea
where $\eta_{i}$ is the time when the initial conditions are imposed (when 
the largest mode of interest is sufficiently inside the Hubble radius during 
inflation), the square brackets imply the commutation of the operators, while 
the angular brackets denote the fact that the expectation values are to be 
evaluated in the initial vacuum state. 
For convenience, we shall set 
\begin{equation}
\cB_{\sf ABC}(\vka,\vkb,\vkc)
= \l(2\,\pi\r)^{-9/2}\, G_{\sf ABC}(\vka,\vkb,\vkc).
\end{equation}

\par

Upon using the above expressions, the mode decompositions~(\ref{eq:st-m-dc}) 
and Wick's theorem, which applies to Gaussian random fields, we obtain that
\bea \label{Grrg}
G_{\sr\sr\gamma}^{m_3n_3}(\vka,\vkb,\vkc)
& = & \sum_{C=1}^{3}\; G_{\sr\sr\gamma\,(C)}^{m_3n_3}(\vka,\vkb,\vkc)\nn\\
& = & \Mp^2\; \sum_{C=1}^{3}\, \sum_{s_3}\
\biggl\{\l[\varepsilon_{m_3n_3}^{s_3}(\vkc)\, f_{\ska}(\ee)\, f_{\skb}(\ee)\,
g_{\skc}(\ee)\r]\nn\\
& & \times\;\cG_{\sr\sr\gamma}^{C}(\vka,\vkb,\vkc) + {\rm complex
~conjugate}\biggr\},
\end{eqnarray}
where the quantities $\cG_{\sr\sr\gamma}^{C}(\vka,\vkb,\vkc)$ are described 
by the integrals
\begin{subequations}\label{eq:cGrrg}
\begin{eqnarray}
\cG_{\sr\sr\gamma}^{1}(\vka,\vkb,\vkc)
&=&-2\; i\; \varepsilon_{ij}^{s_3\ast}(\vkc)\; k_{1i}\,k_{2j}\,
\int_{\ei}^{\ee} \d\eta\; a^2\, \epsilon_1\, f_{\ska}^{\ast}\,
f_{\skb}^{\ast}\,g_{\skc}^{\ast},\label{eq:cGrrg1}\\
\cG_{\sr\sr\gamma}^{2}(\vka,\vkb,\vkc)
&=&\f{i}{2}\;\varepsilon_{ij}^{s_3\ast}(\vkc)\; 
\f{k_{1i}\, k_{2j}\,k_3^2}{k_1^2\,k_2^2}\,
\int_{\ei}^{\ee} \d\eta\; a^2\, \epsilon_1^2\, f_{\ska}^{\prime\ast}\,
f_{\skb}^{\prime\ast}\,g_{\skc}^{\ast},\label{eq:cGrrg2}\\
\cG_{\sr\sr\gamma}^{3}(\vka,\vkb,\vkc)
&=&\f{i}{2}\, \varepsilon_{ij}^{s_3\ast}(\vkc)\, 
\f{k_{1i}\,k_{2j}}{k_1^2\,k_2^2}\,
\int_{\ei}^{\ee} \d\eta\; a^2\, \epsilon_1^2\, 
\l[k_1^2\,f_{\ska}^{\ast}\,f_{\skb}^{\prime\ast}
+k_2^2\,f_{\ska}^{\prime\ast}\,f_{\skb}^{\ast}\r]\,
g_{\skc}^{\prime\ast},\quad\label{eq:cGrrg3}
\end{eqnarray}
\end{subequations}
which, evidently, correspond to the three different terms that constitute
the action~$S_{\cR\cR\g}$ [cf.~Eq.~(\ref{eq:a-sst})].
It is worth mentioning here that, while the first term is of the order of
the first slow roll parameter $\epsilon_1$, the remaining two terms are of 
the order $\epsilon_1^2$~\cite{maldacena-2003}.
In exactly the same way, we can obtain that
\bea \label{Grgg}
G_{\sr\gamma\gamma}^{m_2n_2m_3n_3}(\vka,\vkb,\vkc)
&= & \sum_{C=1}^{3}\;
G_{\sr\gamma\gamma\,(C)}^{m_2n_2m_3n_3}(\vka,\vkb,\vkc)\nn\\
&= & \Mp^2\; \sum_{C=1}^{3}\; \sum_{s_2,s_3}\,
\Biggl\{\l[\vare_{m_{2}n_{2}}^{s_2}(\vkb)\,
\vare_{m_{3}n_{3}}^{s_3}(\vkc)\, f_{\ska}(\ee)\,
g_{\skb}(\ee)\,g_{\skc}(\ee)\r]\nn\\
& &\times\; \cG_{\sr\gamma\gamma}^{C}(\vka,\vkb,\vkc) 
+ {\rm complex~conjugate}\Biggr\},
\end{eqnarray}
with the quantities $\cG_{\sr\g\g}^{C}(\vka,\vkb,\vkc)$ being given by 
\begin{subequations}\label{eq:cGrgg}
\begin{eqnarray}
\cG_{\sr\g\g}^{1}(\vka,\vkb,\vkc)
&=&\f{i}{4}\; \varepsilon_{ij}^{s_2\ast}(\vkb)\;
\varepsilon_{ij}^{s_3\ast}(\vkc)\;
\int_{\ei}^{\ee} \d\eta\; a^2\, \epsilon_1\, f_{\ska}^{\ast}\,
g_{\skb}^{\prime\ast}\,g_{\skc}^{\prime\ast},\label{eq:cGrgg1}\\
\cG_{\sr\g\g}^{2}(\vka,\vkb,\vkc)
&=&-\f{i}{4}\; \varepsilon_{ij}^{s_2\ast}(\vkb)\;
\varepsilon_{ij}^{s_3\ast}(\vkc)\; \l(\vkb\cdot\vkc\r)\,
\int_{\ei}^{\ee} \d\eta\; a^2\, \epsilon_1\, f_{\ska}^{\ast}\,
g_{\skb}^{\ast}\,g_{\skc}^{\ast},\label{eq:cGrgg2}\\
\cG_{\sr\g\g}^{3}(\vka,\vkb,\vkc)
&=&-\f{i}{4}\; \varepsilon_{ij}^{s_2\ast}(\vkb)\;
\varepsilon_{ij}^{s_3\ast}(\vkc)\; 
\int_{\ei}^{\ee} \d\eta\; a^2\, \epsilon_1\, f_{\ska}^{\prime\ast}\,
\biggl[\f{\vka\cdot\vkc}{k_1^2}\,\,g_{\skb}^{\prime\ast}\,
g_{\skc}^{\ast}\nn\\
& &+\,\f{\vka\cdot\vkb}{k_1^2}\,g_{\skb}^{\ast}\,
g_{\skc}^{\prime\ast}\biggr],\label{eq:cGrgg3}
\end{eqnarray}
\end{subequations}
which, again, correspond to the three different terms in the action 
$S_{\sr\gamma\gamma}$ [cf.~Eq.~(\ref{eq:a-stt})].
Note that, in this case, all the terms are of the same order in the
first slow roll parameter $\epsilon_1$.
The hierarchy of the various contributions to the cross-correlations
will also be evident when we discuss specific analytic and numerical
results in the following section.


\subsubsection{The tensor bi-spectrum}

The tensor bi-spectrum  $\cB_{\g\g\g}^{m_1n_1m_2 n_2 m_3 n_3}(\vka,\vkb,\vkc)$
is defined as
\begin{equation}
\langle {\hat \gamma}_{m_1n_1}^{\vka}(\eta _{\rm e})\,
{\hat \gamma}_{m_2n_2}^{\vkb}(\eta _{\rm e})\, 
{\hat \gamma}_{m_3n_3}^{\vkc}(\eta _{\rm e})\rangle 
=\l(2\,\pi\r)^3 \cB_{\g\g\g}^{m_1n_1m_2 n_2 m_3 n_3}(\vka,\vkb,\vkc)\;
\delta^{(3)}\l(\vka+\vkb+\vkc\r)
\end{equation}
and, clearly, in terms of the Hamiltonian ${\hat H}_{\g\g\g}$, it can be 
expressed as 
\begin{equation}
\langle {\hat \g}_{m_1n_1}^{\vka}(\ee)\, {\hat \g}_{m_2n_2}^{\vkb}(\ee)\,
{\hat \gamma}_{m_3n_3}^{\vkc}(\ee)\rangle 
= -i\,\int_{\ei}^{\ee}\d\eta\; 
\langle[ {\hat \g}_{m_1n_1}^{\vka}(\ee)\, {\hat \g}_{m_2n_2}^{\vkb}(\ee)\,
{\hat \gamma}_{m_3n_3}^{\vkc}(\ee),\hat{H}_{\g\g\g}(\eta)]\rangle.
\end{equation}
The corresponding quantity $G_{\g\g\g}^{m_1n_1m_2 n_2 m_3 n_3}(\vka,\vkb,\vkc)$ 
can be arrived in the same fashion as the cross-correlations from the action 
$S_{\g\g\g}$ [cf.~Eq.~(\ref{eq:a-ttt})].
It can be written as~\cite{maldacena-2003,tensor-bs}
\bea \label{Gggg}
G_{\gamma\gamma\gamma}^{m_1n_1m_2n_2m_3n_3}(\vka,\vkb,\vkc)
&= & \sum_{C=1}^{2}\; 
G_{\gamma\gamma\gamma\,(C)}(\vka,\vkb,\vkc)\nn\\
&= & \Mp^2\; \sum_{C=1}^{2}\; \sum_{s_1,s_2,s_3}
\Biggl\{\biggl[\vare_{m_{1}n_{1}}^{s_1}(\vka)\;
\vare_{m_2n_2}^{s_2}(\vkb)\;
\vare_{m_3n_3}^{s_3}(\vkc)\nn\\
& &\times\; g_{\ska}(\ee)\, g_{\skb}(\ee)\,g_{\skc}(\ee)\biggr]\;
\cG_{\gamma\gamma\gamma}^{C}(\vka,\vkb,\vkc)\nn\\
& &+\, {\rm complex~conjugate}\Biggr\},
\end{eqnarray}
where the quantities $\cG_{\gamma\gamma\gamma}^{C}(\vka,\vkb,\vkc)$ are
described by the integrals
\begin{subequations}
\label{eq:cGggg} 
\begin{eqnarray}
\cG_{\g\g\g}^{1}(\vka,\vkb,\vkc)
&=&-\f{i}{4}\; \l[\varepsilon_{ij}^{s_1\ast}(\vka)\;
\varepsilon_{im}^{s_2\ast}(\vkb)\;
\varepsilon_{lj}^{s_3\ast}(\vkc)\; k_{1m}\,k_{1l}
+{\rm five~permutations}\r]\nn\\
& &\times\,\int_{\ei}^{\ee} \d\eta\; a^2\, g_{\ska}^{\ast}\,
g_{\skb}^{\ast}\,g_{\skc}^{\ast},\label{eq:cGggg1}\\
\cG_{\g\g\g}^{2}(\vka,\vkb,\vkc)
&=&\f{i}{8}\; \l[\varepsilon_{ij}^{s_1\ast}(\vka)\;
\varepsilon_{ml}^{s_2\ast}(\vkb)\;
\varepsilon_{ij}^{s_3\ast}(\vkc)\; k_{1m}\,k_{1l}
+{\rm five~permutations}\r]\nn\\
& &\times\,\int_{\ei}^{\ee} \d\eta\; a^2\, g_{\ska}^{\ast}\,
g_{\skb}^{\ast}\,g_{\skc}^{\ast}.\label{eq:cGggg2}
\end{eqnarray}
\end{subequations}
It is evident that the two contributions to the tensor bi-spectrum are of the 
same order in magnitude.
Again, this will be corroborated by explicit analytical and numerical calculations 
in due course.


\subsection{The non-Gaussianity parameters}

Recall that, often, the scalar bi-spectrum is essentially characterized 
by the dimensionless non-Gaussianity parameters $\fnl$.
The basic set of three non-Gaussianity parameters, viz. $(\fnl^{\rm loc},
\fnl^{\rm eq}, \fnl^{\rm ortho})$, do not always capture the complete 
structure of the scalar bi-spectrum, in particular, when there exist 
deviations from the conventional scenario of slow roll inflation driven 
by the canonical scalar field (and, of course, the assumption that the 
perturbations are in the standard Bunch-Davies 
vacuum~\cite{texts,reviews,bunch-1978}).
Nonetheless, they prove to be a convenient tool in understanding the 
amplitude and shape of the scalar bi-spectrum in many situations. 

\par

In a similar manner, one can characterize the cross-correlations and the
tensor bi-spectrum by parameters that are suitable dimensionless ratios 
of the three-point functions and the scalar or the tensor power spectra.
One can generalize the conventional way of introducing the $\fnl$ parameter
to write the scalar and tensor perturbations $\cR$ and $\gamma_{ij}$ as
follows:
\begin{equation}\label{eq:cnlr}
\cR(\eta, {\bm x}) 
= \cR^{\rm G}(\eta, {\bm x}) 
- \f{3\,\fnl}{5}\;[{\cal R}^{\rm G}(\eta, {\bm x})]^2 
- \cnls\; \cR^{\rm G}(\eta,{\bm x})\; 
\gamma_{{\bar m}{\bar n}}^{\rm G}(\eta, {\bm x})
\end{equation}
and
\begin{equation}\label{eq:cnlg}
\gamma_{ij}(\eta, {\bm x}) 
= \gamma_{ij}^{\rm G}(\eta, {\bm x}) 
- \hnl\; \gamma_{ij}^{\rm G}(\eta,{\bm x})\; 
\gamma_{{\bar m}{\bar n}}^{\rm G}(\eta,{\bm x})
- \cnlt\; \gamma_{ij}^{\rm G}(\eta,{\bm x})\;
\cR^{\rm G}(\eta,{\bm x}),
\end{equation}
where $\cR^{\rm G}$ and $\gamma_{ij}^{\rm G}$ denote the Gaussian 
quantities.
Note that the overbars on the indices of the Gaussian tensor perturbation 
imply that the indices should be summed over all allowed values\footnote{It
should be apparent that such a procedure is required to `remove' the additional 
polarization indices that would otherwise occur when the parameters $\cnls$ and 
$\cnlt$ are introduced in the above fashion.
Also, clearly, this procedure is not unique, and there exist other ways of 
`removing' the additional indices.}. 
Upon using the above definitions along with the Wick's theorem to 
calculate the three-point functions (but, retaining terms only to 
the linear order in the non-Gaussianity parameters), we find that 
we can write the parameters $\cnls$, $\cnlt$ and $\hnl$ as follows:
\begin{eqnarray} 
\cnls 
&=& -\f{4}{\l(2\,\pi^2\r)^2}\,
\l[k_1^3\, k_2^3\, k_3^3\; G_{\cR\cR\g}^{m_3n_3}(\vka,\vkb,\vkc)\r]\nn\\
& &\times\; {\l(\Pi_{m_3n_3,{\bar m}{\bar n}}^{\vkc}\r)}^{-1}\;\,
\biggl\{\l[k_1^3\; {\mathcal P}_{_{\rm S}}(k_2)
+k_2^3\; {\mathcal P}_{_{\rm S}}(k_1)\r]\; 
{\mathcal P}_{_{\rm T}}(k_3)\biggr\}^{-1},
\label{eq:cnls}\\
\cnlt 
&=& -\f{4}{\l(2\,\pi^2\r)^2}\,
\l[k_1^3\, k_2^3\, k_3^3\; G_{\cR\g\g}^{m_2n_2m_3n_3}(\vka,\vkb,\vkc)\r]\nn\\
& &\times\,
\biggl\{{\mathcal P}_{_{\rm S}}(k_1)\;
\l[\Pi_{m_2n_2,m_3n_3}^{\vkb}\;k_3^3\; {\mathcal P}_{_{\rm T}}(k_2)
+\Pi_{m_3n_3,m_2n_2}^{\vkc}\; k_2^3\;
{\mathcal P}_{_{\rm T}}(k_3)\r]\biggr\}^{-1},
\label{eq:cnlt}
\end{eqnarray}
and
\begin{eqnarray}
\hnl 
&=&-\l(\f{4}{2\,\pi^2}\r)^2\,
\l[k_1^3\, k_2^3\, k_3^3\; 
G_{\g\g\g}^{m_1n_1m_2n_2m_3n_3}(\vka,\vkb,\vkc)\r]\nn\\
& &\times\; 
\l[\Pi_{m_1n_1,m_2n_2}^{\vka}\,\Pi_{m_3n_3,{\bar m}{\bar n}}^{\vkb}\;
k_3^3\; {\mathcal P}_{_{\rm T}}(k_1)\;{\mathcal P}_{_{\rm T}}(k_2)
+{\rm five~permutations}\r]^{-1},\label{eq:hnl} 
\end{eqnarray}
where the quantity $\Pi_{m_1n_1,m_2n_2}^{\vk}$ is defined 
as~\cite{tensor-bs}
\begin{equation}
\Pi_{m_1n_1,m_2n_2}^{\vk}
=\sum_{s}\;\varepsilon_{m_1n_1}^{s}(\vk)\;
\varepsilon_{m_2n_2}^{s\ast}(\vk).
\end{equation}
While we notice that a parameter such as $\hnl$ to characterize the tensor 
bi-spectrum has been discussed earlier (see, for instance, 
Refs.~\cite{tensor-bs,cc}), to our knowledge, the non-Gaussianity
parameters $\cnls$ and $\cnlt$ describing the cross-correlations do 
not seem to have been considered earlier in the literature.
In retrospect though, the introduction and the utility of these parameters 
in helping to characterize and, eventually, constrain inflationary models
seem evident.


\section{The numerical procedure for evaluating the three-point 
functions}\label{sec:nm}

For a general inflationary model, it proves to be  difficult to analytically 
calculate the scalar-tensor cross-correlations and the tensor bi-spectrum. 
It is therefore useful to develop a numerical approach to evaluate these 
three-point correlations.
It is evident from the discussion in the previous section that the 
three-point functions involve integrals over the background quantities 
as well as the scalar and the tensor modes from the early stages of 
inflation till its very end.
Recently, in the context of the scalar bi-spectrum, it was shown that the 
corresponding super-Hubble contributions prove to be negligible and it 
suffices to carry out the integrals numerically over a suitably smaller 
domain in time~\cite{hazra-2013}. 
We find that similar arguments apply for the other three-point functions too.   
In this section, we shall first show that the super-Hubble contributions to
the three-point functions of our interest here are indeed insignificant and, 
then, based on this result, go on to construct a numerical method to evaluate 
the correlation functions.
We shall also illustrate the accuracy of our numerical procedure by comparing
them with the analytical results that can be obtained in the case of power law 
inflation, the quadratic potential and the non-trivial scenario involving 
departures from slow roll that occurs in the so-called Starobinsky 
model~\cite{starobinsky-1992,martin-2012a,arroja-2011-2012}.


\subsection{Insignificance of the super-Hubble contributions}

To begin with, note that, if we write the scalar and the tensor modes as 
$f_k=v_k/z$ and $g_k={\mathcal U}_k/a$, then the quantities $v_k$ and 
${\mathcal U}_k$ satisfy
the differential equations
\begin{subequations}\label{eq:ms}
\begin{eqnarray}
v_k''+\l(k^2-\f{z''}{z}\r)\,v_k &=& 0,\label{eq:ms-f}\\ 
{\mathcal U}_k''+\l(k^2-\f{a''}{a}\r)\,
{\mathcal U}_k &=& 0,\label{eq:ms-g}  
\end{eqnarray}
\end{subequations}
respectively.
During slow roll inflation, one can show that $z''/z\simeq a''/a \simeq 
2\, {\mathcal H}^2$, where ${\mathcal H}=a\, H$ denotes the conformal Hubble
parameter~\cite{texts,reviews}.
On super-Hubble scales during inflation, i.e. when $k/{\mathcal H}\ll1$, we 
can ignore the $k^2$ term in the above equations in comparison to $z''/z$ and 
$a''/a$, thereby obtaining the following solutions for $f_k$ and $g_k$:
\begin{subequations}\label{eq:shs-fkgk}
\begin{eqnarray}
f_k(\eta) &=& A_k\, 
+ B_k\, \int^{\eta}\f{d{\tilde \eta}}{z^2({\tilde \eta})},
\label{eq:fk-shs}\\
g_k(\eta) &=& \f{\sqrt{2}}{\Mp}\, 
\l(C_k + D_k\, \int^{\eta} \f{\d{\tilde \eta}}{a^2({\tilde \eta})}\r),
\label{eq:gk-shs}
\end{eqnarray}
\end{subequations}
where $A_k$, $B_k$, $C_k$ and $D_k$ are $k$-dependent constants that are
determined by the initial conditions imposed on the modes at early times 
when they are well inside the Hubble radius.
Moreover, the overall factor of $\sqrt{2}/\Mp$ has been introduced in the
solution for $g_k$ by convention, so as to ensure that the resulting tensor 
power spectrum [cf.~Eq.~(\ref{eq:tps})] is dimensionless.
The first terms in the above expressions for $f_k$ and $g_k$ are the growing 
(actually, constant) solutions, while the second represent the decaying (i.e. 
the sub-dominant) ones.
Therefore, at late times, we have 
\begin{equation}
f_k \simeq A_k\quad{\rm and}\quad g_k \simeq \sqrt{2}\, C_k/\Mp
\end{equation}
and, since the derivative of the first terms vanish, we also have, at the
leading order,
\begin{equation}
f_k^{\prime} \simeq B_k/z^{2}= {\bar B}_k/\l(a^{2}\, \epsilon_1\r)
\quad {\rm and}\quad
g_k^{\prime} \simeq \sqrt{2}\,D_k/\l(\Mp\, a^2\r),
\end{equation}
where ${\bar B}_k=B_k/(2\, \Mpl^2)$.
Let us now make use of the above super-Hubble behavior of the modes to
arrive at the corresponding contributions to the three point functions
$G_{\cR\cR\g}^{m_3n_3}$, $G_{\cR\g\g}^{m_2n_2m_3n_3}$ and 
$G_{\g\g\g}^{m_1n_1m_2n_2m_3n_3}$.

\par

Let us first focus on $G_{\cR\cR\g}^{m_3n_3}$. 
Let $\es$ denote the conformal time when the largest of the three wavenumbers 
$k_1$, $k_2$ and $k_3$ is well outside the Hubble radius (in this context, we 
would refer the reader to Fig.~1 of Ref.~\cite{hazra-2013}).  
It is then straightforward to show using the above behavior of the modes
that the super-Hubble contributions to $G_{\cR\cR\g}^{m_3n_3}$ are given 
by
\begin{subequations}
\begin{eqnarray}
G_{\cR\cR\g\,(1)}^{m_3n_3\, ({\rm se})}(\vka,\vkb,\vkc)
&\simeq & -4\, i\; \Pi_{m_3n_3,ij}^{\vkc}\;k_{1i}\, k_{2j}\;
\vert A_\ska\vert^2\,\vert A_\skb\vert^2\,
\vert C_\skc\vert^2\nn\\
& &\times\, \l[I(\ee,\es)-I^{\ast}(\ee,\es)\r],\\
G_{\cR\cR\g\,(2)}^{m_3n_3 ({\rm se})}(\vka,\vkb,\vkc)
&\simeq & i\, \Pi_{m_3n_3,ij}^{\vkc}\, k_{1i}\, k_{2j}\,
\l[k_3^2/\l(k_1^2\,k_2^2\r)\r]\, \vert C_\skc\vert^2\, J(\ee,\es)\nn\\
& &\times\,\l(A_\ska\, A_\skb\, {\bar B}_\ska^{\ast}\,{\bar B}_\skb^{\ast}
-A_\ska^{\ast}\, A_\skb^{\ast}\, {\bar B}_\ska\,{\bar B}_\skb\r),\\
G_{\cR\cR\g\,(3)}^{m_3n_3\, ({\rm se})}(\vka,\vkb,\vkc)
&\simeq & i\, \Pi_{m_3n_3,ij}^{\vkc}\, \l[k_{1i}\, k_{2j}/\l(k_1^2\,k_2^2\r)\r]\,
K(\ee,\es)\nn\\
& &\times\,\biggl[k_1^2\; \vert A_\ska\vert^2\,
\l(A_\skb\,{\bar B}_\skb^{\ast}\, C_\skc\, D_\skc^{\ast}
-A_\skb^{\ast}\,{\bar B}_\skb\, C_\skc^{\ast}\, D_\skc\r)\nn\\
& &+\;k_2^2\; \vert A_\skb\vert^2\,
\l(A_\ska\,{\bar B}_\ska^{\ast}\, C_\skc\, D_\skc^{\ast}
-A_\ska^{\ast}\,{\bar B}_\ska\, C_\skc^{\ast}\, D_\skc\r)\biggr],
\end{eqnarray}
\end{subequations}
where the super-script $({\rm se})$ implies that they correspond 
to the contributions over the time domain $\es$ to $\ee$, and the 
quantities $I(\ee,\es)$, $J(\ee,\es)$ and $K(\ee,\es)$ are described 
by the integrals
\begin{subequations}
\begin{eqnarray}
I(\ee,\es) 
& = & \int_{\es}^{\ee}\,\d\eta\; a^2\, \epsilon_1,\\
J(\ee,\es) 
& = & \int_{\es}^{\ee}\,\f{\d\eta}{a^2},\\
K(\ee,\es) 
& = & \int_{\es}^{\ee}\,\f{\d\eta}{a^2}\;\epsilon_1.
\end{eqnarray}
\end{subequations}
Note that, since $I(\ee,\es)$ is real, the term
$G_{\cR\cR\g\,(1)}^{m_3n_3\, ({\rm se})}$ vanishes identically and, hence,
the super-Hubble contributions arise only due to the other two terms.

\par
In a similar fashion, one can show that the super-Hubble contributions
to $G_{\cR\g\g}^{m_2n_2m_3n_3}$ are given by
\begin{subequations}
\bea
G_{\cR\g\g\,(1)}^{m_2n_2m_3n_3\, ({\rm se})}
&\simeq& \f{i}{\Mpl^2}\; \Pi_{m_2n_2,ij}^{\vkb}\;
\Pi_{m_3n_3,ij}^{\vkc}\,\vert A_\ska\vert^2\, K(\ee,\es)\nn\\
& &\times\,\l(C_\skb\, C_\skc\,D_\skb^{\ast}\, D_\skc^{\ast}
- C_\skb^{\ast}\, C_\skc^{\ast}\,D_\skb\, D_\skc\r),\\
G_{\cR\g\g\,(2)}^{m_2n_2m_3n_3\, ({\rm se})}
&\simeq& -\f{i}{\Mpl^2}\; \Pi_{m_2n_2,ij}^{\vkb}\;
\Pi_{m_3n_3,ij}^{\vkc}\; \l(\vkb\cdot\vkc\r)\,\vert A_\ska\vert^2\,
\vert C_\skb\vert^2\,  \vert C_\skc\vert^2\nn\\                                            
& &\times\;
\l[I(\ee,\es)-I^{\ast}(\ee,\es)\r],\\
G_{\cR\g\g\,(3)}^{m_2n_2m_3n_3\, ({\rm se})}
&\simeq& -\f{i}{\Mpl^2}\; \Pi_{m_2n_2,ij}^{\vkb}\;
\Pi_{m_3n_3,ij}^{\vkc}\; J(\ee,\es)\nn\\
& &\times\,\biggl[\f{\vka\cdot \vkb}{k_1^2}\;\vert C_{k_2}\vert^2\, 
\l(A_{k_1}\, {\bar B}_{k_1}^{\ast}\, C_{k_3}\, D_{k_3}^{\ast} 
- A_{k_1}^{\ast}\, {\bar B}_{k_1}\, C_{k_3}^{\ast}\, D_{k_3}\r)\nn\\
& & +\; \f{\vka\cdot \vkc}{k_1^2}\;\vert C_{k_3}\vert^2\, 
\l(A_{k_1}\, {\bar B}_{k_1}^{\ast}\, C_{k_2}\, D_{k_2}^{\ast} 
- A_{k_1}^{\ast}\, {\bar B}_{k_1}\, C_{k_2}^{\ast}\, D_{k_2}\r)\biggr].
\end{eqnarray}
\end{subequations}
Clearly, the term $G_{\cR\g\g\,(2)}^{m_2n_2m_3n_3\, ({\rm se})}$ 
vanishes for the same reason as $G_{\cR\cR\g\,(1)}^{m_3n_3\, ({\rm se})}$ had 
and, as a result, it is only the remaining two terms that contribute on 
super-Hubble scales to $G_{\cR\g\g}^{m_2n_2m_3n_3}$.

\par

Lastly, let us turn to the tensor bi-spectrum, viz.  
$G_{\g\g\g}^{m_1n_1m_2n_2m_3n_3}$.
In this case, we have, on super-Hubble scales
\begin{subequations}
\begin{eqnarray}
G_{\g\g\g\,(1)}^{m_1n_1m_2n_2m_3n_3\, ({\rm se})}
&=& -\f{2\,i}{\Mpl^4}\; \l(\Pi_{m_1n_1,ij}^{\vka}\;
\Pi_{m_2n_2,im}^{\vkb}\; \Pi_{m_3n_3,lj}^{\vkc}\; k_{1m}\, k_{1l}
+{\rm five~permutations}\r)\nn\label{eq:shc-ggg1}\\
& &\times\,\vert C_\ska\vert^2\,
\vert C_\skb\vert^2\,  \vert C_\skc\vert^2\;
\l[L(\ee,\es)-L^{\ast}(\ee,\es)\r],\\
G_{\g\g\g\,(2)}^{m_1n_1m_2n_2m_2n_3\, ({\rm se})}
&=& \f{i}{\Mpl^4}\; \l(\Pi_{m_1n_1,ij}^{\vka}\;
\Pi_{m_2n_2,ml}^{\vkb}\; \Pi_{m_3n_3,ij}^{\vkc}\; k_{1m}\, k_{1l}
+{\rm five~permutations}\r)\nn\\
& &\times\,\vert C_\ska\vert^2\,
\vert C_\skb\vert^2\,  \vert C_\skc\vert^2\;
\l[L(\ee,\es)-L^{\ast}(\ee,\es)\r],\label{eq:shc-ggg2}
\end{eqnarray}
\end{subequations}
where the quantity $L(\ee,\es)$ is described by the integral
\begin{equation}
L(\ee,\es)= \int_{\es}^{\ee}\,\d\eta\; a^2.\\
\end{equation}
Both of the above expressions obviously vanish since $L(\ee,\es)$ is real.
In other words, the super-Hubble contributions to the tensor bi-spectrum
and the corresponding non-Gaussianity parameter $\hnl$ are identically 
zero.

\par

It is now worthwhile to estimate the extent of the super-Hubble 
contributions to the other two non-Gaussianity parameters $\cnls$ 
and $\cnlt$.
In order to carry out such an estimate, let us focus on power law
inflation wherein the scale factor can be expressed as
\begin{equation}
a(\eta)=a_1\, \l(\f{\eta}{\eta_1}\r)^{\gamma + 1}
\label{eq:a-p-law}
\end{equation}
where $a_1$ and $\eta_1$ are constants, while $\gamma <-2$. 
In such a situation, the first slow roll parameter is a constant,
and is given by $\epsilon_1 = (\gamma + 2)/(\gamma +1)$. 
Also, since $z''/z = a''/a$ in power law inflation, the solutions 
to the scalar and the tensor modes $f_k$ and $g_k$ are exactly the 
same functions, barring overall constants.
In fact, the solutions to the Mukhanov-Sasaki equations~(\ref{eq:ms}) 
can be expressed in terms of the Bessel functions $J_\nu(x)$ as follows 
(see, for instance, Refs.~\cite{p-law,hazra-2012}): 
\begin{equation} 
v_k(\eta) = {\mathcal U}_k(\eta)
=\sqrt{-k\,\eta}\; 
\l[{\mathcal A}_k\, J_\nu(-k\,\eta) 
+ {\mathcal B}_k\, J_{-\nu} (-k\,\eta) \r],
\label{eq:v-p-law}
\end{equation}
where $\nu = \gamma + 1/2$.
The quantities ${\mathcal A}_k$ and ${\mathcal B}_k$ are $k$-dependent 
constants which are determined by demanding that the above solutions 
satisfy the Bunch-Davies initial conditions at early 
times~\cite{texts,reviews,bunch-1978}.
They are found to be
\begin{subequations}
\begin{eqnarray}
{\mathcal A}_k 
&=& -{\mathcal B}_k\; {\rm e}^{-i\,\pi\,(\gamma + 1/2)},\label{eq:cA}\\
{\mathcal B}_k 
&=& \sqrt{\f{\pi}{k}}\;
\f{{\rm e}^{i\,\pi\,\gamma/2}}{2\, {\rm cos}\,(\pi\,\gamma)}.\label{eq:cB}
\end{eqnarray}
\end{subequations}
In the super-Hubble limit, i.e. as $-k\,\eta \rightarrow 0$, the solutions 
for $v_k(\eta)$ and ${\mathcal U}_k(\eta)$ in~(\ref{eq:v-p-law}) can be 
compared with the general solutions~(\ref{eq:shs-fkgk}) to
arrive at the following expressions for the quantities $A_k$, $B_k$,
$C_k$ and $D_k$:
\begin{subequations}
\begin{eqnarray}
A_k &=& \f{C_k}{\sqrt{2\,\epsilon_1}\, \Mpl}
=\f{2^{-(\gamma+1/2)}}{\Gamma(\gamma + 3/2)}\;
\f{(-k\,\eta_1)^{\gamma+1}}{\sqrt{2\,\epsilon_1}\, a_1\, \Mpl}\,
{\mathcal A}_k,\label{eq:A-pli}\\
B_k &=& \sqrt{2\,\epsilon_1}\, \Mpl\; D_k
=-\,\f{(2\,\gamma + 1)\; 2^{\gamma + 1/2}}{\Gamma(-\gamma +1/2)}\;
\f{\sqrt{2\,\epsilon_1}\,a_1\,\Mpl}{\eta_1}\;(-k\,\eta_1)^{-\gamma}\;
{\mathcal B}_k.
\end{eqnarray}
\end{subequations}
Moreover, the scalar and tensor power spectra in power law inflation, 
evaluated in the super-Hubble limit, can be shown to be
\begin{equation}
{\mathcal P}_{_{\rm S}}(k)
=\f{k^3}{2\,\pi^2}\; \vert A_k\vert^2
= \f{{\mathcal P}_{_{\rm T}}(k)}{16\, \epsilon_1},
\end{equation}
a well known result that is also valid in slow roll 
inflation~\cite{texts,reviews}.

\par

We now have all the quantities required to arrive at an estimate for
the super-Hubble contributions to the parameters $\cnls$ and $\cnlt$
[cf. Eqs.~(\ref{eq:cnls}) and (\ref{eq:cnlt})] in power law inflation.
Let us restrict ourselves to the equilateral limit, i.e. $k_1 = k_2 = 
k_3$, for simplicity.
In such a case, upon using the results we have obtained above, one 
can show, after a bit of algebra~\cite{hazra-2012,hazra-2013}, that
\begin{eqnarray}
C_{_{\rm NL}}^{\cR\, ({\rm se})}
&=& \f{3}{4}\,C_{_{\rm NL}}^{\g\, ({\rm se})}\nn\\
&=& \frac{3}{16\,\pi}\; \Gamma^2\l(\gamma +\f{1}{2}\r)\,
2^{2\,\gamma+1}\, \l(2\,\gamma+1\r)\, (\gamma+2)\,
\l\vert \gamma+1\r\vert^{-2\,(\gamma+1)}\,\sin\,(2\,\pi\,\gamma)\nn\\
& &\times\, \l[1-\f{H_{\rm s}}{H_{\rm e}}\;
{\rm e}^{-3\,(N_{\rm e}-N_{\rm s})}\r]\;
\l(\f{k}{a_{\rm s}\, H_{\rm s}}\r)^{-(2\,\gamma+1)},
\end{eqnarray}
where $(N_{\rm s}, N_{\rm e})$ and $(H_{\rm s},H_{\rm e})$ denote the
number of e-folds and the Hubble parameter at the conformal times
$(\es,\ee)$. 
We should also add that, in arriving at the above expression, we have 
ignored overall factors involving $\Pi_{mn,ij}^{\vk}$, which 
can be assumed to be of order unity without any loss of generality.
Further, we have set the constant $a_1$ to be $a_{\rm s}$, viz. the
scale factor at the time $\es$.
If we now choose $\gamma\simeq -(2+\delta)$, where $\delta\ll 1$, then, 
we obtain that 
\begin{equation}
C_{_{\rm NL}}^{\cR\, ({\rm se})}
= \f{3}{4}\,C_{_{\rm NL}}^{\g\, ({\rm se})}
\simeq -\,\delta^2\; 
\l(\f{k_{\rm s}}{a_{\rm s}\, H_{\rm s}}\r)^3
\lesssim 10^{-19},
\end{equation} 
where $k_{\rm s}$ is the largest wavenumber of interest and, in arriving 
at the final inequality, we have assumed that  $k_{\rm s}/(a_{\rm s}\,
H_{\rm s})=10^{-5}$ and $\delta\simeq 10^{-2}$. 
As we shall see later, this value always proves to be considerably smaller 
than the corresponding values generated as the modes leave the Hubble radius 
during inflation.
This implies that we can safely ignore the super-Hubble contributions 
to the scalar-tensor cross correlations and the tensor bi-spectrum as
well as the corresponding non-Gaussianity parameters.


\subsection{Details of the numerical method}

Let us now turn to discuss the numerical procedure for evaluating the 
three-point functions. 
It should by now be clear that evaluating the three-point functions and 
the non-Gaussianity parameters involves solving for the evolution of the 
background and the perturbations and, eventually, computing the integrals 
involved.
Given the inflationary potential $V(\phi)$ that describes the scalar 
field and the values for the parameters, the background evolution is 
completely determined if the initial conditions on the scalar field are
specified.
Typically, the initial value of the scalar field is chosen so that one
achieves about $60$ or so e-folds of inflation (as is required to overcome 
the horizon problem) before the accelerated expansion is terminated as
the field approaches a minima of the potential.
Further, the initial velocity of the field is often chosen such that 
the field starts on the inflationary attractor (in this context, see, 
for example, Refs.~\cite{ne-ps}).

\par

Once the background has been solved for, the scalar and the tensor 
perturbations are evolved from the standard Bunch-Davies initial 
conditions using the governing 
equations~(\ref{eq:de-p})~\cite{texts,reviews,bunch-1978}.
Then, in order to arrive at the three-point functions, it is a matter 
of being able to carry out the various integrals involved.
Recall that, when calculating the power spectra, the initial conditions
are imposed on the modes when they are sufficiently inside the Hubble 
radius, typically, when $k/(a\, H)\simeq 10^2$.
The spectra are evaluated in the super-Hubble domain, when the amplitudes
of the modes have reached a constant value, which often occurs when 
$k/(a\, H)\simeq 10^{-5}$ (see, for instance, 
Refs.~\cite{e-dfsr,ne-ps}).
Since the super-Hubble contributions to the three-point functions are
negligible, it suffices to carry out the integrals from the earliest
time $\ei$ when the smallest of the three wavenumbers $(k_1,k_2,k_3)$ 
is well inside the Hubble radius to the final time $\es$ when the 
largest of them is sufficiently outside.
However, there is one point that needs to be noted though.
In the extreme sub-Hubble domain, the modes oscillate rapidly and, 
theoretically, a cut-off is required in order to identify the correct 
perturbative vacuum~\cite{maldacena-2003,ng-ncsf,ng-reviews}.  
This proves to be handy numerically, as the introduction of a cut-off 
ensures that the integrals converge quickly (for the original 
discussion on this point, see Refs.~\cite{ng-ne}).
Motivated by the consistent results arrived recently in the case of 
the scalar bi-spectrum~\cite{hazra-2013}, we introduce a cut-off of
the form $\exp-\l[\kappa\, k/(a\, H)\r]$, where $\kappa$ is a small
parameter.
As we shall discuss below, a suitable combination of $\kappa$, $\ei$
and $\es$ (or, $N_{\rm i}$ and $N_{\rm s}$, in terms of e-folds) ensure 
that the final results are fairly robust against changes in their values. 
  
\par

\begin{figure*}[!htb]
\begin{center}
\includegraphics[width=15.0cm]{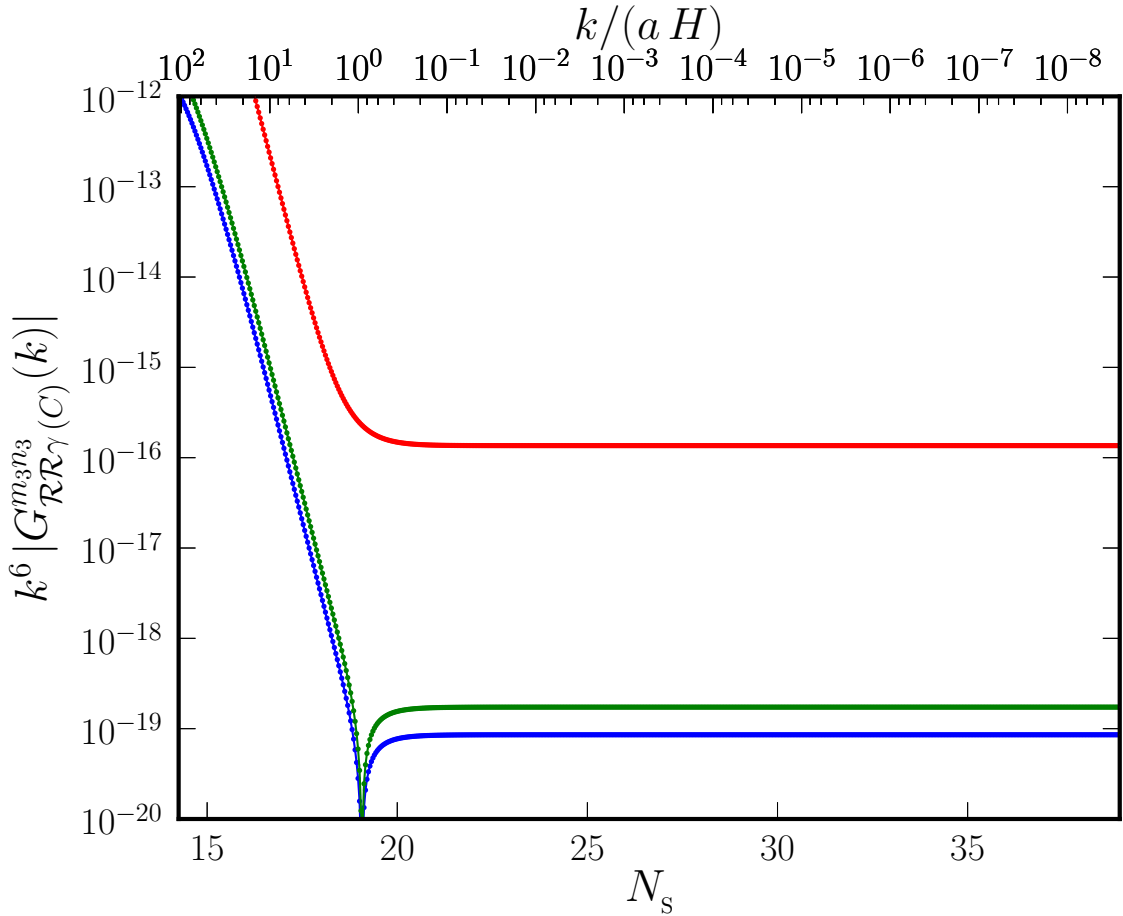}
\end{center}
\vskip -15pt
\caption{\label{fig:qp-v-ns} 
The absolute value of the different contributions to the scalar-scalar-tensor 
cross-correlation evaluated in the equilateral limit, 
i.e. $k^6\, G_{\cR\cR\g}^{m_3n_3}$, have been plotted for a specific mode 
(which leaves the Hubble radius at about $40$ e-folds before the end of 
inflation), evolving in the background driven by the conventional quadratic 
potential, as a function of the upper limit of integration~$\Ns$.
In this figure and in the ones that follow, we shall adopt the following choice 
of colors to represent the different contributions to the three-point functions.
The first, the second and the third terms of the three-point functions will 
always be represented by red, green and blue curves, in that order, respectively. 
We should also mention here that we shall ignore factors such as 
$\Pi_{mn,ij}^{\bf k}$ in plotting these quantities.
It is clear from the above figure that the different contributions settle down 
to their final value soon after the mode has emerged from the Hubble radius 
[say, by $k/(a\,H)\simeq 10^{-2}$].
We find that all the contributions to the other three-point functions too 
exhibit the same behavior.}
\end{figure*}

\begin{figure*}[!htb]
\begin{center}
\includegraphics[width=15.0cm]{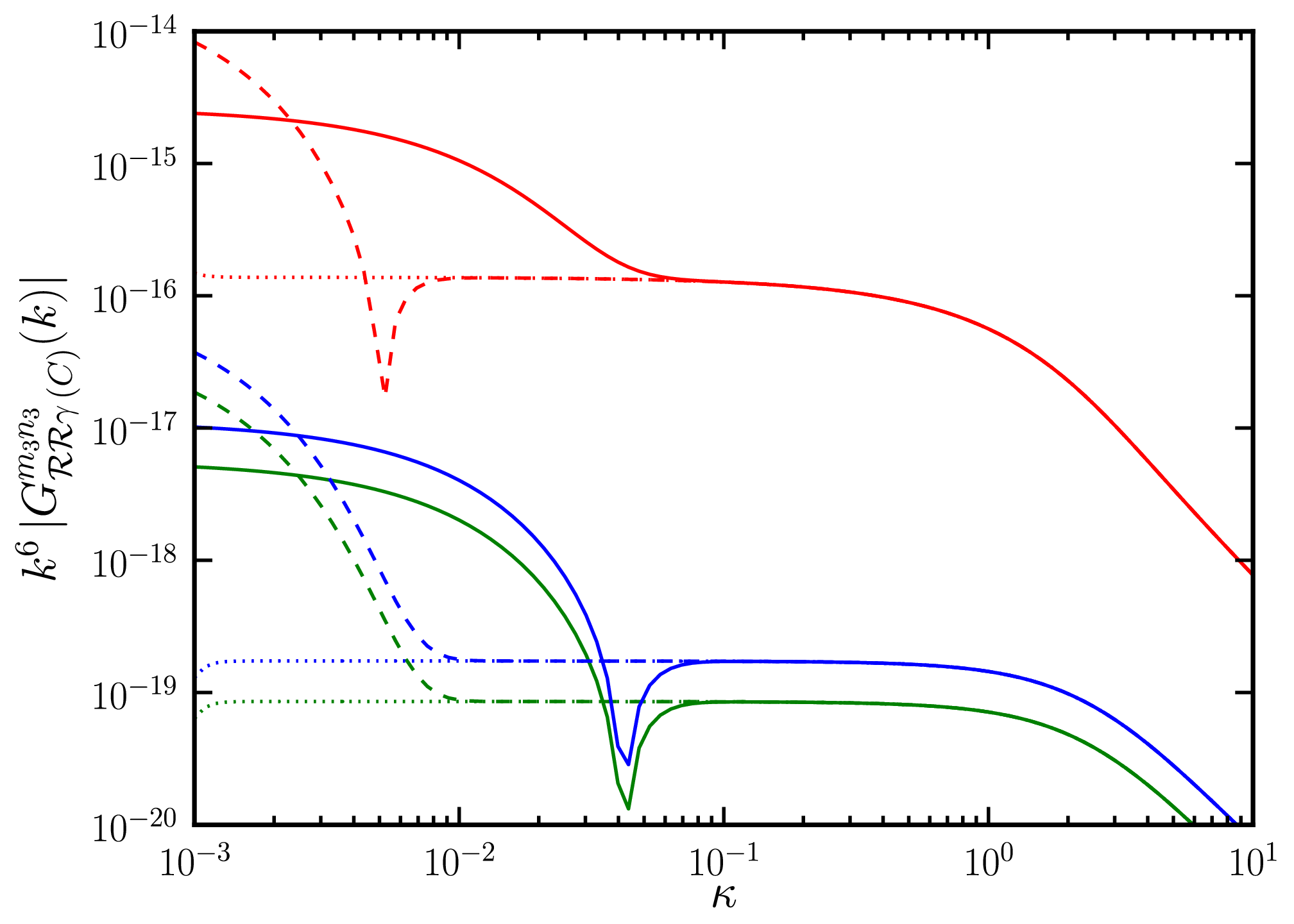}
\end{center}
\vskip -15pt
\caption{\label{fig:qp-v-k-ni} 
The absolute value of the different contributions to the scalar-scalar-tensor 
cross-correlation in the equilateral limit, viz. $k^6\, G_{\cR\cR\g}^{m_3n_3}$, 
discussed in the previous figure, have been plotted for the same model and mode 
for a few different combinations of $\Ni$ and $\kappa$, but with a fixed $\Ns$
(corresponding to $k/(a\,H)$ of $10^{-5}$).
The solid, the dashed and the dotted lines correspond to integrals evaluated 
from different $\Ni$, corresponding to $k/(a\,H)$ of $10^2$, $10^3$ and 
$10^4$, respectively.
We should point out that too large a value of $\kappa$ (say, much beyond 
$\kappa\simeq 0.1$) brings down the values of the integrals, as it then
essentially kills the contributions that occur as the modes leave the 
Hubble radius. 
It is also evident that the choice of $\kappa=0.1$ and an $\Ni$ corresponding
to $k/(a\,H)=10^2$ leads to consistent results (as all the curves converge
over this domain).
Again, we find that the same conclusions apply to all the contributions
to the other three-point functions as well.}
\end{figure*}

We solve the background and the perturbation equations using the fifth 
order Runge-Kutta algorithm (see, for instance, Ref.~\cite{nr}), with 
e-folds as the independent variable.
We carry out the integrals involved using the so-called Bode's 
rule to arrive at the three-point functions and the non-Gaussianity
parameters\footnote{There seems to be some confusion in the literature 
regarding whether it is the Bode's or the Boole's rule!
Following Ref.~\cite{nr}, we have called it the Bode's rule.}.
In Figs.~\ref{fig:qp-v-ns} and~\ref{fig:qp-v-k-ni}, with the help of an 
example (viz. the three different contributions to the cross-correlation
$G_{\cR\cR\g}^{m_3n_3}$, evaluated in the equilateral limit), we demonstrate 
the robustness of the procedure we have described above for a specific mode 
evolving in the popular quadratic potential.
In arriving at the first figure, we have fixed the values of $\Ni$ and 
$\kappa$, and vary $\Ns$.
Whereas, the second figure corresponds to a few different values of $\Ni$,
but a fixed value of $\Ns$.
It is clear from the figures that the choices of $\Ns$ corresponding to
$k/(a\, H)$ of $10^{-5}$, and the combination of $\Ni$ corresponding to
$k/(a\, H)$ of $10^{2}$ and $\kappa$ of $0.1$ leads to consistent
results.
We have carried out similar exercises for all the models that we shall discuss 
in this paper, and we have found that the above set of values for $\Ni$, $\Ns$
and $\kappa$ lead to robust results in all the cases. 
Also, as we shall illustrate in the following sub-section, the numerical results 
arrived at in such a fashion are consistent with the various analytical results
that are available.
Actually, we find that, the numerical results obtained with a $\kappa$ of $0.1$ 
and an $N_i$ corresponding to $k/(a\, H)$ of $10^2$ matches the analytical 
results at the level of $5\%$, just as it had in the case of the scalar
bi-spectrum~\cite{hazra-2013}.
The match improves to $1$--$2\%$ if we work with a $\kappa$ of, say, $0.02$, and 
simultaneously integrate from an $N_i$ corresponding to $k/(a\,H)$ of $10^3$.
We should emphasize here that we have worked with these set of values in arriving 
at all the latter figures (i.e. Fig.~\ref{fig:pl-sm-el} and thereafter).


\subsection{Comparison with the analytical results}

In this section, as it was done in the context of the scalar bi-spectrum
(see Ref.~\cite{hazra-2013}), we shall compare the numerical results for the
three-point functions (or, equivalently, for the non-Gaussianity parameters)
with the spectral dependence that can be arrived at in power-law inflation 
in the equilateral and the squeezed limits and the results for an arbitrary 
triangular configuration that can be obtained in the slow roll scenario (as 
applied to the case of the quadratic potential).
With the motivation to consider a non-trivial situation involving departures
from slow roll, we shall also evaluate the three-point functions for the case 
of the Starobinsky model analytically and compare them with the corresponding 
numerical results.
We shall relegate some of the details of the calculation in the case of the
Starobinsky model to the appendix.


\subsubsection{The case of power law inflation}

As we have already discussed, power law inflation is described by the scale
factor~(\ref{eq:a-p-law}).
Also, in such a scenario, the scalar and the tensor modes $v_k$ and 
${\mathcal U}_k$ can be obtained analytically [cf. Eq.~(\ref{eq:v-p-law})]. 
Note that these modes depend only on the combination $k\, \eta$.
Due to this reason, interestingly, one finds that, with a simple rescaling
of variables, the spectral dependence (but, not the amplitudes) of all the 
contributions to the scalar-tensor cross correlations as well as the tensor 
bi-spectrum can be arrived at without actually having to evaluate the integrals 
involved~\cite{hazra-2013}.
Since the solutions to the scalar as well as the tensor modes are of the 
same form, in the equilateral limit, i.e. when $k_1=k_2=k_3=k$, one finds 
that all the contributions to the three-point functions have the same spectral 
dependence, viz. $k^{6}\, G_{{\sf ABC}\,(C)}(k) \propto k^{4\,(\gamma+2)}$. 

\par

\begin{figure}[!h]
\begin{center}
\vskip -20pt
\begin{tabular}{cc}
\includegraphics[width=7.40cm]{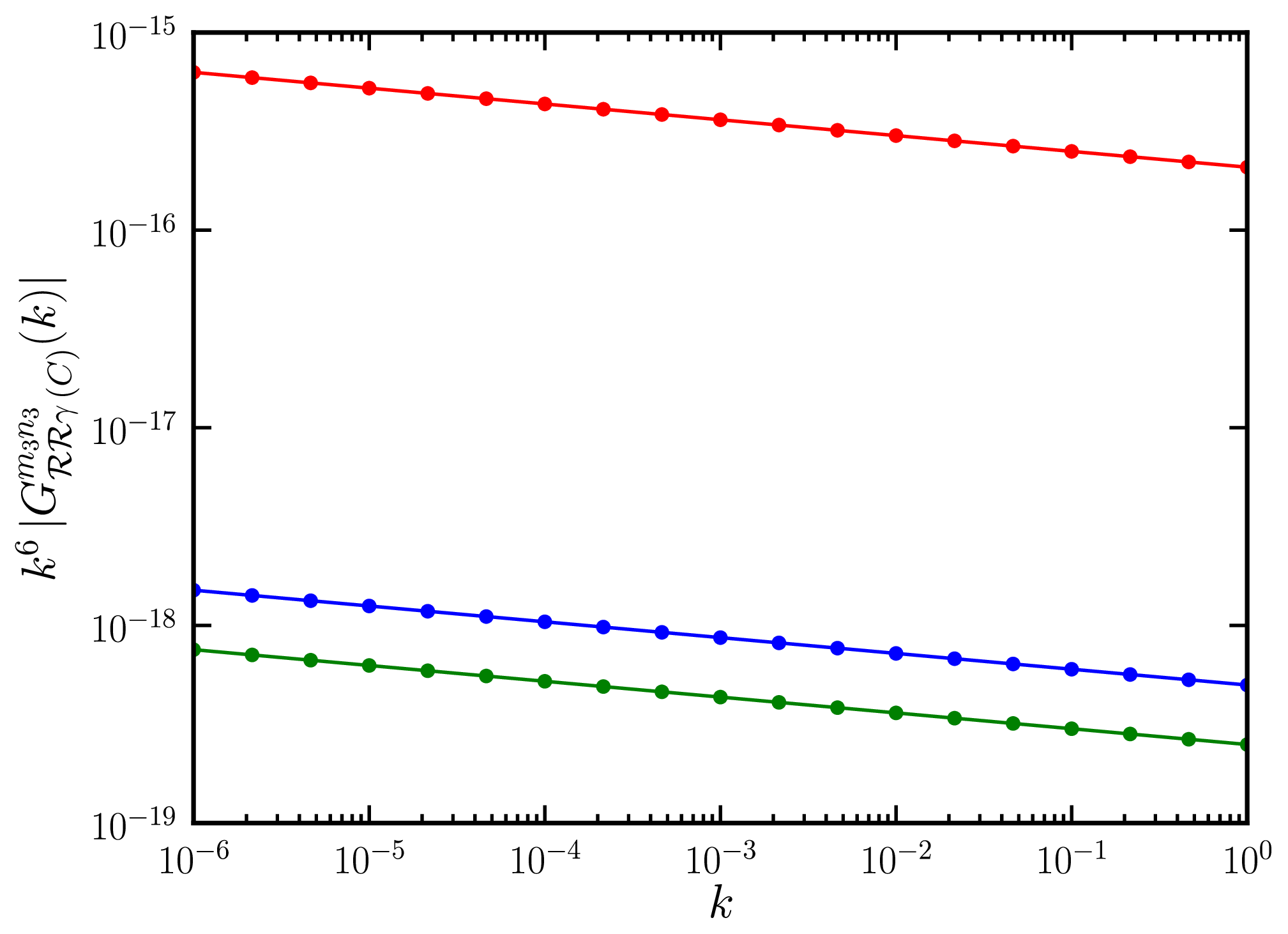} &
\hskip-5pt
\includegraphics[width=7.40cm]{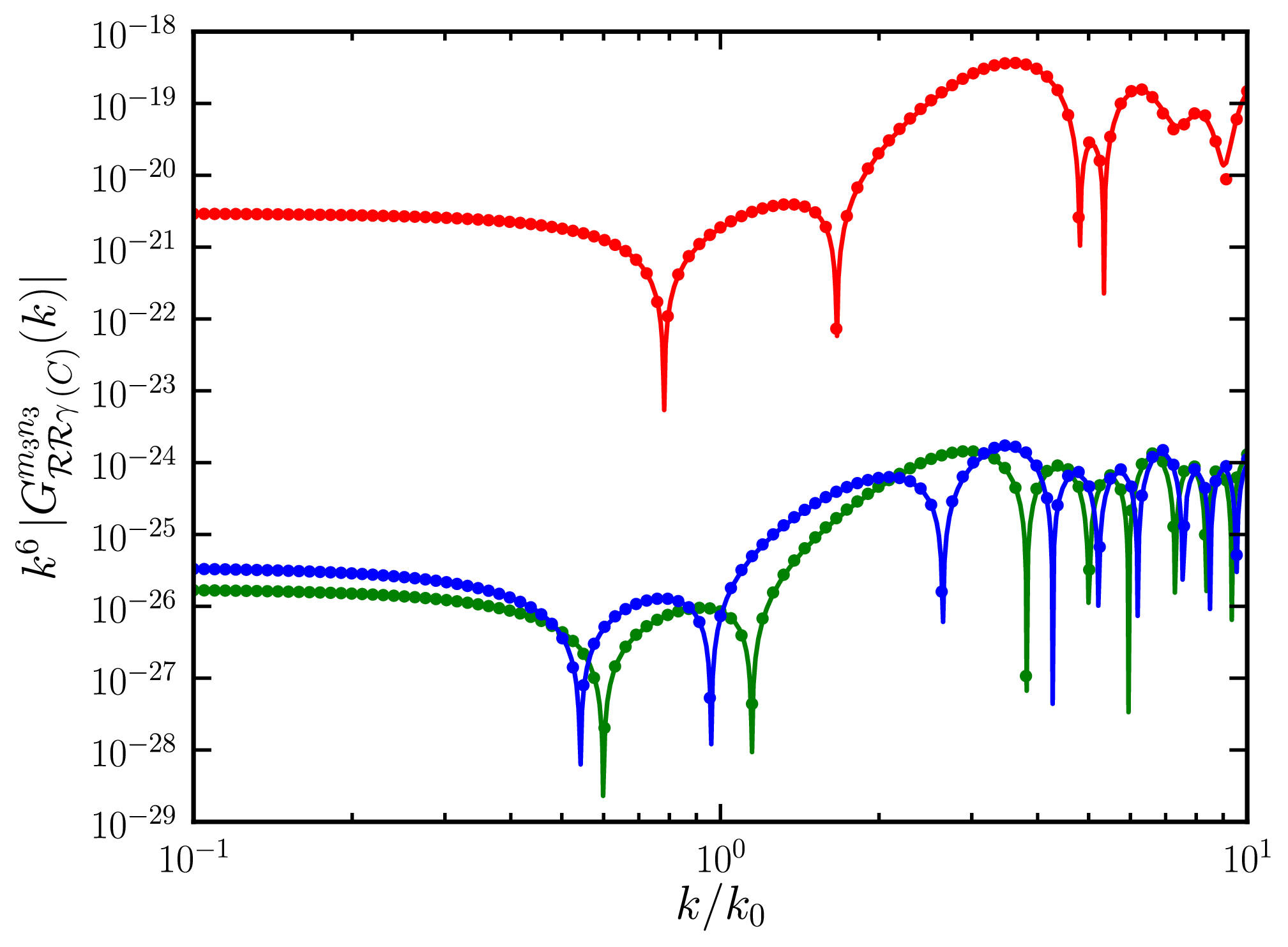} \\
\includegraphics[width=7.40cm]{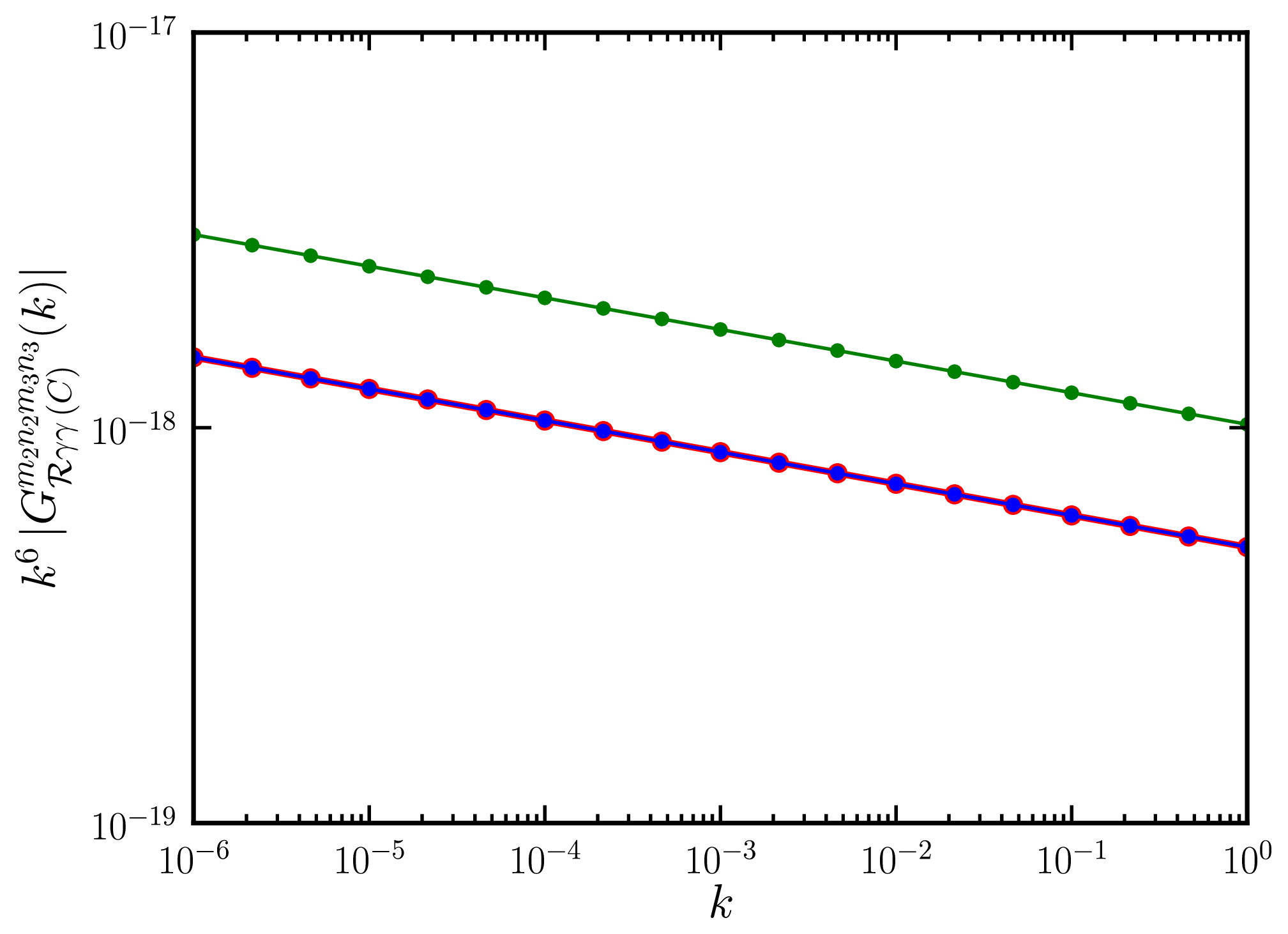} &
\hskip-5pt
\includegraphics[width=7.40cm]{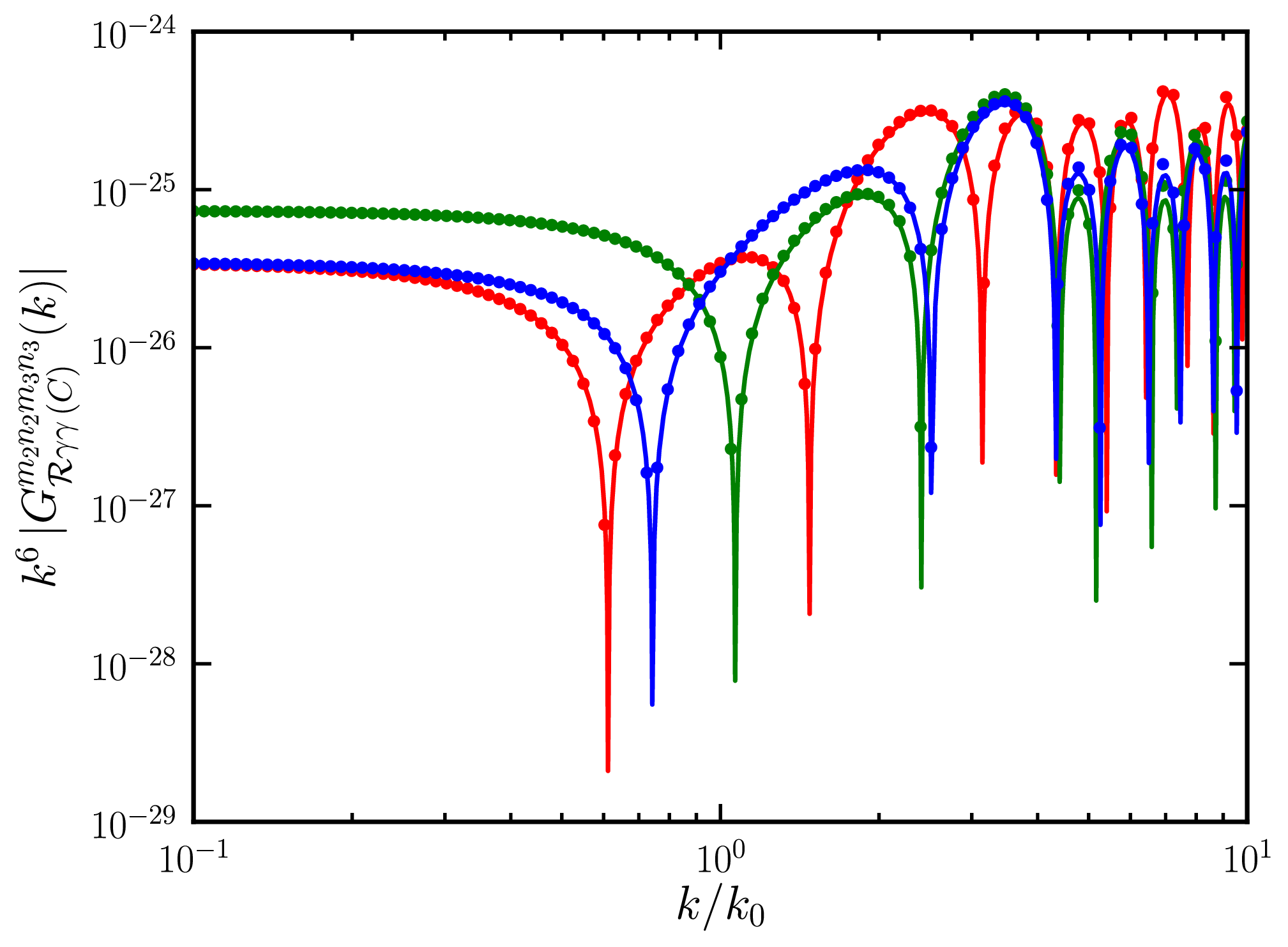} \\
\includegraphics[width=7.40cm]{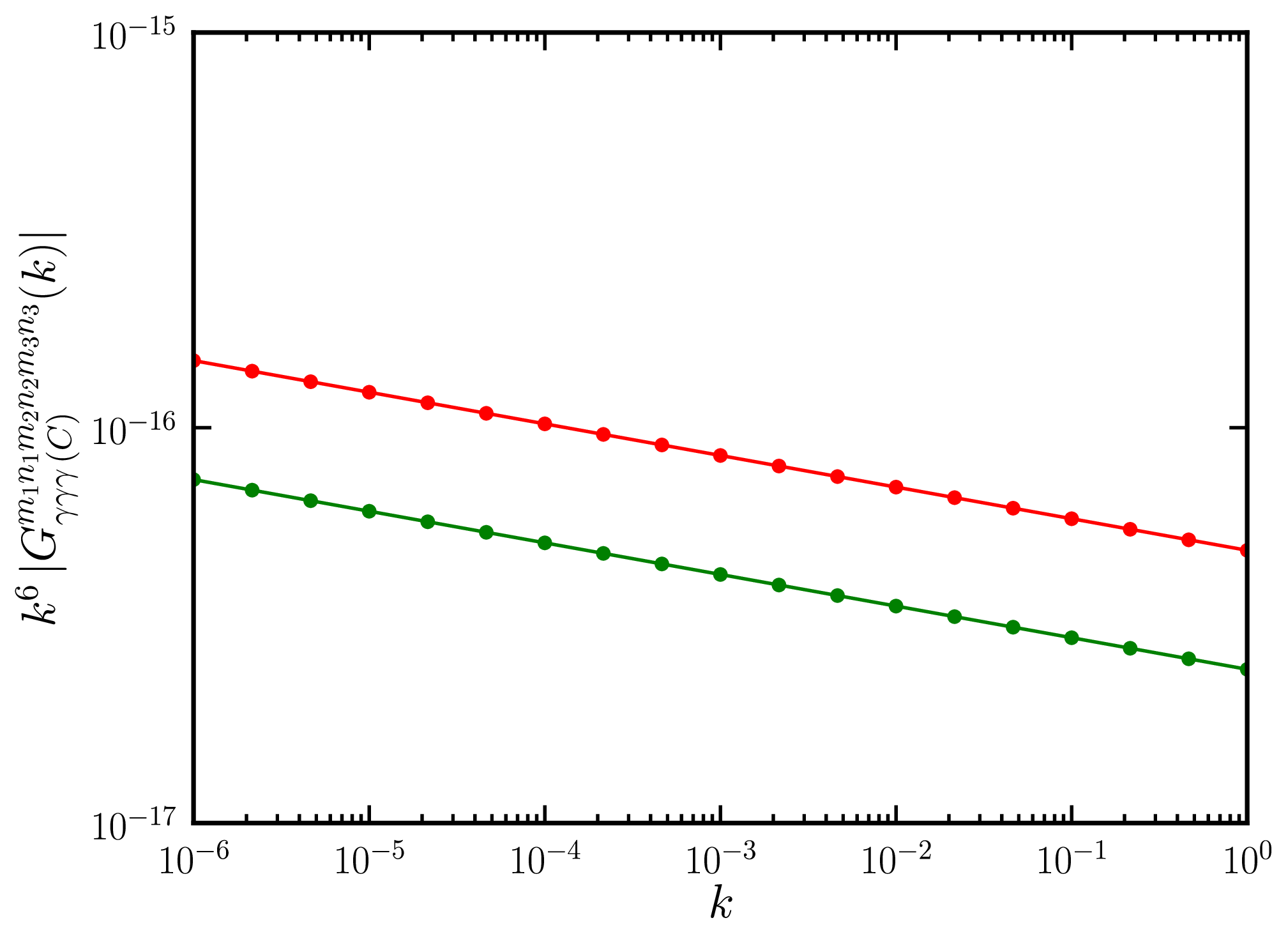} &
\hskip-5pt
\includegraphics[width=7.40cm]{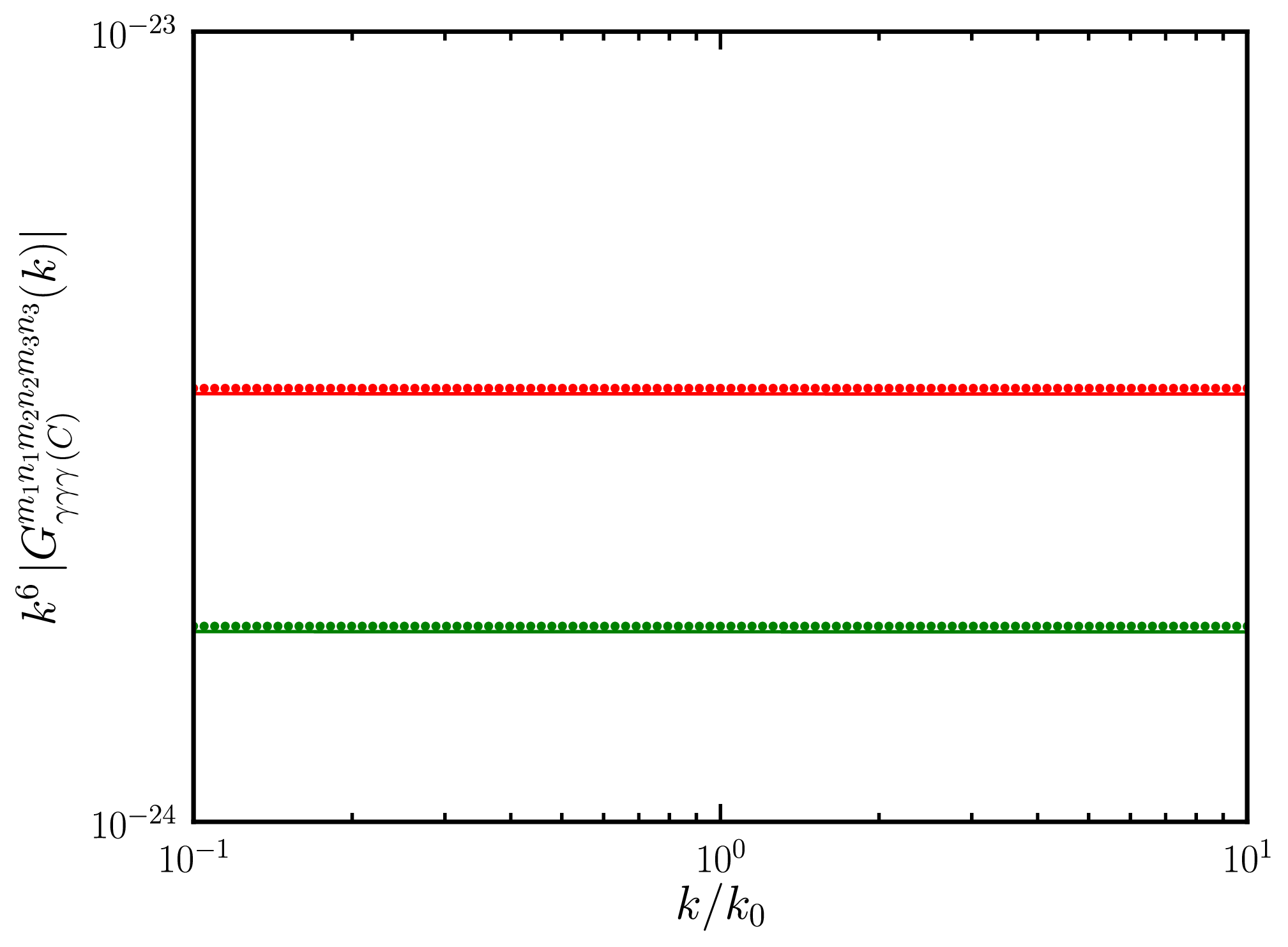} \\
\end{tabular}
\caption{\label{fig:pl-sm-el}
A comparison of the numerical results (plotted as solid lines) with the 
analytical results (marked with dots) for the various contributions to
the three-point functions in the equilateral limit, viz. $k^6$ times the 
absolute values of $G_{\cR\cR\g\, (C)}^{m_3n_3}$ (on top), 
$G_{\cR\g\g\, (C)}^{m_2n_2m_3n_3}$ (in the middle) and
$G_{\gamma\gamma\gamma\, (C)}^{m_1n_1m_2n_2m_3n_3}$ (at the bottom), for 
power law inflation (on the left) and the Starobinsky model (on the right).
In the case of power law inflation, in plotting the analytical, spectral 
dependences, we have chosen the amplitude by hand so that they match the 
numerical result at a specific wavenumber.
The hierarchy of the different contributions are clear from the above 
figure.
Note that, in the cases of the scalar-tensor-tensor cross-correlation and 
the tensor bi-spectrum, as is expected from their dependence on the first slow 
roll parameter $\epsilon_1$, the different contributions to these quantities 
prove to be of the same order.
Whereas, in the case of the scalar-scalar-tensor cross-correlation, the
second and the third terms are of the same order, but are sub-dominant to
the first term.}
\end{center}
\end{figure}

\begin{figure}[!t]
\begin{center}
${}$\vskip -40pt
\begin{tabular}{cc}
\includegraphics[width=7.40cm]{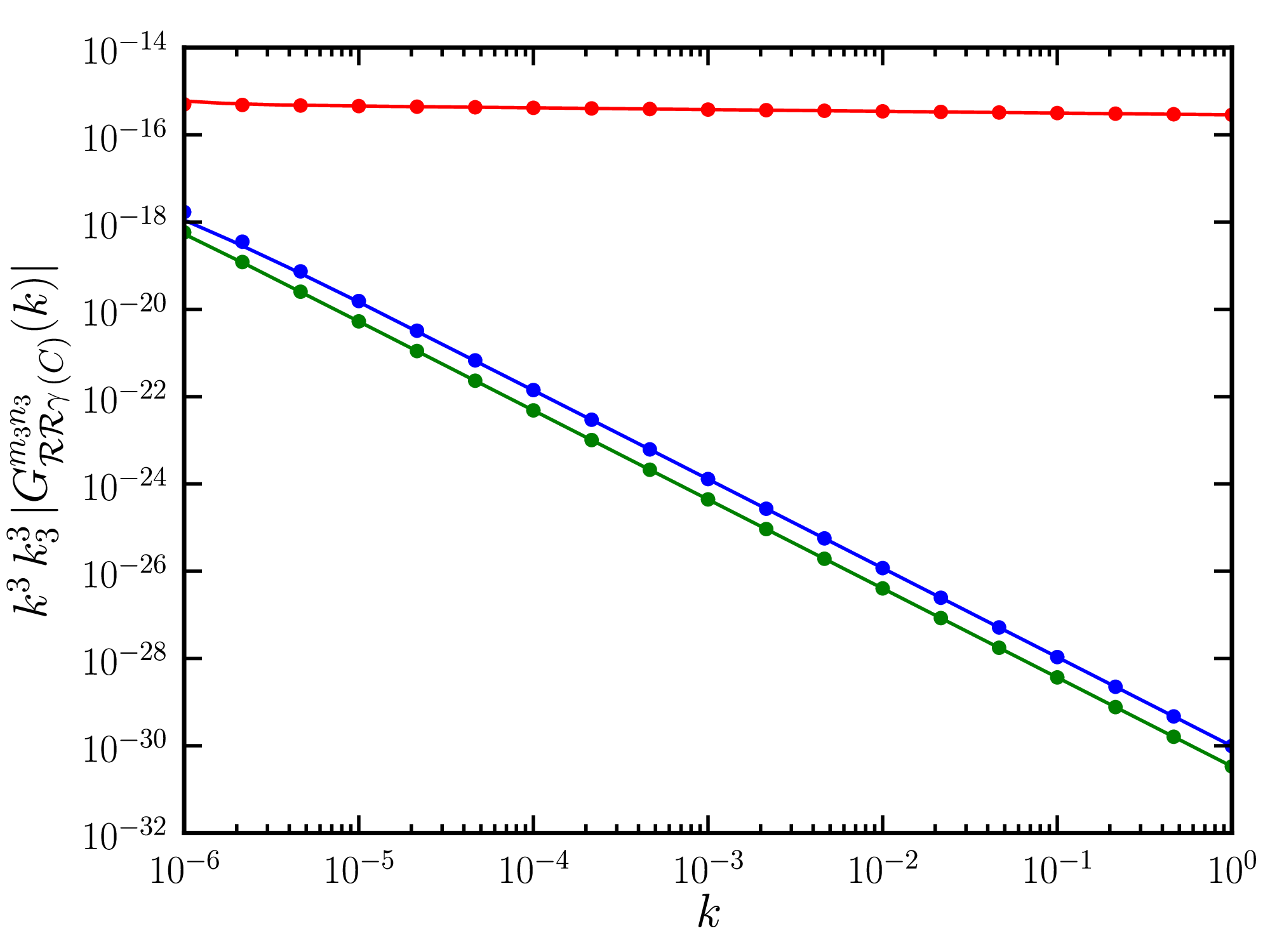} &
\hskip-5pt
\includegraphics[width=7.40cm]{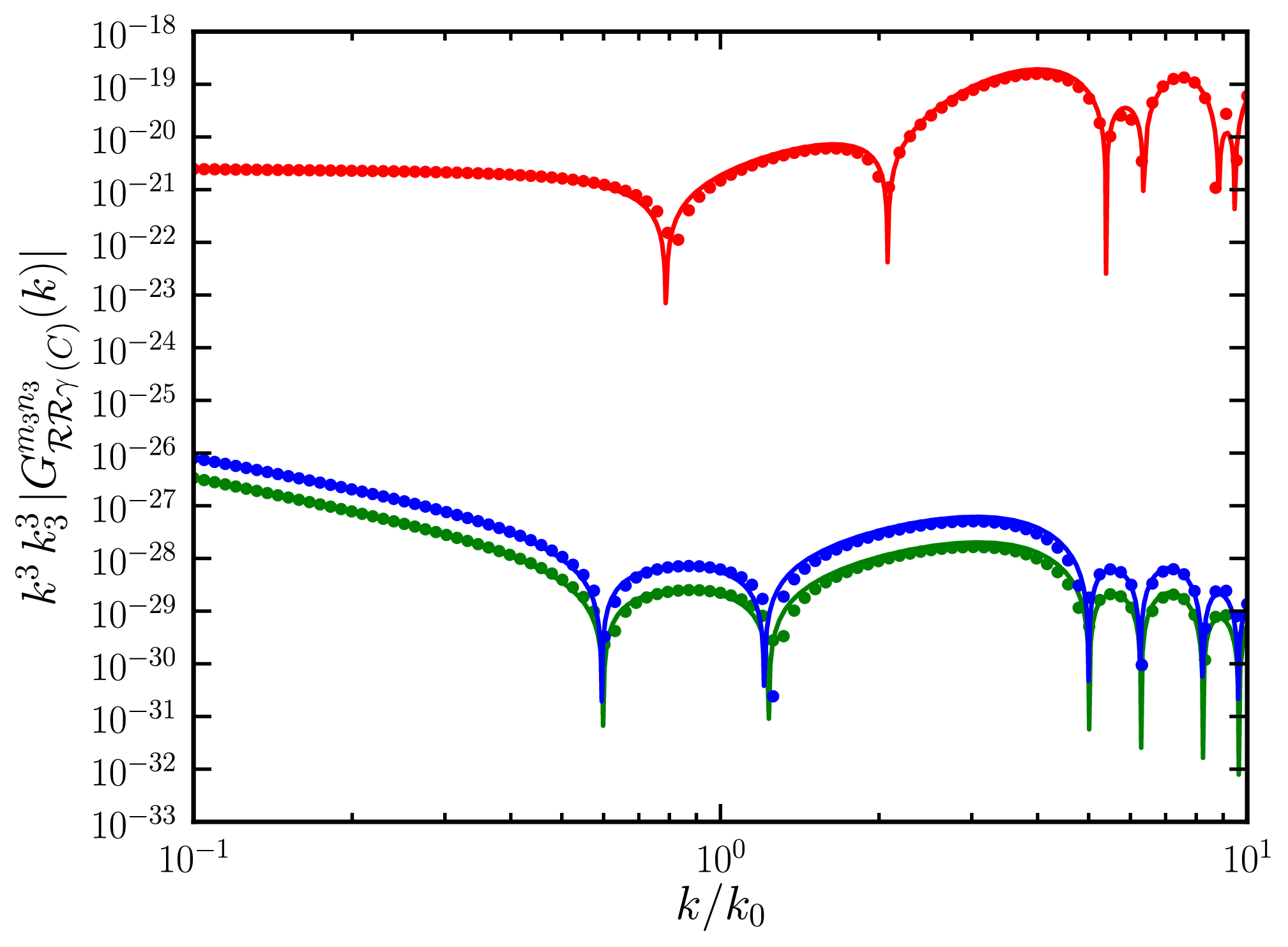}\\
\includegraphics[width=7.40cm]{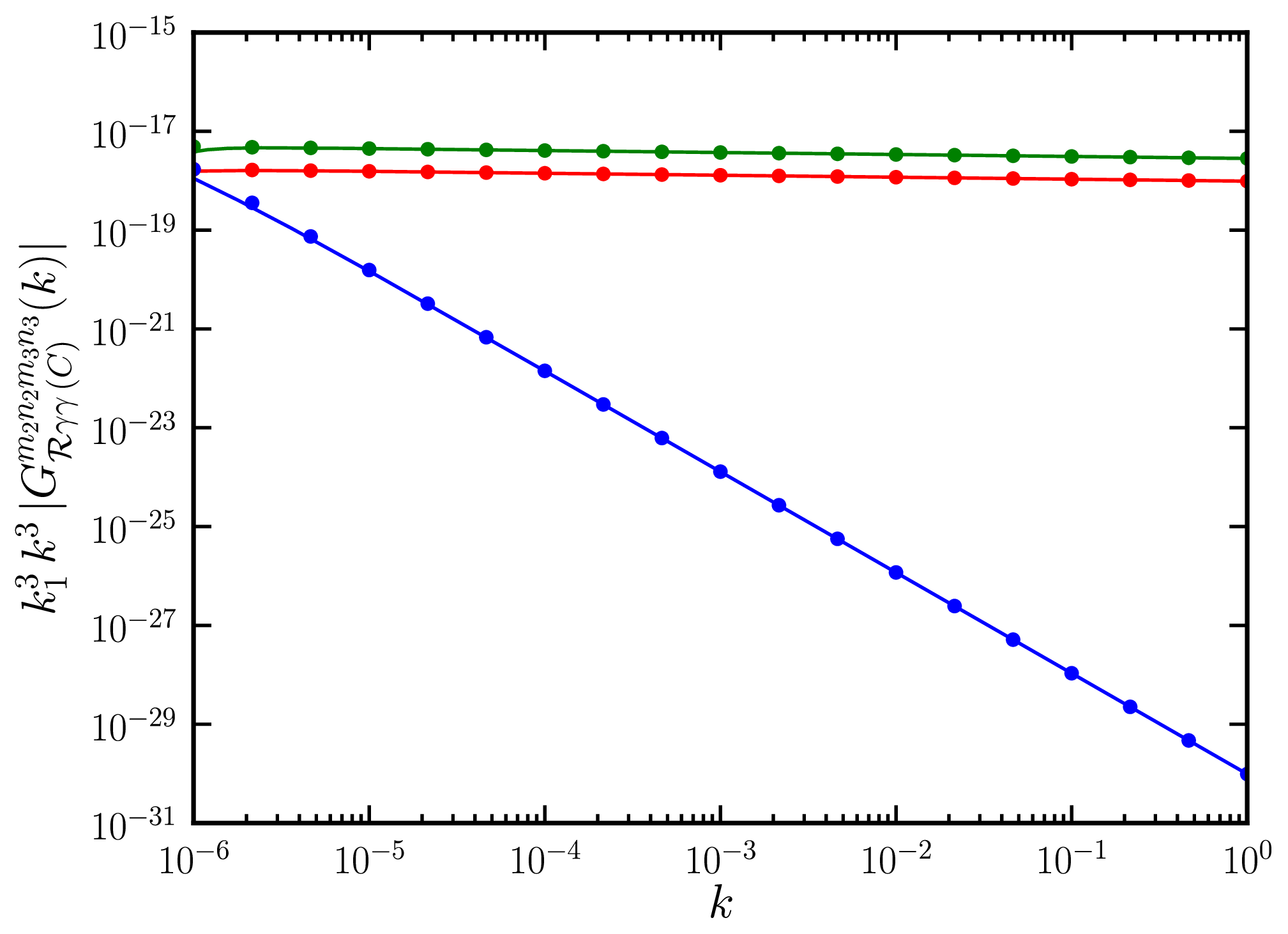} &
\hskip-5pt
\includegraphics[width=7.40cm]{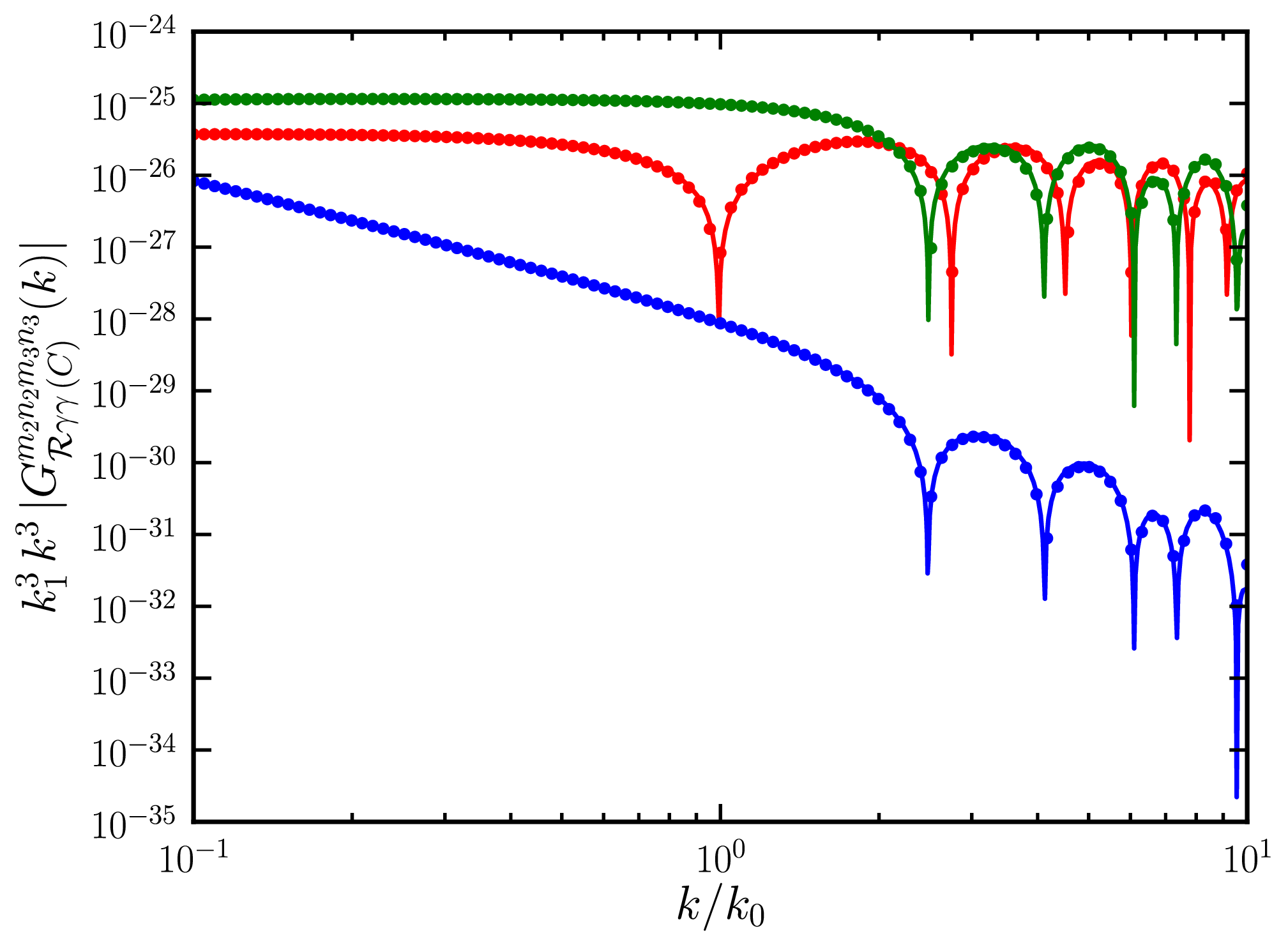}\\
\includegraphics[width=7.40cm]{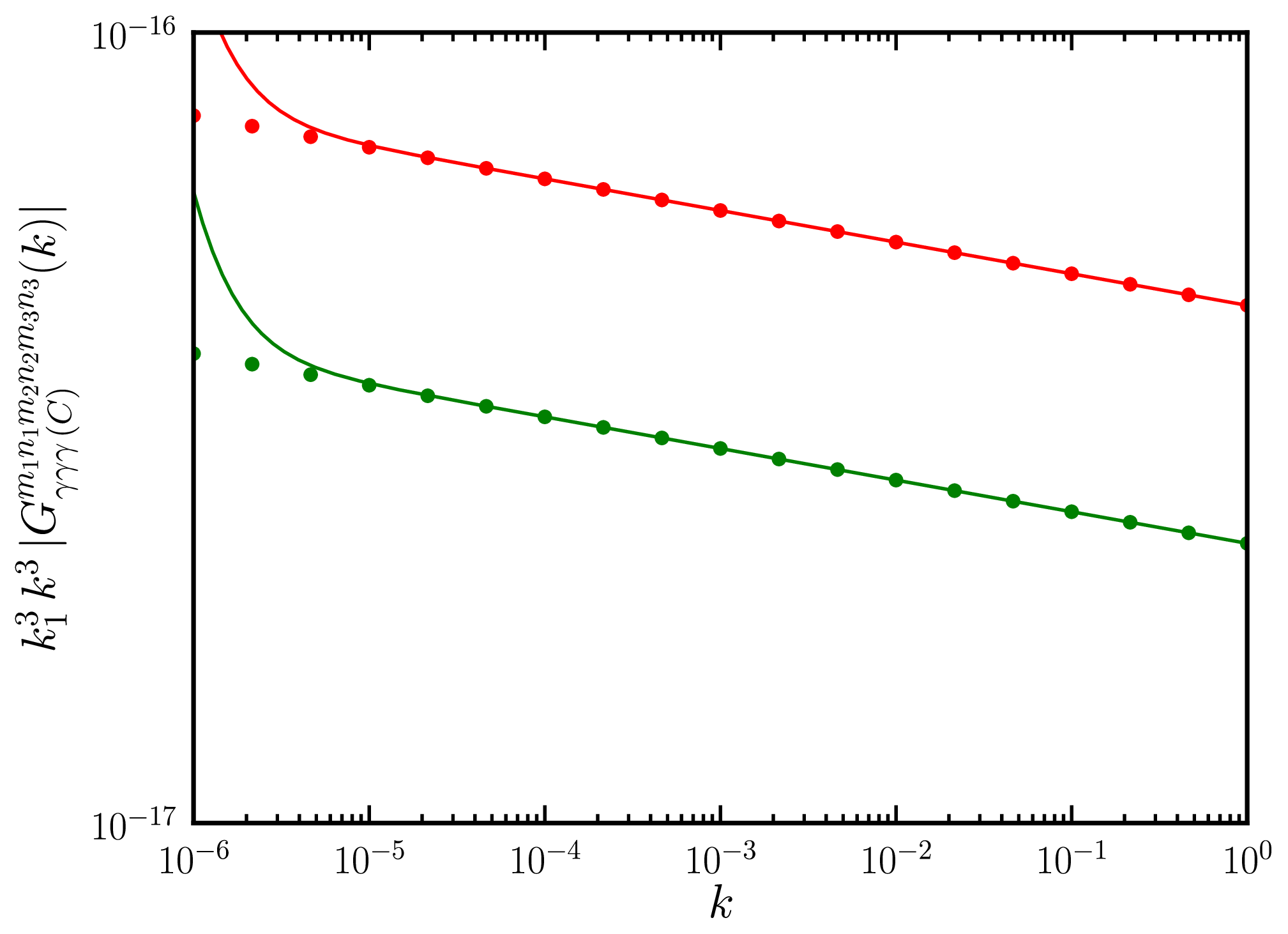} &
\hskip-5pt
\includegraphics[width=7.40cm]{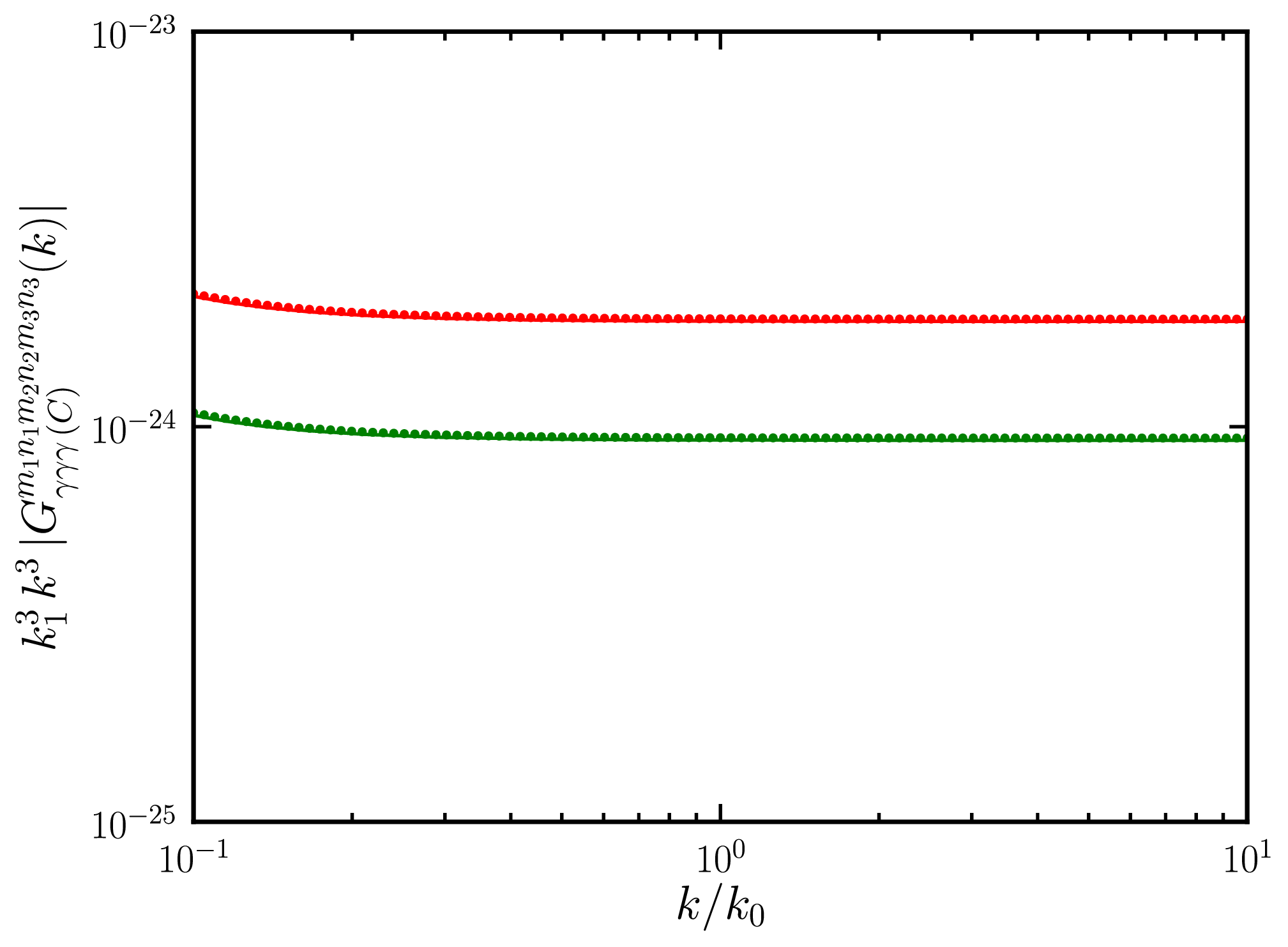}
\end{tabular}
\vskip -10pt
\caption{\label{fig:pl-sm-sl}
A comparison of the analytic and the numerical results in the
squeezed limit for the same set of quantities and models as in 
the previous figure.
Note that, in arriving at the theoretical spectral dependences in
the squeezed limit, we have taken the wavenumber of the tensor mode 
to zero in the case of $G_{\cR\cR\g}^{m_3n_3}$ and we have assumed 
that the wavenumber of the scalar mode vanishes in the case of 
$G_{\cR\g\g}^{m_2n_2m_3n_3}$.
Clearly, the numerical results match the analytical results quite 
well in the equilateral limit.
However, in the squeezed limit, while the match is good at large $k$, 
there is a noticeable difference between the theoretical and the 
numerical results at small $k$ in some cases.
This difference essentially arises due to the fact that, while the 
theoretical results have been arrived at by assuming that one of the
wavenumbers (either $k_1$ or $k_3$) vanishes, it is impossible 
to set a wavenumber to be zero numerically and one has to work with 
a suitably small value that permits the evolution of the modes as well 
as the evaluation of the integrals involved.
We have chosen the value of the large scale mode $k_1$ or $k_3$ to be
$8.3\times 10^{-7}$ and $4.3\times10^{-2}\, k_0$ in the power law case and the 
Starobinsky model, respectively.
If needed, the match can be improved by working with a smaller wavenumber, 
but the effort can become numerically taxing.}
\end{center}
\end{figure}

In fact, in power law inflation, we find that the spectral dependence of all 
the contributions can also be arrived at in the squeezed limit, which corresponds 
to setting two of the wavenumbers to be the same, while allowing the third to
vanish. 
Note that, as far as the cross-correlations go, in the squeezed limit, there 
exist two possibilities.
We can either consider the limit wherein the wavenumber of a scalar mode goes 
to zero or we can consider the situation wherein the wavenumber of a tensor 
mode vanishes. 
We obtain the following behavior for $G_{\cR\cR\g\, (C)}^{m_3n_3}(\vka,\vkb,\vkc)$
when $k_1=k_2=k$ and $k_3\to 0$ (i.e. when the wavenumber of the tensor mode 
vanishes):
\begin{subequations}
\begin{eqnarray}
k^{3}\,k_3^{3}\,G_{\cR\cR\g\, (1)}^{m_3n_3}(k,k_3)
&\propto& k^{2\,(\gamma+2)}\, k_3^{2\,(\gamma+2)},\\
k^{3}\,k_3^{3}\,G_{\cR\cR\g\, (2)}^{m_3n_3}(k,k_3)
&\propto& k^{2\,(\gamma+1)}\, k_3^{2\,(\gamma+3)},\\ 
k^{3}\,k_3^{3}\,G_{\cR\cR\g\, (3)}^{m_3n_3}(k,k_3)
&\propto& k^{2\,(\gamma+1)}\, k_3^{2\,(\gamma+3)},
\end{eqnarray}
\end{subequations}
whereas we find that all the terms have the following spectral dependence 
as $k_1\to 0$ (i.e. as the wavenumber of a scalar mode goes to zero) and 
$k_2=k_3=k$:
\begin{equation}
k_1^{3}\,k^{3}\,G_{\cR\cR\g\, (C)}^{m_3n_3}(k_1,k)
\propto k_1^{2\,\gamma+5}\, k^{2\,\gamma+3}.
\end{equation}
Similarly, in the case of $G_{\cR\g\g\, (C)}^{m_2n_2m_3n_3}(\vka,\vkb,\vkc)$,
when $k_2=k_3=k$ and $k_1\to 0$ (i.e. when the wavenumber of the scalar mode 
vanishes), we obtain that
\begin{subequations}
\begin{eqnarray}
k_1^{3}\,k^{3}\, G_{\cR\g\g\, (1)}^{m_2n_2m_3n_3}(k_1,k)
&\propto& k_1^{2\,(\gamma+2)}\, k^{2\,(\gamma+2)},\\ 
k_1^{3}\,k^{3}\, G_{\cR\g\g\, (2)}^{m_2n_3m_3n_3}(k_1,k)
&\propto& k_1^{2\,(\gamma+2)}\, k^{2\,(\gamma+2)},\\
k_1^{3}\,k^{3}\, G_{\cR\g\g\, (3)}^{m_2n_3m_3n_3}(k_1,k)
&\propto& k_1^{2\,(\gamma+3)}\, k^{2\,(\gamma+1)},
\end{eqnarray}
\end{subequations}
whereas we find that all the terms have the following spectral dependence 
when $k_1=k_2=k$ and $k_3\to 0$ (i.e. as the wavenumber of the tensor mode 
goes to zero):
\begin{equation}
k^{3}\,k_{3}^{3}\,G_{\cR\g\g\, (C)}^{m_2n_2m_3n_3}(k,k_3)
\propto k^{2\,(\gamma+1)}\, k_3^{2\,(\gamma+3)}.
\end{equation}
Lastly, one can show that, in power law inflation, in the squeezed limit, say, 
when $k_2=k_3=k$ and $k_1 \to 0$, the two contributions to the tensor bi-spectrum 
behave as
\begin{equation}
k_1^{3}\,k^{3}\,G_{\g\g\g\, (C)}^{m_1n_1m_2n_2m_3n_3}(k_1,k) 
\propto k_1^{2\,(\g+2)}\,k^{2\,(\gamma+2)}.
\end{equation}
In Figs.~\ref{fig:pl-sm-el} and~\ref{fig:pl-sm-sl}, we have compared the 
spectral dependences we have obtained above in the equilateral and the 
squeezed limits for all the different contributions to the three-point 
functions of interest with the corresponding numerical results. 
We find the agreement between the analytical and the numerical results 
to be quite good (about $1$-$2\%$, as we have alluded to before).


\subsubsection{Comparison in the case of the Starobinsky 
model}\label{subsec:sm}

The Starobinsky model is characterized by a linear potential with a sharp 
change in slope at a specific point~\cite{starobinsky-1992}.
The potential that governs the model is given by
\begin{equation} \label{starobinsky potential}
V(\phi)
=\left\{ \begin{array}{rcl}                  
V_0 + A_+(\phi -\phi_0) &{\rm for}& \phi > \phi_0,\\  
V_0 + A_-(\phi -\phi_0) &{\rm for}& \phi < \phi_0,
\end{array}\right.
\end{equation}
where $V_0$, $A_+$, $A_-$ and $\phi_0$ are constants.
Evidently, the derivative of the potential contains a discontinuity 
at $\phi_0$.
The discontinuity leads to a brief period of fast roll as the field 
crosses the point, before slow roll is restored again at late times.
It is assumed that the constant term $V_0$ in the potential is dominant 
as the transition across the discontinuity takes place.
Hence, the scale factor always behaves as that of de Sitter with the 
constant Hubble parameter, say, $H_0$, being given by $H_0^2\simeq 
V_0/(3\, \Mp^2)$ (for recent discussions on the evolution of the 
background as well as the perturbations, see 
Refs.~\cite{martin-2012a,arroja-2011-2012}).

\par 

Note that the only background quantity required to evaluate the tensor 
bi-spectrum is the scale factor $a$ [cf.~Eqs.~(\ref{eq:cGggg})]. 
Since, in the Starobinsky model, the scale factor is always of the de 
Sitter form, i.e. $a(\eta)=-1/(H_0\, \eta)$, the tensor modes remain
unaffected by the transition.
As a result, the tensor bi-spectrum that one arrives at in this case 
is essentially the same as the one obtained in the slow roll 
approximation (to be precise, in the de Sitter limit). 
As far as the three-point cross-correlations are concerned, we require, 
apart from the scale factor, the behavior of the first slow roll parameter 
$\epsilon_1$ as well [cf.~Eqs.~(\ref{eq:cGrrg}) and (\ref{eq:cGrgg})].
Let us denote the various quantities before and after the transition by 
the sub-scripts (or super-scripts, as is convenient) plus and minus, 
respectively.
One finds that the behavior of the first slow roll parameter $\epsilon_1$ 
can be expressed as~\cite{martin-2012a,arroja-2011-2012} 
\begin{subequations}
\begin{eqnarray}\label{eq:srp1}
\epsilon_{1+} 
& \simeq & \f{A_+^2}{18\,\Mp^2\,H_0^4},\label{srpm}\\ 
\epsilon_{1-} 
& \simeq & \f{A_-^2}{18\,\Mp^2\,H_0^4}\,
\l[1+\rho^3\,\eta^3\r]^2,\label{eq:e1m}
\end{eqnarray}
\end{subequations}
where $\rho^3= -(\Delta A/A_-)\,\,(1/\eta_0)^3$, with $\Delta A = A_- - A_+$ 
and $\eta_0$ being the conformal time at the transition.
Actually, as we shall see below, the derivative of the scalar modes which
are required to evaluate the three-point functions also involve the second
slow roll parameter $\epsilon_2=\d\, {\rm ln}\, \epsilon_1/\d N$.
It can be shown that the second slow roll parameter behaves as 
follows~\cite{martin-2012a,arroja-2011-2012}:
\begin{subequations}
\begin{eqnarray}\label{eq:srp2} 
\epsilon_{2+} &\simeq & 4\,\epsilon_{1+},\\ 
\epsilon_{2-} &\simeq & \f{-\,6\,\rho^3\,\eta^3}{1+\rho^3\,\eta^3}.
\label{eq:e2m}
\end{eqnarray}
\end{subequations} 

\par

Evidently, we also require the scalar and the tensor modes, $f_k$ and $g_k$,
as well as their derivatives with respect to the conformal time, in order to 
arrive at the three-point functions.
As we have already mentioned, since the scale factor remains unaffected by
the transition, the tensor modes are given by the standard Bunch-Davies
solutions in the de Sitter spacetime, viz.
\begin{equation}
g_k(\eta) = \f{i\,\sqrt{2}\;H_0}{\Mpl\,\sqrt{2\,k^3}}\,
\l(1+i\,k\,\eta\r)\, {\rm e}^{-i\,k\,\eta},\label{eq:gk-ds}
\end{equation}
the time derivative of which is straightforward to evaluate.
The scalar modes $f_k$ before and after the transition can be 
expressed as~\cite{martin-2012a,arroja-2011-2012}:
\begin{subequations}
\begin{eqnarray}
f_k^{+}(\eta)
&=&\frac{i\, H_0}{2\, \Mp\, \sqrt{{k^3}\,\epsilon_{1+}}}\,
\l(1+i\,k\,\eta\right)\,{\rm e}^{-i\,k\,\eta},\label{eq:fk-bt}\\
f_k^{-}(\eta)
&=&\frac{i\,H_0\,\alpha_k}{2\,\Mp\,\sqrt{{k^3}\,\epsilon_{1-}}}\,
\l(1+i\,k\,\eta\r)\, {\rm e}^{-i\,k\,\eta}
-\frac{i\,H_0\,\beta_k}{2\,\Mp\,\sqrt{{k^3}\,\epsilon_{1-}}}
\l(1-i\,k\,\eta\right)\, {\rm e}^{i\,k\,\eta}.\label{eq:fk-at}
\end{eqnarray}
\end{subequations}
The derivatives of $f_k$ can be obtained to be, at the level of
the approximation one works in,
\begin{eqnarray}
f_k^{+}{}'(\eta)
&=&\frac{i\, H_0}{2\, \Mp\, \sqrt{{k^3}\,\epsilon_{1+}}}\;
k^2\,\eta\; {\rm e}^{-i\,k\,\eta},\label{eq:fkp-bt}\\
f_\vk^{-}{}'(\eta)
&=&\frac{i\,H_0\,\alpha_k}{2\,\Mp\,\sqrt{{k^3}\epsilon_{1-}}}
\l[\f{\epsilon_{2-}}{2\, \eta}\,\l(1+i\,k\,\eta\r)
+k^2\,\eta\r]\, {\rm e}^{-i\,k\,\eta}\nn\\
&-&\frac{i\,H_0\,\beta_k}{2\,\Mp\,\sqrt{{k^3}\epsilon_{1-}}}
\l[\f{\epsilon_{2-}}{2\,\eta}\,\l(1-i\,k\,\eta\right)
+k^2\,\eta\r]{\rm e}^{i\,k\,\eta}.\label{eq:fkp-at}
\end{eqnarray}
The quantities $\alpha_k$ and $\beta_k$ that appear in the above
expressions are the standard Bogoliubov coefficients, which are 
obtained by matching the modes $f_k$ and their derivatives $f_k'$ 
at the transition.
They are found to be~\cite{starobinsky-1992,martin-2012a,arroja-2011-2012}
\begin{subequations}
\begin{eqnarray}
\alpha_\vk 
&=& 1+\frac{3\,i\,\Delta A}{2\,A_{+}}\;\frac{k_0}{k}\,
\left(1+\frac{k_0^2}{k^2}\right),
\label{eq:alphak-sm}\\
\beta_\vk 
&=& -\frac{3\,i\,\Delta A}{2\,A_+}\;\f{k_0}{k}\,
\l(1+\frac{i\, k_0}{k}\r)^2\, {\rm e}^{2\,i\,k/k_{0}},
\label{eq:betak-sm}
\end{eqnarray}
\end{subequations}
with $k_0 = -1/\eta_0 = a_0\,H_0$ and $a_0$ being the value of the
scale factor at the transition.

\par 

We have already mentioned that, in the Starobinsky model, the tensor 
bi-spectrum will essentially be the same as the one arrived at in the 
slow roll approximation (in this context, see, for instance, 
Refs.~\cite{maldacena-2003,tensor-bs}).
Note that since the scalar modes (and the first two slow roll parameters) 
behave differently before and after the transition, while evaluating the 
scalar-tensor cross-correlations, one needs to divide the integrals 
involved into two, and carry out the integrals before and after the 
transition separately, just as it was done in the context of the scalar 
bi-spectrum~\cite{martin-2012a,arroja-2011-2012}.
We find that the cross-correlations can be evaluated completely analytically
for an arbitrary triangular configuration of the wavenumbers (which, in fact, 
proves to be difficult to carry out for the scalar bi-spectrum).
Since the calculations and the expressions involved prove to be rather 
long and cumbersome, we have relegated the calculations to the appendix.
In Figs.~\ref{fig:pl-sm-el} and~\ref{fig:pl-sm-sl}, we have compared the 
analytic results we have obtained with the corresponding numerical results 
for the cross-correlations and the tensor bi-spectrum in the equilateral 
and the squeezed limits.
We should mention here that, in order to solve the problem numerically, the 
discontinuity in the potential of the Starobinsky model has been suitably 
smoothened~\cite{hazra-2013}.
The figures suggest that the match between the analytic and the numerical
results is very good. 


\subsubsection{The case of the quadratic potential}

\begin{figure}[!htb]
\begin{center}
\begin{tabular}{cc}
\hskip -10pt
\includegraphics[width=7.750cm,angle=0]{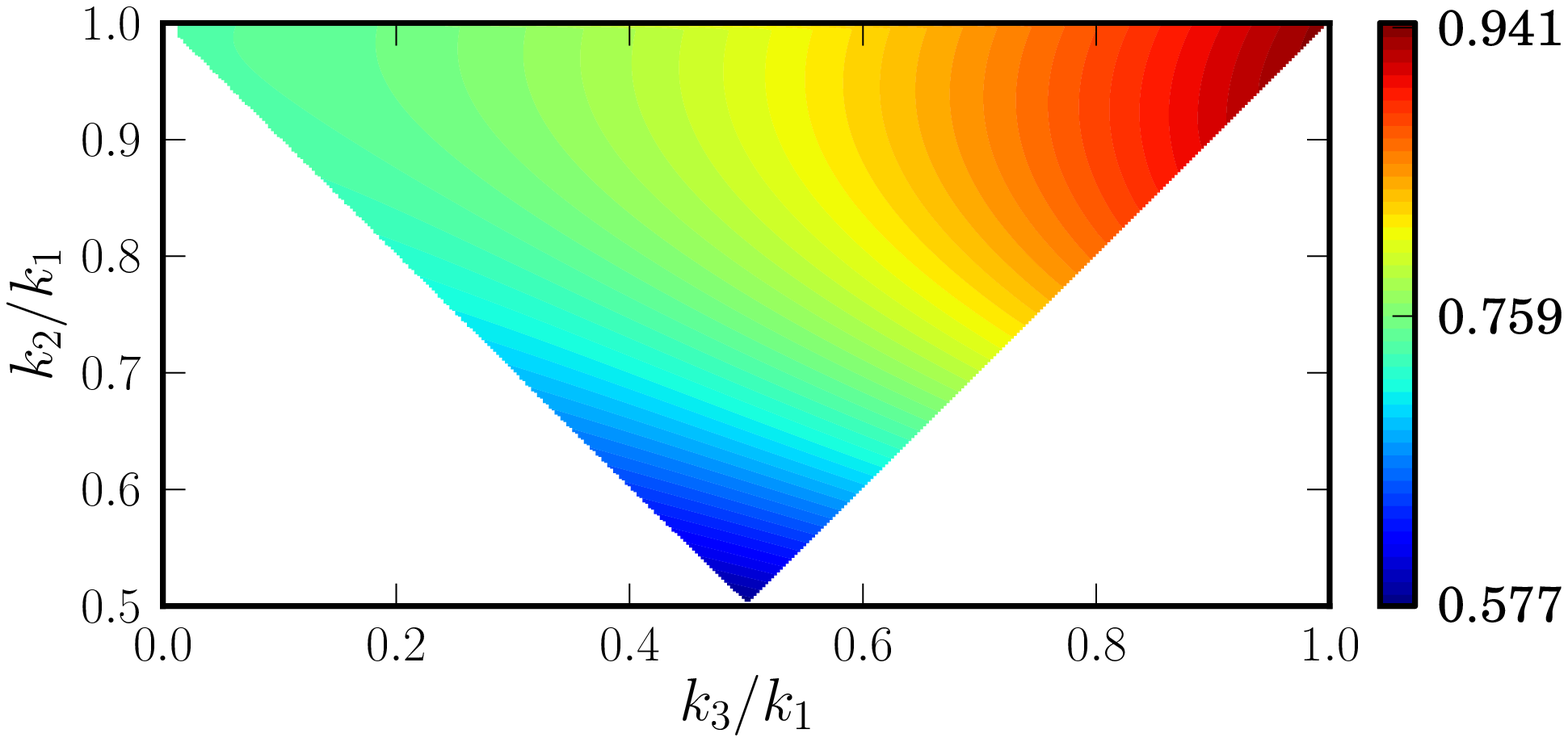} &
\hskip -10pt
\includegraphics[width=7.750cm,angle=0]{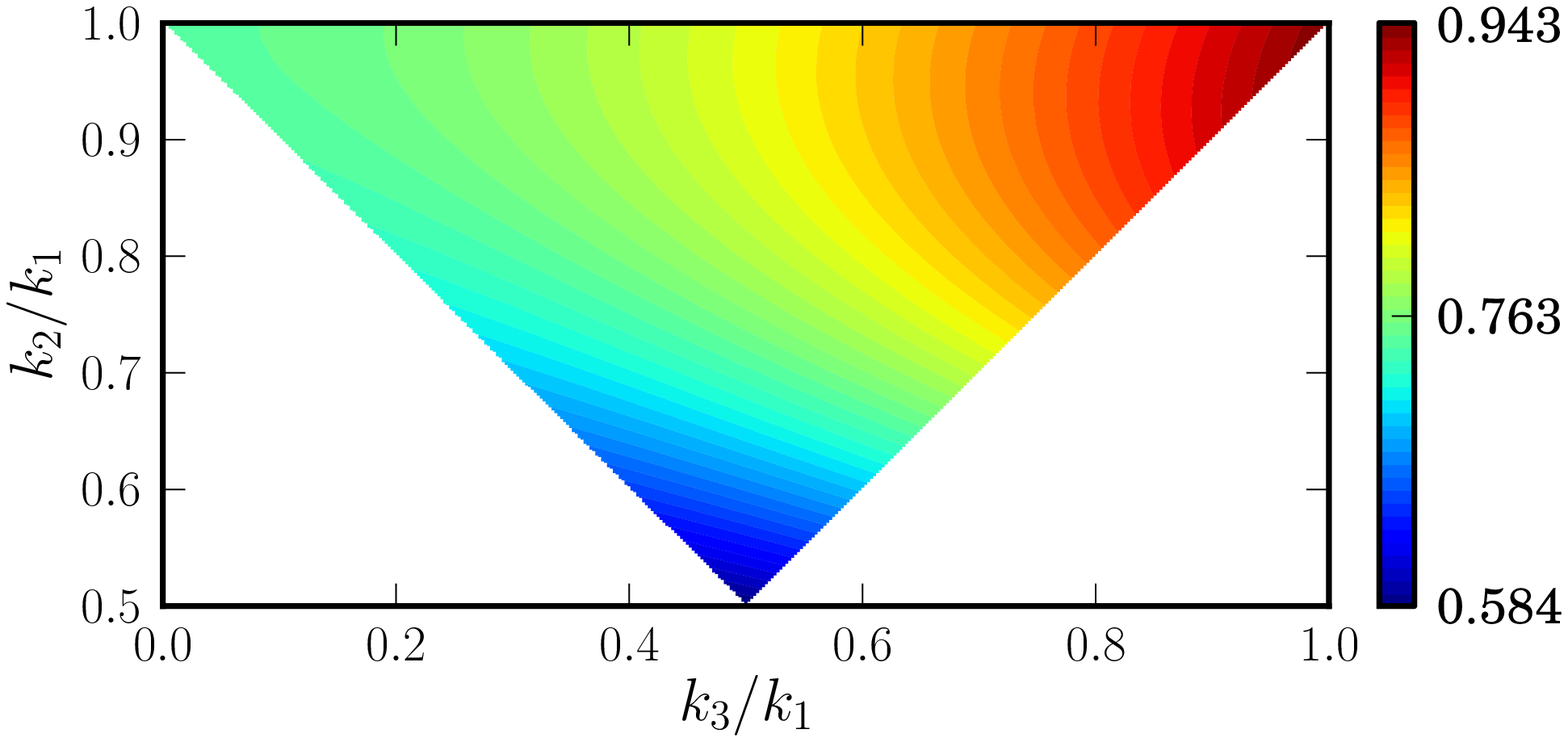} \\
\hskip -10pt
\includegraphics[width=7.750cm,angle=0]{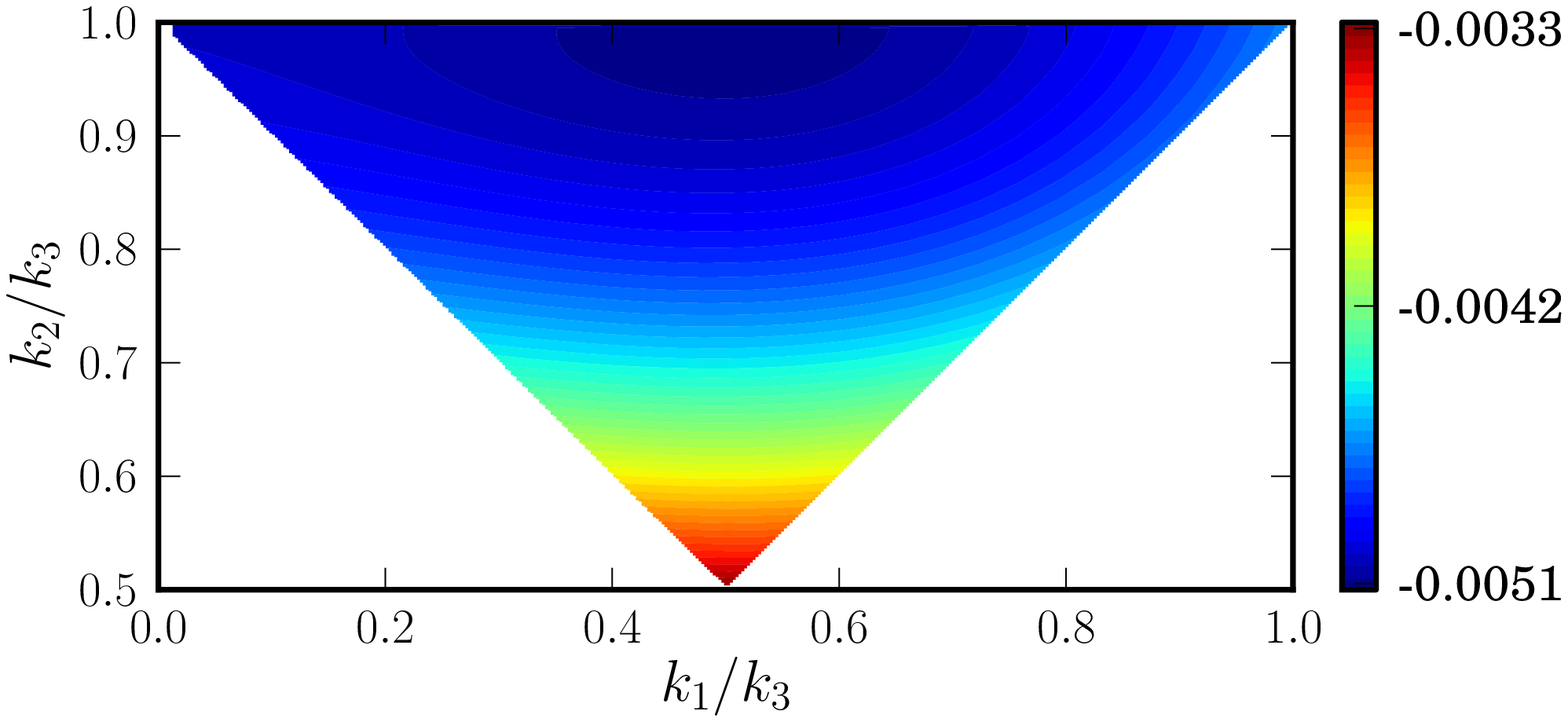} &
\hskip -10pt
\includegraphics[width=7.750cm,angle=0]{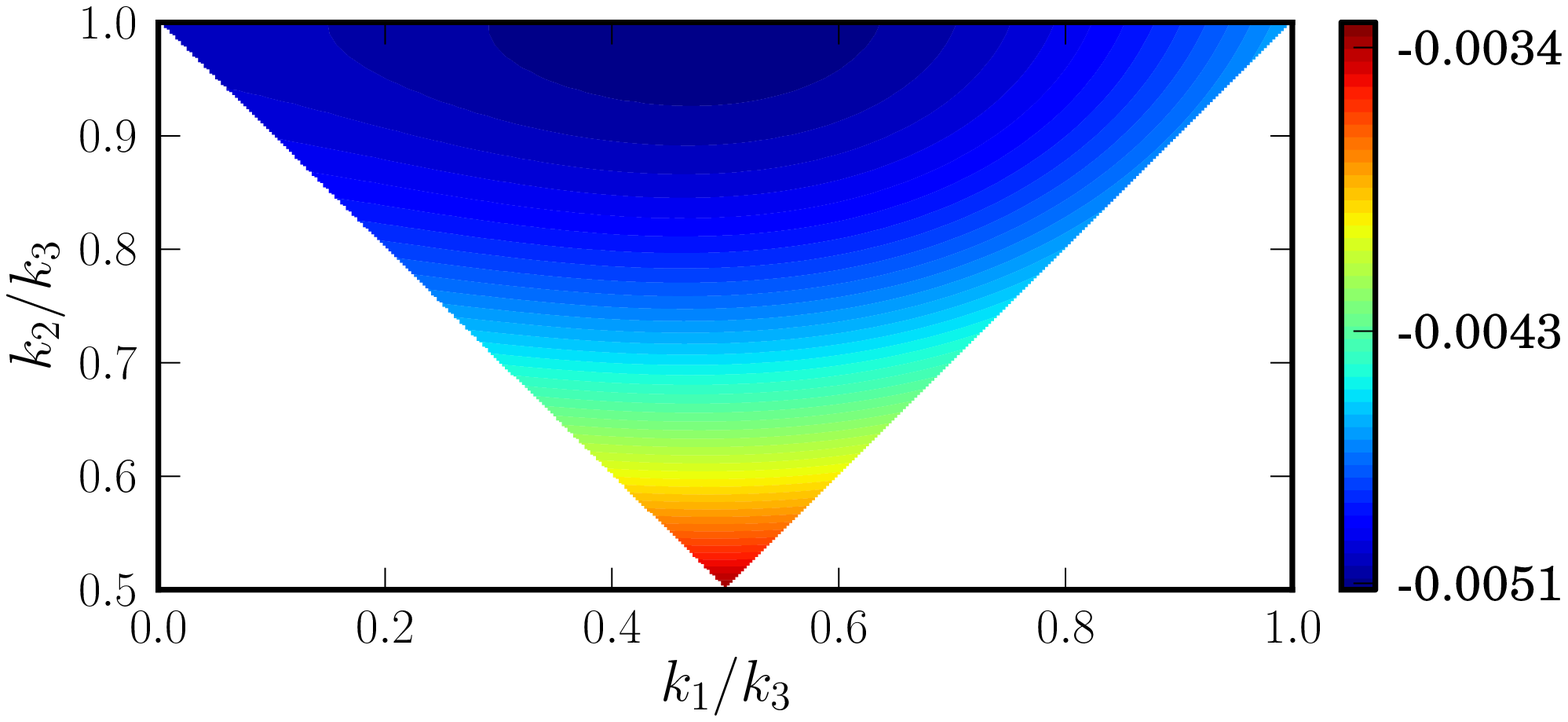} \\
\hskip -10pt
\includegraphics[width=7.750cm,angle=0]{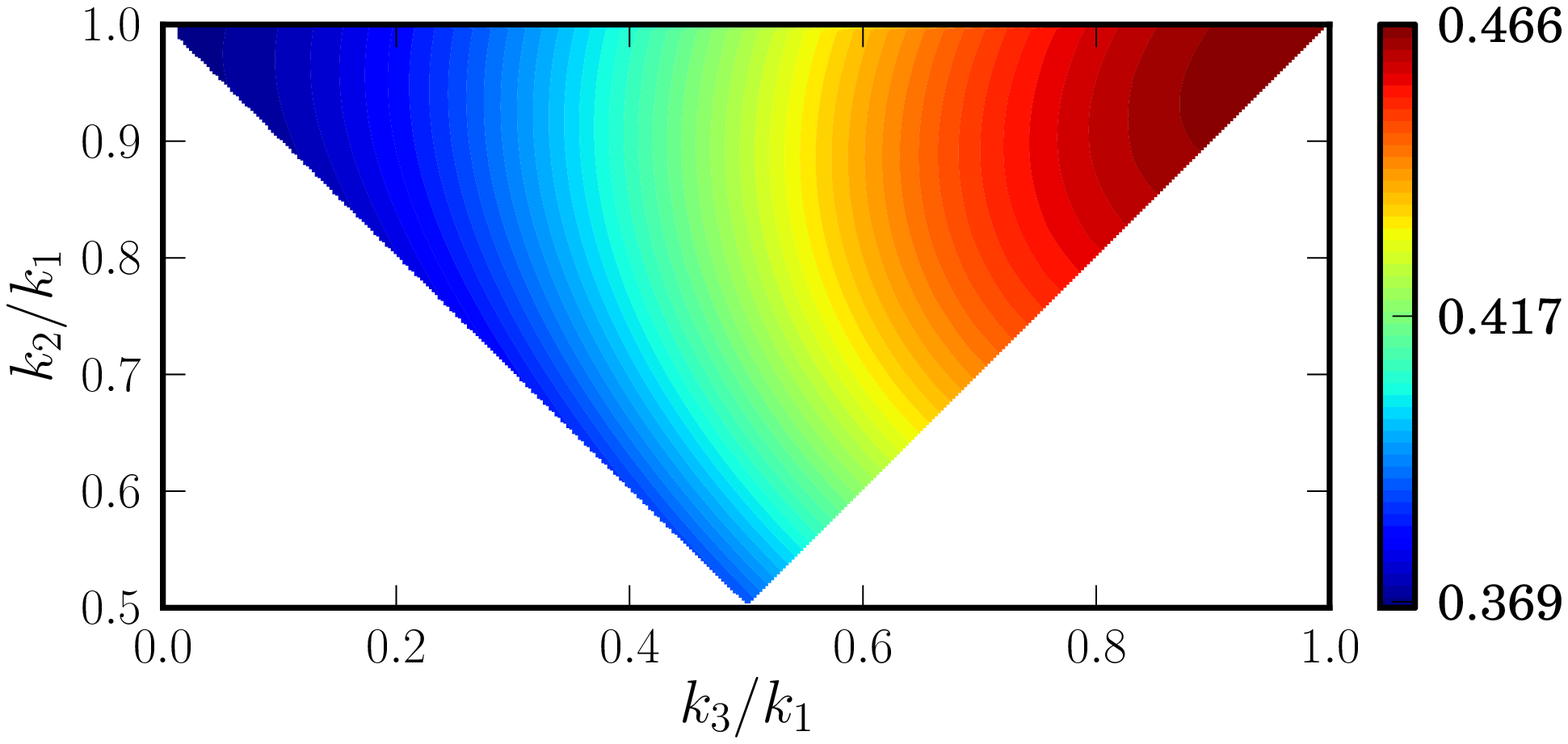} &
\hskip -10pt
\includegraphics[width=7.750cm,angle=0]{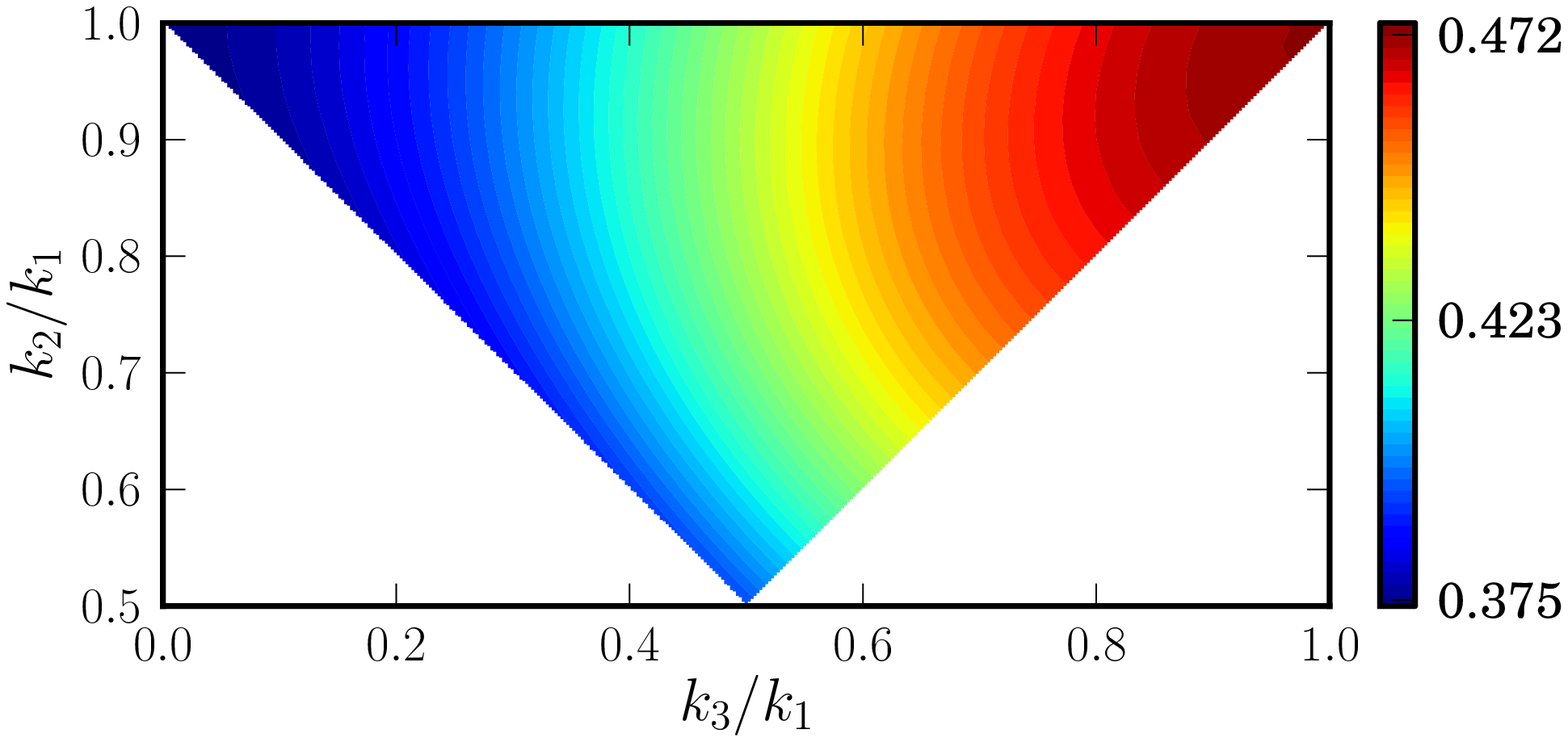} 
\end{tabular}
\caption{\label{fig:qp-atc-anr}
Density plots of the non-Gaussianity parameters $\cnls$ (on top), $\cnlt$
(in the middle) and $\hnl$ (at the bottom) for an arbitrary triangular 
configuration of the wavenumbers for the case of the conventional, quadratic 
potential.
In arriving at the above figures, when $k_1$ and $k_3$ appear in the denominators
of the two axes, we have chosen them to be $k_{\ast}$.
Evidently, the strong similarity between the numerical results (on the left)
and the corresponding quantities arrived at using the slow roll approximation
(on the right) indicates the robustness of the numerical procedure we have
adopted to compute the three-point functions.
We find that the numerical results match the analytical estimates to better 
than $1\%$ over a large domain of the wavenumbers of interest.}
\end{center}
\end{figure}

As is well known, the conventional quadratic potential leads to slow 
roll and, hence, in this case, one can utilize the three-point functions 
evaluated in the slow roll limit to compare with the numerical results.
For the sake of completeness, we shall write down here the entire expressions
for the non-Gaussianity parameters evaluated in the slow roll approximation.
We find that, if we ignore factors involving $\Pi_{mn,ij}^{\vk}$, they are 
given by
\begin{subequations}
\begin{eqnarray}
\cnls 
&=& \l(k_{2|\ast}^{n_{_{\rm S}}-1}\, k_{3|\ast}^{n_{_{\rm T}}} 
+ k_{2|1}^3\, k_{1|\ast}^{n_{_{\rm S}}-1}\, k_{3|*}^{n_{_{\rm T}}}\r)^{-1}\nn\\
& &\times\,\l[k_{2|1}\,k_{{_{\rm T}}|1}\,
\l(1 - \f{k_{2|1} + k_{3|1} + k_{2|1}\, k_{3|1}}{k_{{_{\rm T}}|1}^2}
- \f{k_{2|1}\,k_{3|1}}{k_{{_{\rm T}}|1}^3}\r)
- \epsilon_1\,\f{k_{2|1}\, k_{3|1}^2}{k_{{_{\rm T}}|1}}\r],\\
&=& \l(k_{1|3}^3\,k_{2|\ast}^{n_{_{\rm S}}-1}\, k_{3|\ast}^{n_{_{\rm T}}} 
+ k_{2|3}^3\, k_{1|\ast}^{n_{_{\rm S}}-1}\, k_{3|*}^{n_{_{\rm T}}}\r)^{-1},\nn\\
& &\times\,\l[k_{1|3}\,k_{2|3}\,k_{{_{\rm T}}|3}\,
\l(1 - \f{k_{1|3} + k_{2|3} + k_{1|3}\, k_{2|3}}{k_{{_{\rm T}}|3}^2}
- \f{k_{1|3}\,k_{2|3}}{k_{{_{\rm T}}|3}^3}\r)
- \epsilon_1\,\f{k_{1|3}\, k_{2|3}}{k_{{_{\rm T}}|3}}\r],\quad\\
\cnlt 
&=& \f{\epsilon_1}{4}\,
\l(k_{3|1}^3\,k_{1|\ast}^{n_{_{\rm S}}-1}\, k_{2|\ast}^{n_{_{\rm T}}} 
+ k_{2|1}^3\, k_{1|\ast}^{n_{_{\rm S}}-1}\, k_{3|*}^{n_{_{\rm T}}}\r)^{-1}\,
\l(1-k_{2|1}^2-k_{3|1}^2-\f{8\,k_{2|1}^2\, k_{3|1}^2}{k_{{_{\rm T}}|1}}\r),\\
&=& \f{\epsilon_1}{4}\,
\l(k_{1|\ast}^{n_{_{\rm S}}-1}\, k_{2|\ast}^{n_{_{\rm T}}} 
+ k_{2|3}^3\, k_{1|\ast}^{n_{_{\rm S}}-1}\, k_{3|*}^{n_{_{\rm T}}}\r)^{-1}\,
\l(k_{1|3}^3-k_{1|3}\, \l(1+k_{2|3}^2\r)-\f{8\,k_{2|3}^2}{k_{{_{\rm T}}|3}}\r),\\
\hnl
&=& \l(k_{2|*}^{n_{_{\rm T}}}\, k_{3|*}^{n_{_{\rm T}}}\,
+ k_{2|1}^3\, k_{1|*}^{n_{_{\rm T}}}\, k_{3|*}^{n_{_{\rm T}}}
+ k_{3|1}^3\, k_{1|*}^{n_{_{\rm T}}}\, k_{2|*}^{n_{_{\rm T}}}\r)^{-1}\,
\l(1+k_{2|1}+k_{3|1}\r)\,\l(1+k_{2|1}^2+k_{3|1}^2\r)\nn\\
& &\times\,
\l(1-\f{k_{2|1}+k_{3|1}+k_{2|1}\, k_{3|1}}{k_{{_{{\rm T}}|1}^2}}
-\f{k_{2|1}\,k_{3|1}}{k_{{_{\rm T}}|1}^3}\r),
\end{eqnarray}
\end{subequations}
where $k_{i|j}=k_i/k_j$, $k_{i|\ast}=k_i/k_\ast$, $k_{{_{\rm T}}|1}=1
+k_{2|1}+k_{3|1}$ and $k_{_{{\rm T}}|3}=k_{1|3}+k_{2|3}+1$.
Recall that, in the slow roll approximation, $n_{_{\rm S}}=1-2\, \epsilon_1
-\epsilon_2$, while $n_{_{\rm T}}=-2\, \epsilon_1$.
In Fig.~\ref{fig:qp-atc-anr}, we have plotted the above analytical results
for the non-Gaussianity parameters and the corresponding numerical results
for an arbitrary triangular configuration of the wavenumbers for the case
of the quadratic potential.
There is clearly a striking similarity between the structure of the numerical 
results and the corresponding analytical estimates.
We find that the numerical and analytical results match to better than $1\%$
over a large region of the wavenumbers involved.


\section{The three-point functions in models leading to features 
in the scalar power spectrum}\label{sec:mwf}

As we had discussed in the introduction, there has been considerable interest
in studying the possibility of features in the scalar power spectrum over the
last decade.
Specifically, a large amount of attention has been focused on models leading
to three types of features, viz. a sharp cut-off on large scales, a burst of
oscillations over an intermediate range of scales and small but repeated 
oscillations over a wide range of scales (in this context, see 
Refs.~\cite{pi,l2240,hazra-2010,benetti-2011,pso,pahud-2009,flauger-2010,kobayashi-2011,aich-2013,peiris-2013}).
And, not surprisingly, it is exactly such classes of models that have been 
considered by the Planck team~\cite{planck-2013-ci}.

\par

In this section, we shall utilize our code to study the behavior of the 
three-point functions of interest in models leading to deviations from 
slow roll.
We shall consider three different models that lead to features in the 
scalar power spectrum of the three types  mentioned above (see, in this 
context, Fig.~9 of Ref.~\cite{hazra-2013}).
The first of the models that we shall consider is the model described by 
the following potential:
\begin{equation}
V(\phi) = \f{m^2}{2}\,\phi^2 
-\f{\sqrt{2\,\lambda\,(n-1)}\, m}{n}\, \phi^n
+\f{\lambda}{4}\,\phi^{2\,(n-1)}.
\end{equation}
For suitable values of the parameters, this model leads to a brief period
of departure from inflation before slow roll is restored again, a scenario
that has been dubbed punctuated inflation~\cite{pi}.
Due to the sudden deviation from slow roll that one encounters, this model 
leads to sharp features in the scalar power as well as 
bi-spectra~\cite{pi,hazra-2013}.

\par

The second model that we shall consider is the one described by the
popular quadratic potential, but with an additional step that has been
introduced by hand.
The complete potential is given by the 
expression~\cite{l2240,hazra-2010,benetti-2011}
\begin{equation}
V(\phi) = \f{m^2}{2}\,\phi^2\,
\l[1 + \alpha\,\tanh\,\l(\f{\phi -\phi_0}{\Delta\phi}\r)\r],
\end{equation}
where, clearly, $\alpha$ and $\Delta \phi$ denote the height and the
width of the step, respectively, while $\phi_0$ represents its location.

\par

The last model that we shall consider is the so-called axion monodromy 
model that consists of a linear potential with super-imposed oscillations.
The potential is motivated by string theory and is given 
by~\cite{flauger-2010,aich-2013,peiris-2013}
\begin{equation}
V(\phi) = \lambda\, \l[\phi 
+ \alpha\, \cos\,\l(\f{\phi}{\beta} +\delta\r)\r].
\end{equation}

\par

\begin{figure}[!htb]
\begin{center}
\begin{tabular}{ccc}
\hskip -10pt
\includegraphics[width=5.25cm]{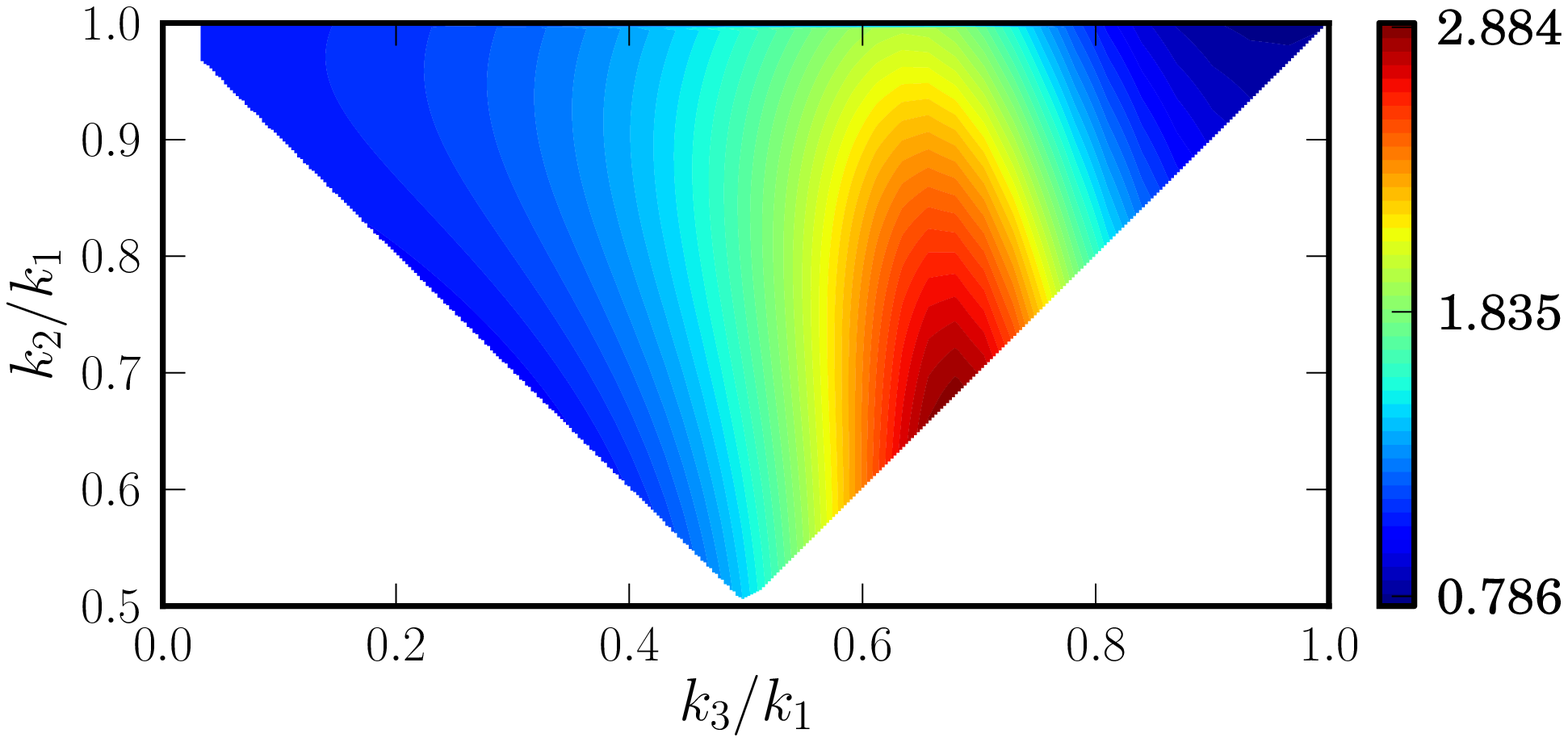} & 
\hskip -10pt
\includegraphics[width=5.25cm]{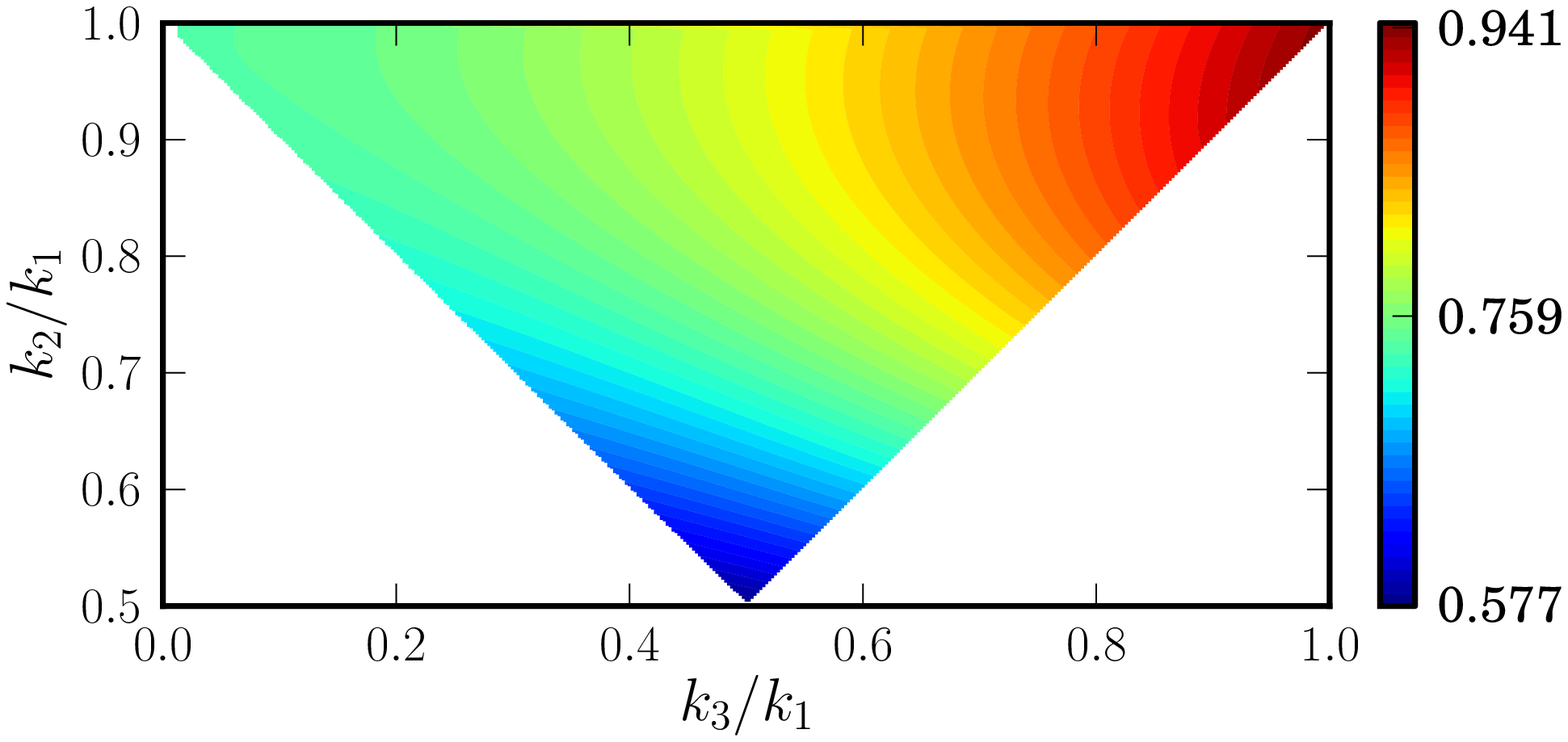} &
\hskip -10pt
\includegraphics[width=5.25cm]{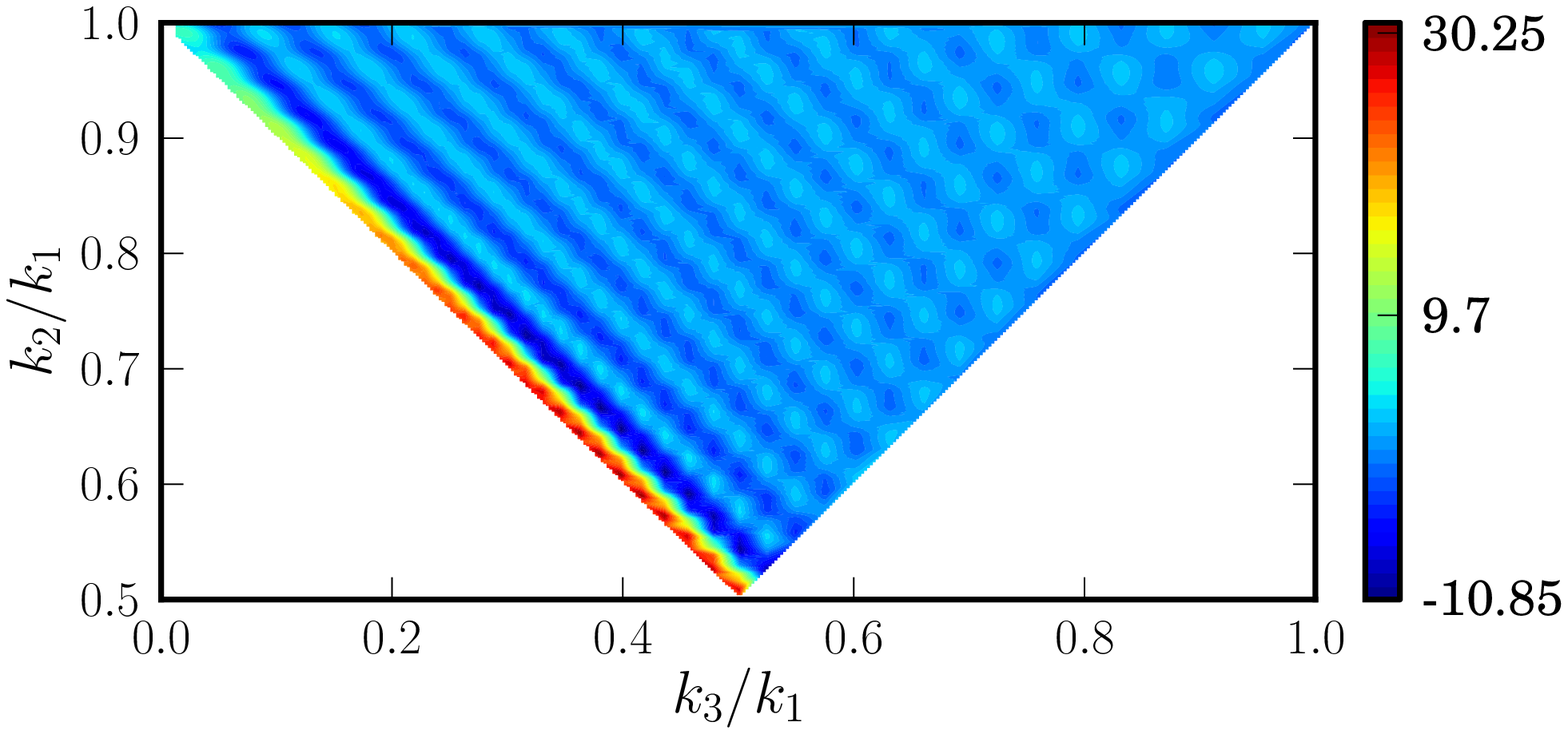} \\
\hskip -10pt
\includegraphics[width=5.25cm]{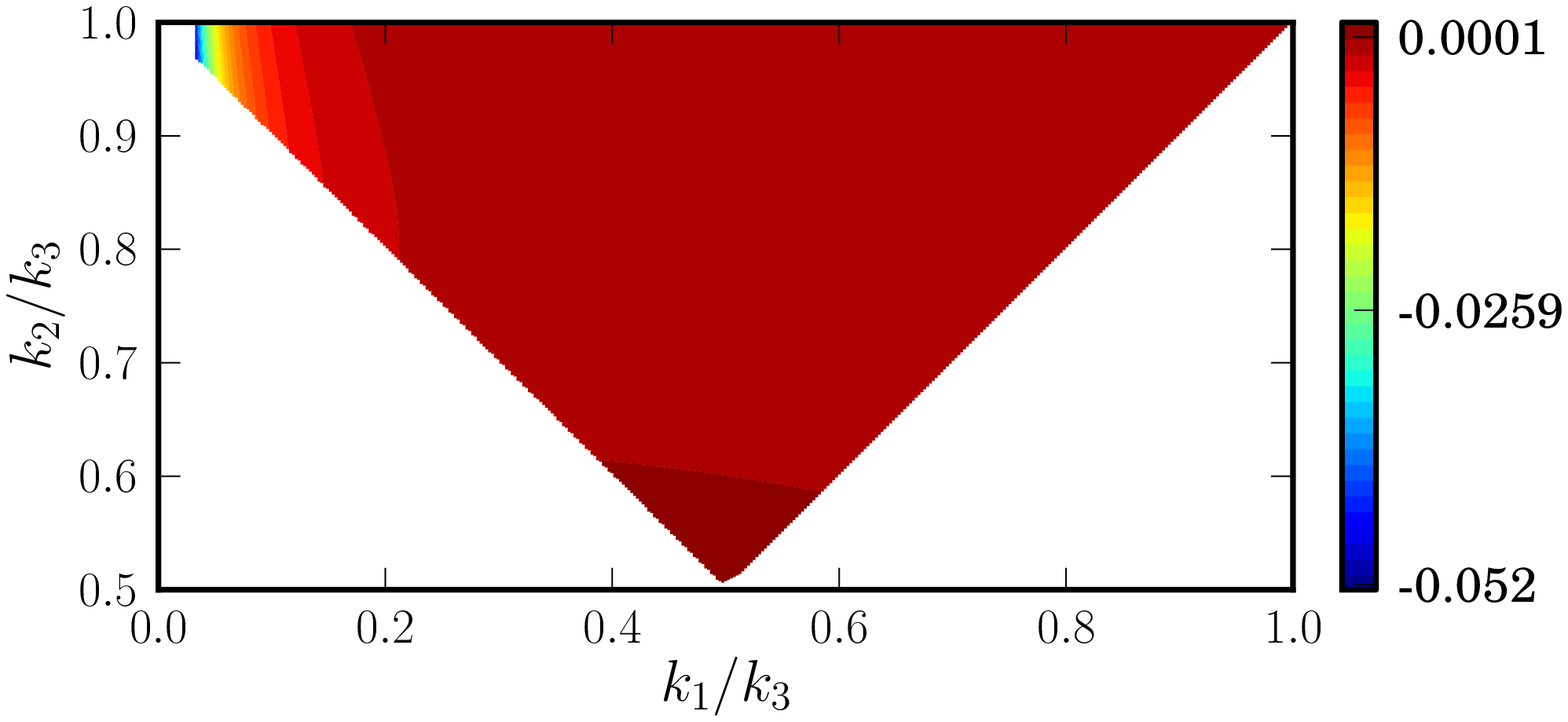} &
\hskip -10pt
\includegraphics[width=5.25cm]{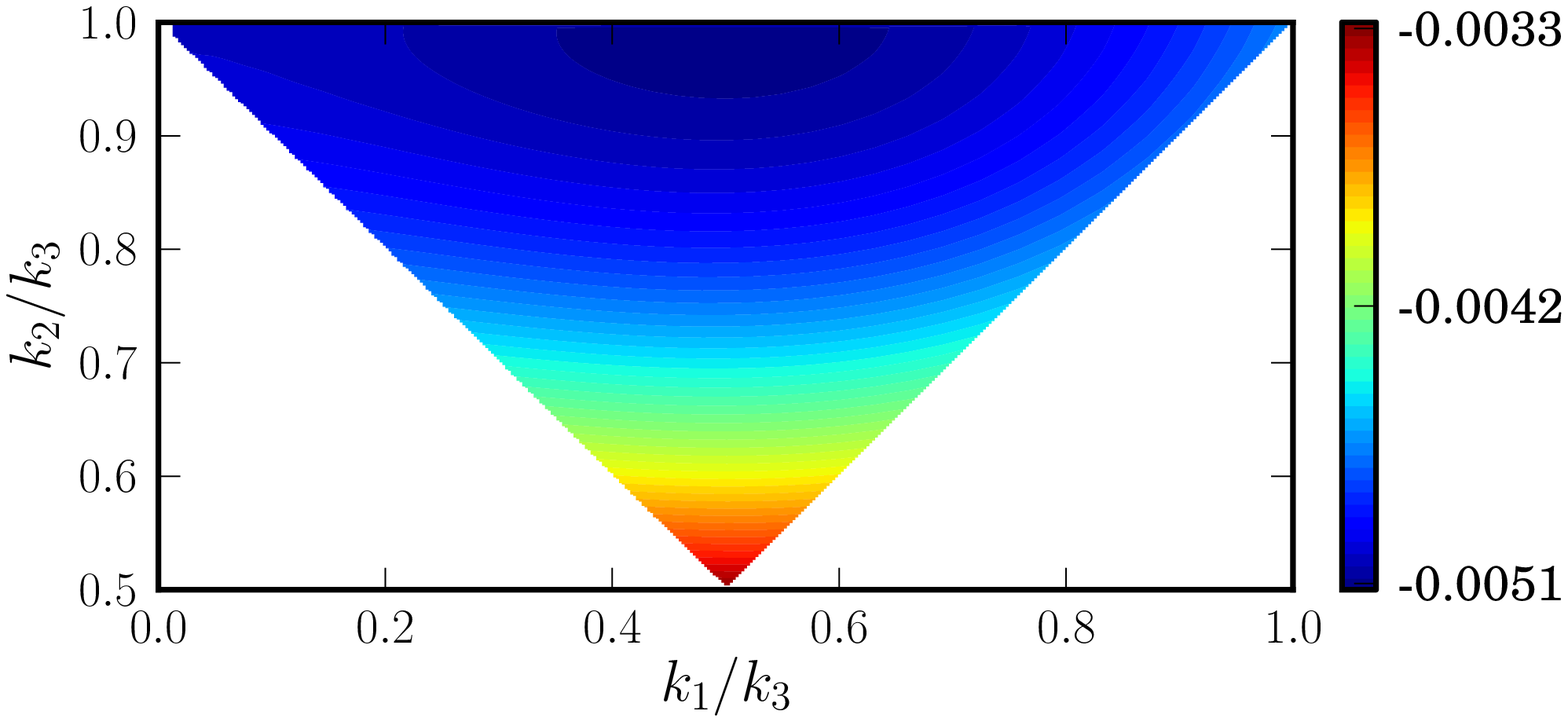} &
\hskip -10pt
\includegraphics[width=5.25cm]{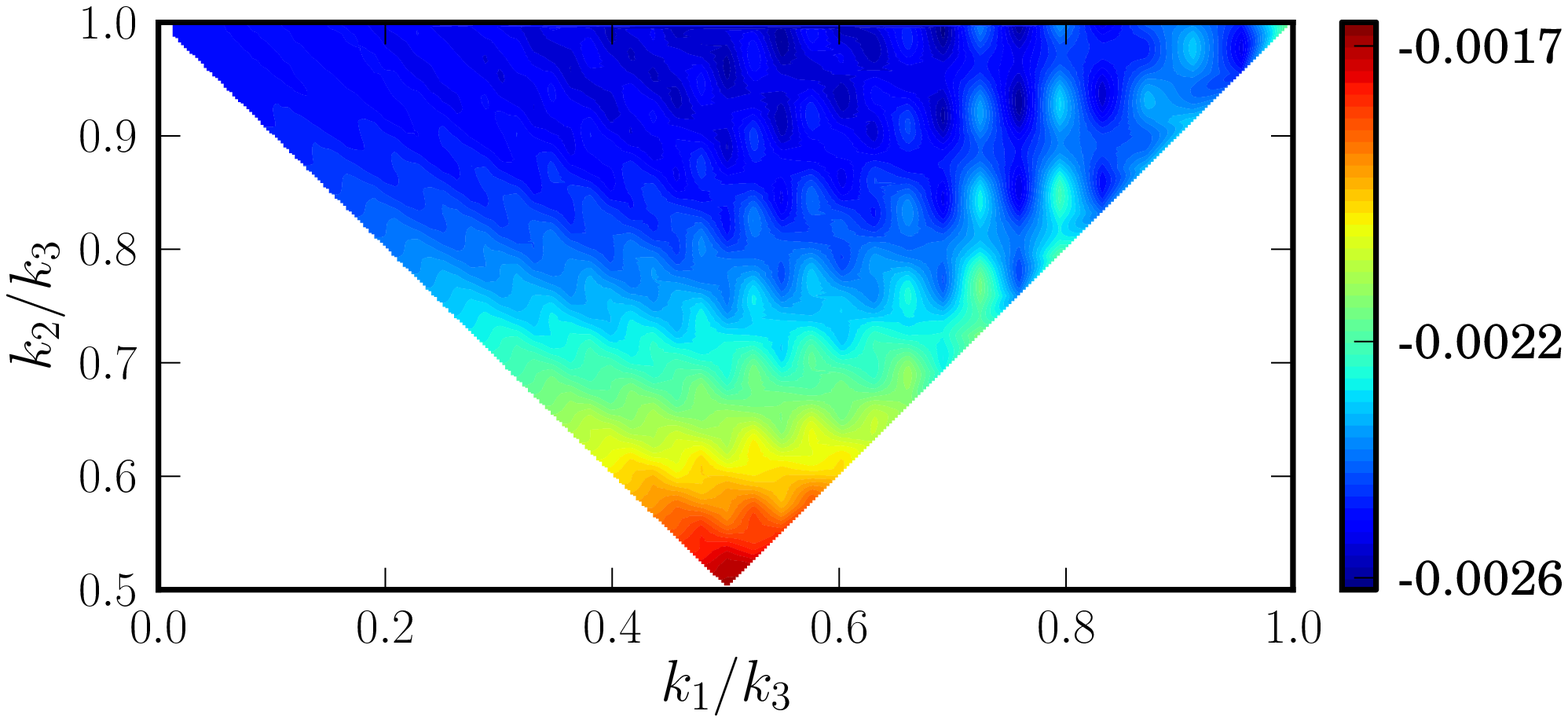} \\
\hskip -10pt
\includegraphics[width=5.25cm]{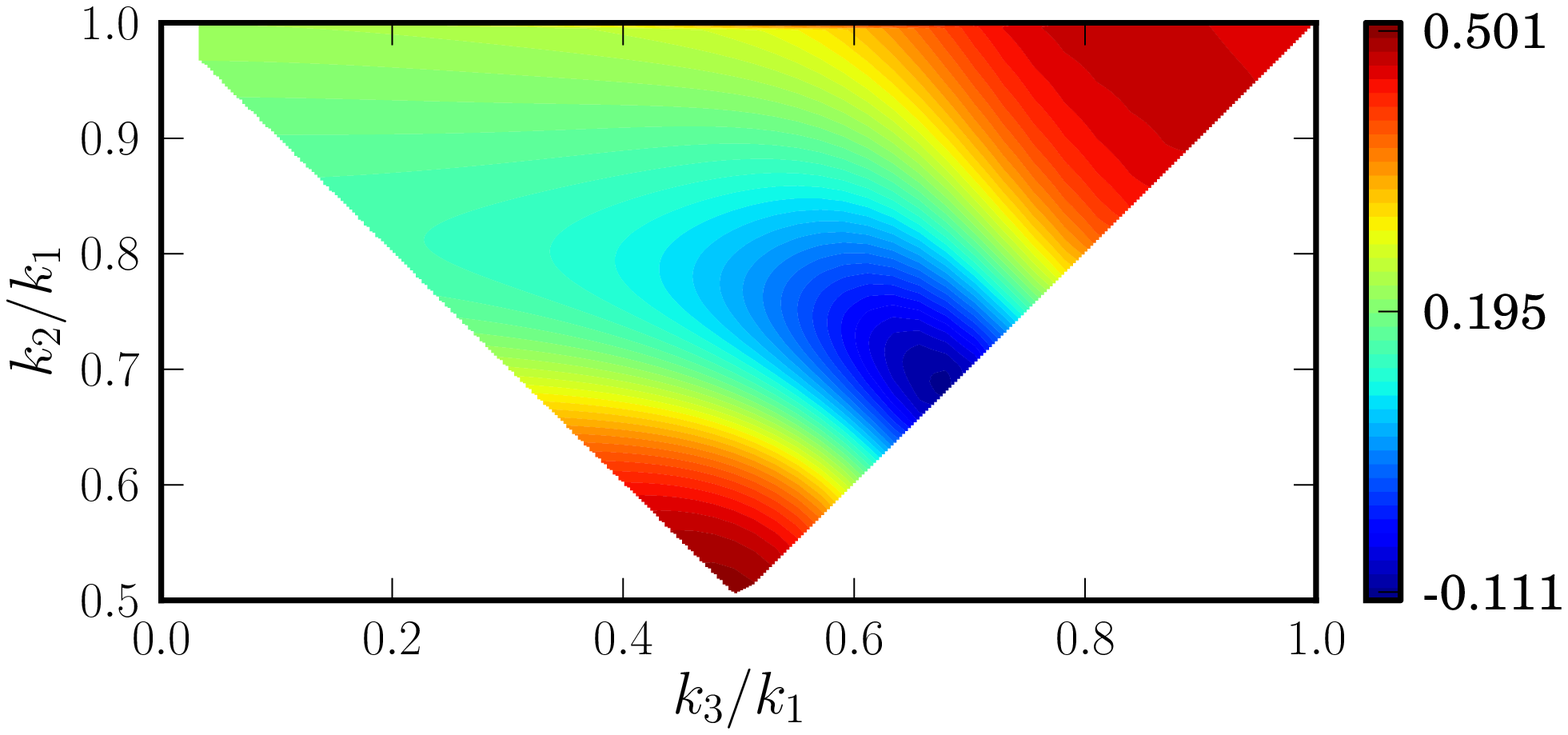} &
\hskip -10pt
\includegraphics[width=5.25cm]{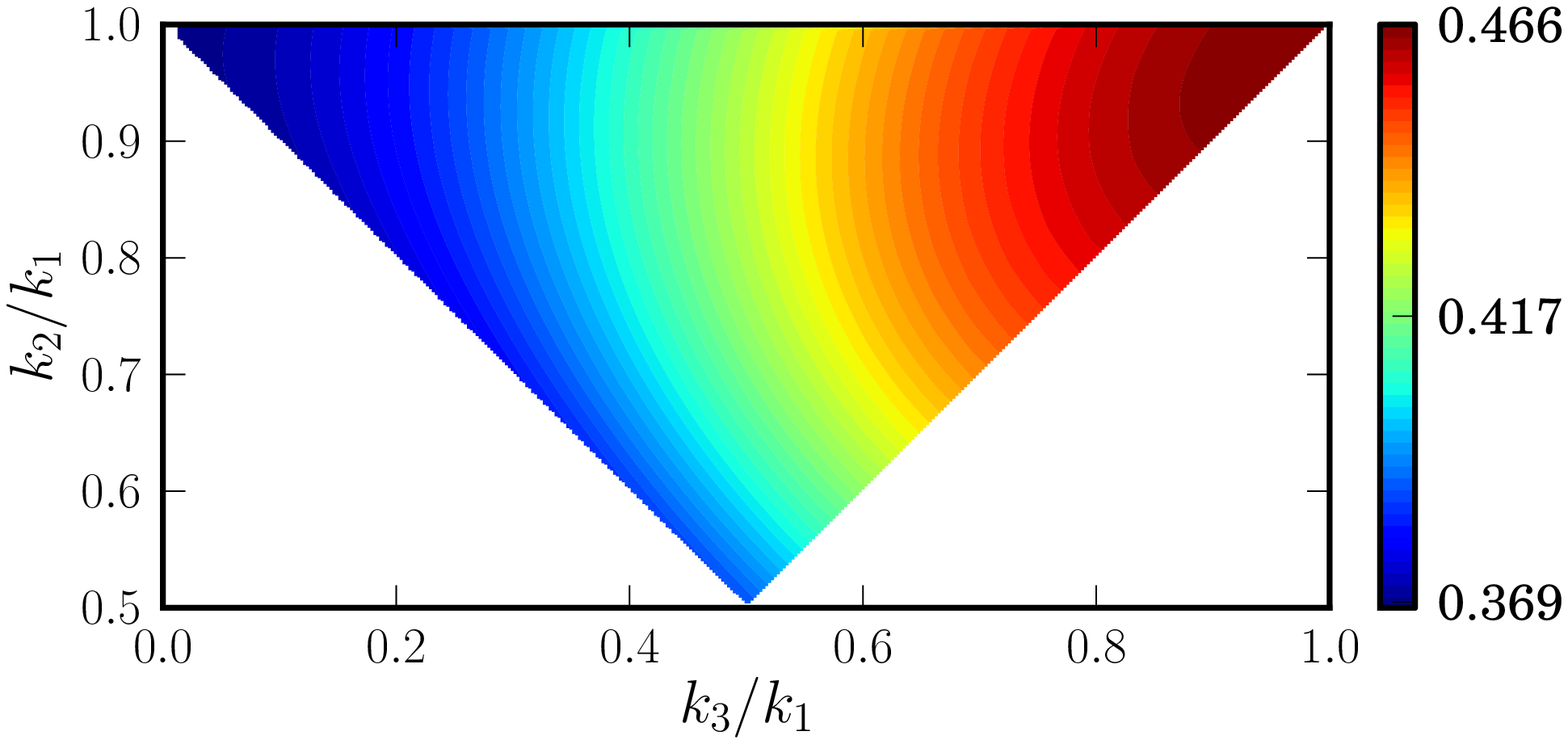} &
\hskip -10pt
\includegraphics[width=5.25cm]{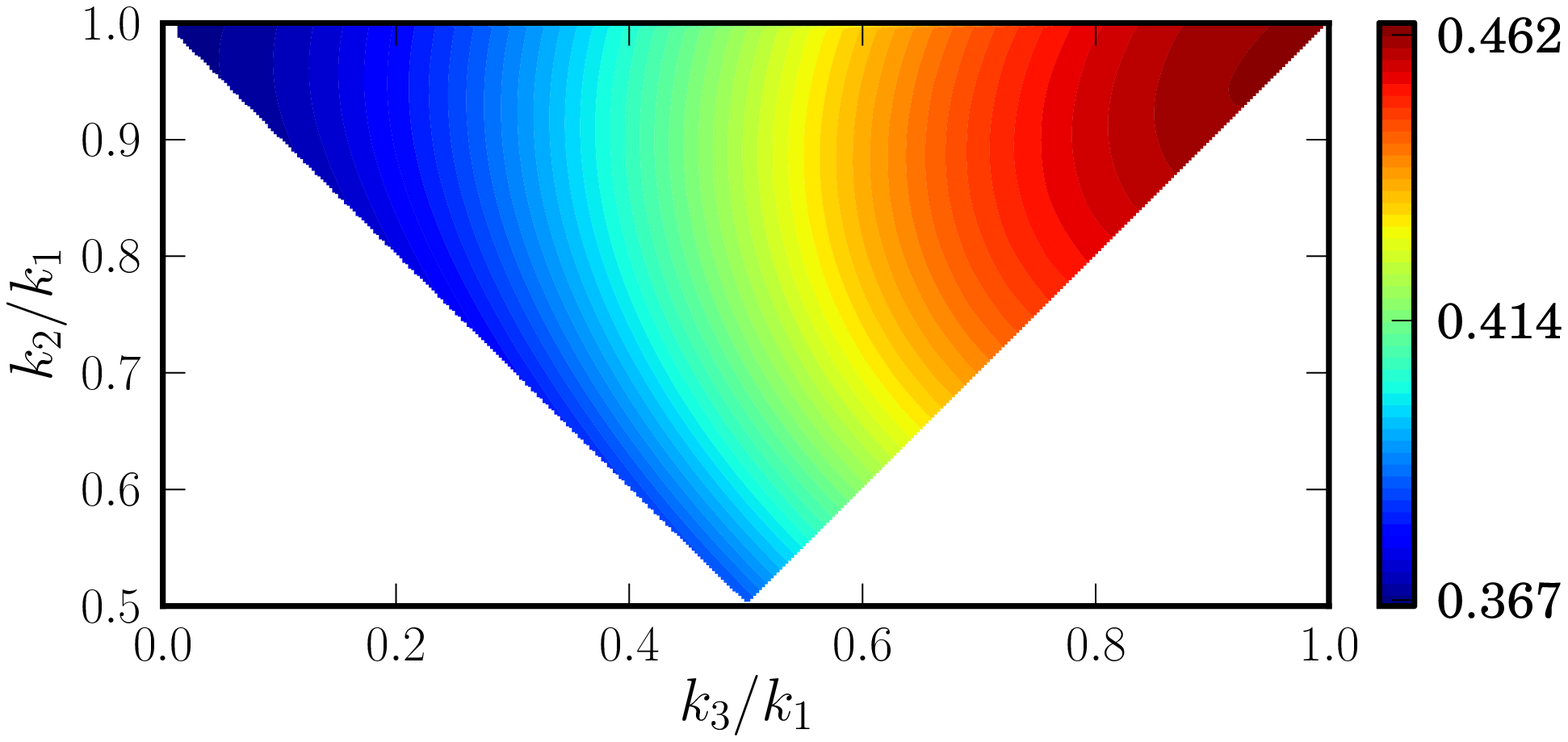} 
\end{tabular}
\caption{\label{fig:atc-pi-qpws-amm}
Density plots of the non-Gaussianity parameters $\cnls$ (on top), $\cnlt$
(in the middle) and $\hnl$ (at the bottom) evaluated numerically for an 
arbitrary triangular configuration of the wavenumbers for the case of the 
punctuated inflationary scenario (the left column), the quadratic potential 
with the step (the middle column) and the axion monodromy model (the right
column).
It should be evident that, in models leading to features, the shape of the
three-point functions as plotted against $k_3/k_1$ and $k_2/k_1$ in the 
cases of $\cnls$ and $\hnl$ or, against $k_1/k_3$ and $k_2/k_3$ in the
case of $\cnlt$, will depend on the choices of denominators $k_1$ and $k_3$.
In order to capture the non-trivial shapes that these models lead to, we have 
fixed $k_1$ and $k_3$ (when they appear in the denominators) to be $10^{-3}$, 
$2\times 10^{-3}$ and $5\times10^{-2}$ in the cases of punctuated inflation, 
the quadratic potential with a step and the monodromy model, respectively.}
\end{center}
\end{figure}

We have evaluated the scalar-tensor cross-correlations and the tensor
bi-spectrum numerically for the three models listed above.
We should mention here that we have worked with parameters for the 
models that lead to an improved fit to the WMAP seven~\cite{wmap-2011} 
or nine-year data~\cite{wmap-2012}.
(We would refer the reader to the earlier effort~\cite{hazra-2013} to 
calculate the scalar bi-spectrum in these models for the values of the 
potential parameters, including that of the Starobinsky model which we 
had discussed before. 
We would also refer the reader to Fig.~8 of the work for a plot of the 
various potentials.)
It is important that we add here that models such as the quadratic potential
with the step and the axion monodromy model have very recently been compared 
with the Planck data (see Refs.~\cite{benetti-2013} 
and~\cite{meerburg-2013a,meerburg-2013b,easther-2013}).
These investigations suggest that the resulting features lead to an improved 
fit to the Planck data too. 
Moreover, models similar to punctuated inflation, which lead to suppression of
power on large scales continue to attract attention as well (in this context,
see Refs.~\cite{lp-ls}).
In Fig.~\ref{fig:atc-pi-qpws-amm}, we have plotted the three non-Gaussianity
parameters, viz. $\cnls$, $\cnlt$ and $\hnl$, for the above three models
for an arbitrary triangular configuration of the wavenumbers.

\par

Let us now highlight certain aspects of the results that we have obtained.
We had earlier pointed out (see the caption of Fig.~\ref{fig:pl-sm-el})
the hierarchy of the various contributions to the three-point functions.
We find that the hierarchy is maintained even when deviations from slow
roll occurs.
This is not surprising because the tensor bi-spectrum is independent of the
slow roll parameters, whereas the cross-correlations at the most depend on 
the first slow roll parameter $\epsilon_1$. 
Since the first slow roll parameter cannot remain large for an extended period
without completely terminating inflation, the hierarchy of the different 
contributions is preserved even in situations involving departures from 
slow roll.

\par

It is clear from Fig.~\ref{fig:atc-pi-qpws-amm} that the tensor bi-spectrum 
in the cases of the quadratic potential with the step and the axion monodromy 
model resemble each other very closely.
In fact, they have virtually the same amplitude and shape as in the slow 
roll case illustrated in Fig.~\ref{fig:qp-atc-anr}.
This should not be surprising.
After all, since the deviations from slow roll are rather minimal in these 
models, the tensors are hardly affected.
In contrast, punctuated inflation, because of the brief departure from 
accelerated expansion that occurs, leads to a rather large effect on 
the tensors, with the tensor amplitude being considerably suppressed on 
small scales~\cite{pi}.
This is reflected in the non-trivial shape of the associated $\hnl$ parameter.
The ringing effects on the scalars that arises due to the resonance encountered
in the monodromy model (see, for example, Refs.~\cite{flauger-2010,aich-2013}) 
is clearly reflected in the amplitudes and shapes of the corresponding $\cnls$ 
and the $\cnlt$ parameters.
It is this resonance that leads to a substantially larger value for the
$\cnls$ parameter, as it does to the scalar non-Gaussianity parameter 
$\fnl$ (in this context, see, for instance, Ref.~\cite{hazra-2013}). 
Note that, apart from the ringing, the shape of the $\cnlt$ parameter is 
somewhat similar in the cases of the quadratic potential with the step and 
the monodromy model.
In the case of punctuated inflation, the shape of the $\cnls$ and $\cnlt$
parameters are considerably influenced by the contrasting fall and rise of
the scalar and the tensor powers at large scales.
This behavior results in a larger value for the $\cnls$ parameter than the 
corresponding value encountered in, say, the case of the model with the step.
 

\section{The contributions during preheating}\label{sec:c-d-prh}

In most models of inflation, the scalar field rolls down the potential 
and inflation is terminated when the field is close to a mimina of the 
potential.
Thereafter, typically, the scalar field oscillates at the bottom of the 
potential. 
During this epoch, the inflaton, due to its coupling to the matter fields,
is expected to decay and thermalize, thereby leading to the conventional
radiation dominated era~\cite{prh}.    

\par

Immediately after inflation and before the inflaton starts decaying, there 
exists a brief domain when the scalar field is oscillating at the bottom 
of the potential and continues to dominate the background evolution. 
This brief epoch is referred to as preheating~\cite{prh}. 
Since the scalar field is the dominant source of the background, the
perturbations (both scalar and tensor) continue to be governed by the
same actions and equations of motion as they are during inflation.
It is interesting to then investigate the contributions to the three-point 
functions that we have considered due to this epoch.
In fact, the contributions to the scalar bi-spectrum during preheating in
single field inflationary models was evaluated recently~\cite{hazra-2012}.
Our aim in this section is to extend the analysis to the case of the
other three-point functions.

\par 

In order to do so, as should be clear by now, we require the behavior 
of the background as well as the perturbations during the epoch of
preheating.
If one considers single field inflationary models with quadratic minima, 
say, $V(\phi) \simeq m^2\,\phi^2/2$, then it can be shown that, during 
the epoch of preheating, the first slow roll parameter behaves 
as~\cite{prh,ep-d-prh,hazra-2012}
\begin{equation}
\epsilon_1 \simeq 3\, \cos^2(m\,t + \Delta),
\end{equation}
where $t$ is the cosmic time (measured since the end of inflation), while
$\Delta$ is an arbitrary phase, chosen suitably to match the transition
from inflation to preheating.
The average value of the above slow roll parameter is $3/2$, which corresponds
to a matter dominated era.
Note that, all perturbations of cosmological interest are on super-Hubble 
scales during the domain of preheating.
Naively, one may imagine that the super-Hubble solutions for the 
scalar and the tensor perturbations during inflation, as given by 
Eqs.~(\ref{eq:shs-fkgk}), will continue to hold during the
epoch of preheating too.
The tensor modes are governed by the quantity $a''/a$, which behaves
monotonously during inflation as well as preheating.
Therefore, the $k^2$ term in Eq.~(\ref{eq:ms-g}) can indeed be ignored 
when compared to $a''/a$ even during preheating, so that the super-Hubble 
solutions to the tensor modes, viz. Eq.~(\ref{eq:gk-shs}), continue to be
applicable~\cite{ep-d-prh}. 
However, the quantity $z''/z$, as it involves the scalar field, behaves
differently during inflation and preheating.
While it grows monotonically during the latter stages of inflation, the 
quantity can even vanish during preheating (since the scalar field is 
oscillating at the minimum of the potential).
Hence, it is not a priori clear that the inflationary, super-Hubble, solutions 
will remain valid once the accelerated expansion has terminated.
A careful analysis however illustrates that, under certain conditions which 
are easily achieved in quadratic minima (for details, see, for instance, 
Refs.~\cite{ep-d-prh,hazra-2012}), the inflationary super-Hubble solutions 
for the scalar modes continue to be applicable during preheating.

\par 

Recall that the contributions to the tensor bi-spectrum (and, hence, to 
the corresponding non-Gaussianity parameter $\hnl$) on super-Hubble scales
during inflation is strictly zero.
This is true even during the epoch of preheating.
For simplicity, let us ignore the oscillations at the bottom of the
quadratic minima and use the average value of the first slow roll 
parameter, viz. that $\epsilon_1\simeq 3/2$.
In such a case, if one focuses on the equilateral limit, one can show that 
the contribution to the non-Gaussianity parameters $\cnls$ and $\cnlt$ arising 
due to the evolution from the end of inflation to the e-fold, say, $\Nf$, 
during preheating can be expressed as
\begin{equation}
C_{_{\rm NL}}^{\cR\, ({\rm ef})}(k)
= \l(\f{4\,\g+5}{5\,\g+7}\r)\; C_{_{\rm NL}}^{\g\, ({\rm ef})}(k)
=\f{12}{115}\, \l(\f{4\,\g+5}{\g+2}\r)\;\fnl^{\,({\rm ef})}(k),
\end{equation}
where $\fnl^{\,({\rm ef})}$ is the contribution due to preheating to the
non-Gaussianity parameter associated with the scalar bi-spectrum and is 
given by~\cite{hazra-2013}
\begin{eqnarray}
\fnl^{\,({\rm ef})}(k) 
&=& \frac{115\, (\g+2)}{288\,\pi\,(\g+1)}\;\Gamma^2\l(\gamma +\f{1}{2}\r)\,
2^{2\,\gamma+1}\, \l(2\,\gamma+1\r)^2\,\sin\,(2\,\pi\,\gamma)\nn\\
& &\times\, \l\vert \gamma+1\right\vert^{-2\,(\gamma+1)}\,
\l[1-{\rm e}^{-3\,(\Nf-\Ne)/2}\r]\nn\\
& & \times\, \left[\left(\frac{\pi^2\,g_{\rm eff}}{30}\right)^{-1/4}\,
\left(1+z_{\rm eq}\right)^{1/4}\,\frac{\rho_{\rm cri}^{1/4}}{T_{\rm rh}}
\right]^{-(2\,\gamma +1)}\, 
\l(\frac{k}{a_{\rm now}\, H_{\rm now}}\r)^{-(2\,\gamma+1)}.
\end{eqnarray}
We should mention here that we have arrived at this expression assuming
inflation to be of the power law form, with the scale factor being given 
by Eq.~(\ref{eq:a-p-law}) and with $\epsilon_1=(\gamma+2)/(\gamma+1)$, as
we have pointed out earlier.
In the above expression, the quantity $g_{\rm eff}$ denotes the effective 
number of relativistic degrees of freedom at reheating, $T_{\rm rh}$ the 
reheating temperature and $z_{\rm eq}$ the redshift at the epoch of equality.
Also, $\rho_{\rm cri}$, $a_{\rm now}$ and $H_{\rm now}$ represent the critical 
energy density, the scale factor and the Hubble parameter {\it today},\/ 
respectively\footnote{Since we have already made use of $a_0$ and $H_0$ 
to denote the scale factor at the transition and the Hubble parameter 
around the transition in the Starobinsky model, to avoid degeneracy, we 
have used the less conventional $a_{\rm now}$ and $H_{\rm now}$ to denote 
the scale factor and the Hubble parameter today!}.
It should be clear from the above expression that the contributions
due to preheating is mainly determined by the quantity 
$\rho_{\rm cri}^{1/4}/T_{\rm rh}$. 
For an inflationary model wherein $\gamma \simeq -2$ and a reheating 
temperature of $T_{\rm rh} \simeq 10^{10}\, \mbox{GeV}$, one obtains 
that $\fnl^{\,({\rm ef})}\sim C_{_{\rm NL}}^{\cR\, ({\rm ef})}
\sim C_{_{\rm NL}}^{\g\, ({\rm ef})} \sim 10^{-60}$ for the 
modes of cosmological interest (i.e. for wavenumbers such that 
$k\simeq a_{\rm now}\,H_{\rm now}$). 
Needless to add, these values are simply unobservable (also see, 
Ref.~\cite{ng-prh1}; in this context, however, see Ref.~\cite{ng-prh2}). 
In other words, as in the case of the scalar parameter $\fnl$, the contribution 
to the other non-Gaussianity parameters due to the epoch of preheating is 
completely insignificant.


\section{Discussion}\label{sec:c}

In this work, based on the Maldacena formalism and extending the recent
effort towards calculating the scalar bi-spectrum, we have developed a
numerical procedure for calculating the other three-point functions of
interest.
Motivated by the parameters often introduced to characterize the scalar
and the tensor bi-spectra, we have introduced dimensionless non-Gaussianity
parameters to describe the scalar-tensor cross-correlations.
We have compared the performance of the code with the analytical results
that are available in different situations and have utilized the code to
calculate the three-point functions and the corresponding non-Gaussianity
parameters in a class of models that lead to features in the scalar
power spectrum. 
We have also shown that, as in the case of the scalar bi-spectrum, the 
contributions to the other three-point functions during the epoch of 
preheating proves to be completely negligible.
In fact, we have made available a sample of the numerical 
code that we have worked with to arrive at the results discussed in this 
paper at the following URL: 
{\it https://www.physics.iitm.ac.in/{\textasciitilde{}}sriram/tpf-code/registration.html}.
The sample code corresponds to the specific case of the quadratic potential 
with the step that we have considered.
The code can be easily extended to other inflationary models.

\par

Before we conclude, we would like to make a couple of clarifying remarks 
concerning the status of models leading to features in the scalar
power spectrum in the light of the Planck data.
While, as we had mentioned in the introduction, the Bayesian evidence for 
features do not seem substantial~\cite{planck-2013-ci}, model independent 
reconstruction efforts seem to consistently point to the possibility of 
scale-dependent power spectra (in this context, see the recent efforts, 
Refs.~\cite{rc}).
Importantly, the Planck team finds that the constraints on the scalar
non-Gaussianity parameters $\fnl$ (that we had quoted in the introductory
section) turn less stringent when one permits features (contrast, for
instance, Table~8 with Tables~12 and~13 of Ref.~\cite{planck-2013-cpng}). 
This is an aspect that seems to deserve closer examination.
 
\par

We believe that the non-Gaussianity parameters $\cnls$ and $\cnlt$ which
we have introduced here provide additional quantities to characterize an
inflationary model.
It will be interesting to arrive at constraints on these parameters as
well from the observational data and understand its implications.
We are currently investigating these issues.


\section*{Acknowledgements}

RT's work is supported under the DST-Max Planck India Partner Group in Gravity 
and Cosmology. 
RT wishes to thank the Indian Institute of Technology Madras, Chennai, India,
for support and hospitality during a visit, where this work was initiated.
VS would like to acknowledge the hospitality provided by Dr.~S.~Shankaranarayanan 
at the Indian Institute of Science Education and Research, Thiruvananthapuram, 
India, where part of this work was carried out.
The authors wish to thank J\'er\^ome Martin and Dhiraj Hazra for discussions as 
well as detailed comments on the manuscript.
We acknowledge the use of the high performance computing facility at the
Indian Institute of Technology Madras, Chennai, India.


\appendix

\section{Three-point functions in the Starobinsky model}\label{sec:sm}

In this appendix, we shall provide some of the essential details for arriving
at the analytical results for the three-point functions of our interest in the 
case of the Starobinsky 
model~\cite{starobinsky-1992,martin-2012a,arroja-2011-2012}.
In Subsec.~\ref{subsec:sm}, we have already discussed the behavior of the 
background as well as the perturbations in the model.
It is just a matter of substituting the various quantities in the integrals
that describe the three-point functions and being able to carry out the integrals
involved.
As we had pointed out earlier, due to the transition at the discontinuity in 
the potential, the integrals need to be divided into two.
The integrals up to the transition essentially lead to the slow roll results,
but with suitable modifications that arise because of the reason that the 
integrals are not to be carried out until late times.
Though slow roll is violated briefly due to the discontinuity, we find that all 
the integrals can be evaluated in terms of simple functions to arrive at the 
three-point correlations.
Since the scale factor is always of the de Sitter form, as we had mentioned,
the tensor bi-spectrum proves to be the same as the one arrived at in the slow 
roll approximation~\cite{maldacena-2003,tensor-bs}.
Therefore, we do not discuss it here.
In what follows, we shall list out the results of the integrals involved in arriving 
at the two cross-correlations.


\subsection{Calculation of ${\mathcal G}_{\cR\cR\gamma}^C$}

Evidently, the quantities $\cG_{\cR\cR\gamma}^{C}$, with $C=(1,2,3)$
[cf.~Eqs.~(\ref{eq:cGrrg1})--(\ref{eq:cGrrg3})], need to be first 
evaluated in order to arrive at the cross-correlation
$G_{\cR\cR\gamma}^{m_3n_3}$.
Upon dividing the integrals into two, we find that the contributions 
before the transition are given by the following expressions:
\begin{subequations}
\begin{eqnarray}
\sg_{\sr\sr\g}^{1+}(\vka,\vkb,\vkc) 
&=& \f{H_0}{2\,\Mpl^3\,\sqrt{k_1^3\,k_2^3\,k_3^3}}\;
\vare_{ij}^{s_3\ast}(k_3)\;k_{1i}\,k_{2j}\;\biggl[k_0 +
\f{i\,\l(k_1\,k_2 +k_1\,k_3 + k_2\,k_3\r)}{\kT}\nn\\ 
& &-\, \f{k_1\,k_2\,k_3}{\kT\, k_0} + \f{i\,k_1\,k_2\,k_3}{\kT^2}\biggr]\,
{\rm e}^{-i\,\kT/k_0},\\
\sg_{\sr\sr\g}^{2+}(\vka,\vkb,\vkc)  
& = & \f{A_+^2}{144\, H_0^3\,\Mpl^5\,\sqrt{k_1^3\,k_2^3\,k_3^3}}\;
\vare_{ij}^{s_3\ast}(k_3)\;k_{1i}\,k_{2j}\,k_3^2\nn\\
& &\times\,\l[\f{i}{\kT} - \f{k_3}{\kT\, k_0} 
+ \f{i\,k_3}{\kT^2}\r]\,{\rm e}^{-i\,\kT/k_0},\\
\sg_{\sr\sr\g}^{3+}(\vka,\vkb,\vkc)  
&=& -\,\f{A_+^2}{144\,H_0^3\,\Mpl^5\,\sqrt{k_1^3\,k_2^3\,k_3^3}}\;
\vare_{ij}^{s_3\ast}(k_3)\;k_{1i}\,k_{2j}\,k_3^2\;
\bigg[\l(-\f{i}{\kt}+\f{k_1}{\kt\,\ko} -\f{i\,k_1}{\kt^2}\r)\nn\\
& & +\, \l(-\f{i}{\kt} +\f{k_2}{\kt\,\ko} -\f{i\,k_2}{\kt^2}\r)\biggr]\, 
{\rm e}^{-i\,\kt/\ko},
\end{eqnarray}
\end{subequations}
where, as we have indicated earlier, $k_{_{\rm T}}=k_1+k_2+k_3$.
Similarly, after the transition, upon substituting the corresponding modes 
describing the perturbations, we find that, we can write
\begin{subequations}
\begin{eqnarray}
\sg_{\sr\sr\g}^{1-}(\vka,\vkb,\vkc) 
&=& \f{H_0}{2\,\Mpl^3\,\sqrt{k_1^3\,k_2^3\,k_3^3}}\;
\vare_{ij}^{s_3\ast}(k_3)\;k_{1i}\,k_{2j}\,
\biggl[\alpha_{k_1}^\ast\,\alpha_{k_2}^\ast\, 
\cI_{\sr\sr\g}^{11}(k_1,k_2,k_3)\nn\\ 
& &-\,\alpha_{k_1}^\ast\,\beta_{k_2}^\ast\, \cI_{\sr\sr\g}^{12}(k_1,k_2,k_3)
-\beta_{k_1}^\ast\,\alpha_{k_2}^\ast\, \cI_{\sr\sr\g}^{13}(k_1,k_2,k_3)\nn\\
& & +\,\beta_{k_1}^\ast\,\beta_{k_2}^\ast\, \cI_{\sr\sr\g}^{14}(k_1,k_2,k_3)\biggr],\\
\sg_{\sr\sr\g}^{2-}(\vka,\vkb,\vkc)  
&=& -\,\f{H_0}{8\,\Mpl^3\, \sqrt{k_1^3\,k_2^3\,k_3^3}}\,
\vare_{ij}^{s_3\ast}(k_3)\; \f{k_3^2\, k_{1i}\, k_{2j}}{k_1^2\,k_2^2}\,
\biggl[\alpha_{k_1}^\ast\,\alpha_{k_2}^\ast\, \cI_{\sr\sr\g}^{21}(k_1,k_2,k_3)\nn\\
& &-\, \alpha_{k_1}^\ast\,\beta_{k_2}^\ast\,\cI_{\sr\sr\g}^{22}(k_1,k_2,k_3)
-\beta_{k_1}^\ast\,\alpha_{k_2}^\ast\, \cI_{\sr\sr\g}^{23}(k_1,k_2,k_3)\nn\\
& & +\,\beta_{k_1}^\ast\,\beta_{k_2}^\ast\,
\cI_{\sr\sr\g}^{24}(k_1,k_2,k_3)\biggr],\\
\sg_{\sr\sr\g}^{3-}(k_1,k_2,k_3)
&=&-\,\f{H_0}{8\,\Mpl^3\, \sqrt{k_1^3\,k_2^3\,k_3^3}}\, 
\vare_{ij}^{s_3\ast}(k_3)\;k_{1i}\,k_{2j}\,
\biggl\{\f{k_3^2}{k_2^2}\, \biggl[\alpha_{k_1}^\ast\,\alpha_{k_2}^\ast\, 
\cI_{\sr\sr\g}^{31}(k_1,k_2,k_3)\nonumber \\
& &-\,\alpha_{k_1}^\ast\,\beta_{k_2}^\ast\,\cI_{\sr\sr\g}^{32}(k_1,k_2,k_3)
-\beta_{k_1}^\ast\,\alpha_{k_2}^\ast\, \cI_{\sr\sr\g}^{33}(k_1,k_2,k_3)\nn\\
& &+\,\beta_{k_1}^\ast\,\beta_{k_2}^\ast\, \cI_{\sr\sr\g}^{34}(k_1,k_2,k_3)\biggr]\nn\\
& &+\,\f{k_3^2}{k_1^2}\, \biggl[\alpha_{k_1}^\ast\,\alpha_{k_2}^\ast\,
\cJ_{\sr\sr\g}^{31}(k_1,k_2,k_3)
-\alpha_{k_1}^\ast\,\beta_{k_2}^\ast\,\cJ_{\sr\sr\g}^{32}(k_1,k_2,k_3)\nn\\
& &-\,\beta_{k_1}^\ast\, \alpha_{k_2}^\ast\, 
\cJ_{\sr\sr\g}^{33}(k_1,k_2,k_3) 
+\beta_{k_1}^\ast\,\beta_{k_2}^\ast\, 
\cJ_{\sr\sr\g}^{34}(k_1,k_2,k_3)\biggr]\biggr\}.
\end{eqnarray}
\end{subequations}
The  expressions for the functions $\cI_{\sr\sr\g}^{1i}(k_1,k_2,k_3)$, 
$\cI_{\sr\sr\g}^{2i}(k_1,k_2,k_3)$ and $\cI_{\sr\sr\g}^{3i}(k_1,k_2,k_3)$ 
as well as $\cJ_{\sr\sr\g}^{3i}(k_1,k_2,k_3)$, where $i=1,2,3,4$, are
furnished in the last sub-section. 


\subsection{Calculation of ${\mathcal G}_{\cR\g\g}^C$}

In this case, the contributions before the transition are given by
\begin{subequations}
\begin{eqnarray}
\mathcal{G}^{1+}_{R\gamma\gamma}(k_1,k_2,k_3) 
&=& \frac{i\,A_{+}}{24\,H_{0}\,\Mpl^{4}\,\sqrt{2\,k_1^3\, k_2^3\,k_3^3}}\; 
\epij^{s_2\ast}(k_2)\;\epij^{s_3\ast}(k_3)\, k_2^{2}\,k_{3}^{2}\,\nn\\
& &\times\, \l(\frac{1}{\kt}+\frac{i\,k_{1}}{\kt\, k_{0}}
+\frac{k_{1}}{\kt^{2}}\r)\,{\rm e}^{-i\,\kt/k_{0}},\\
\mathcal{G}^{2+}_{R\gamma\gamma}(k_1,k_2,k_3) 
&=&\frac{i\,A_{+}}{24\,H_{0}\,\Mpl^{4}\, \sqrt{2\,k_1^3\,k_2^3\,k_3^3}}\;
\epij^{s_2\ast}(k_2)\;\epij^{s_3\ast}(k_3)\, \l(\vkb\cdot\vkc\r)\nn\\ 
& &\times\,\biggl(-i\,k_{0}+\frac{k_1\,k_2+k_1\,k_3+k_2\,k_3}{\kt}
+\frac{i\,k_1\,k_2\,k_3}{\kt\, k_{0}} 
+ \frac{k_1\,k_2\,k_3}{\kt^{2}}\biggr)\nn\\
& &\times\,{\rm e}^{-i\,\kt/k_{0}},\\
\mathcal{G}^{3+}_{R\gamma\gamma}(k_1,k_2,k_3) 
&=&-\,\frac{i\,A_{+}}{24\,H_{0}\,\Mpl^{4}\,\sqrt{2\,k_1^3\,k_2^3\,k_3^3}}\;
\epij^{s_2\ast}(k_2)\;\epij^{s_3\ast}(k_3)\nn\\
& &\times\,\biggl[\l(\vka\cdot \vkb\r)\,k_{3}^{2}\,
\l(\f{1}{\kt}+\frac{i\,k_{2}}{\kt\, k_{0}}+\frac{k_{2}}{\kt^{2}}\r)\nn\\
& & +\,\l(\vka\cdot\vkc\r)\, k_{2}^{2}\,
\l(\f{1}{\kt} +\frac{i\,k_{3}}{\kt\,k_{0}}
+\f{k_{3}}{\kt^{2}}\r)\biggr]\; {\rm e}^{-i\,\kt/k_{0}}.
\end{eqnarray}
\end{subequations}
The corresponding quantities after the transition are found to be
\begin{subequations}
\begin{eqnarray}
\mathcal{G}^{1-}_{R\gamma\gamma}(k_1,k_2,k_3) 
&=& \frac{i\,A_{-}}{24\,H_{0}\,\Mpl^4\,\sqrt{2\,k_1^3\,k_2^3\,k_3^3}}\;
\epij^{s_2\ast}(k_2)\;\epij^{s_3\ast}(k_3)\;k_{2}^{2}\,k_{3}^{2}\nn\\
& &\times\,\l[\alpha_{k1}^{\ast}\, \cM_{1}(k_{1},k_{2},k_{3}) 
-\beta_{k1}^{\ast}\, \cM_{1}(-k_{1},k_{2},k_{3})\r],\\
\mathcal{G}^{2-}_{R\gamma\gamma}(k_1,k_2,k_3)  
&=& -\frac{i\,A_-}{24\,H_{0}\,\Mpl^4\,\sqrt{2\,k_3\,k_2^3\,k_3^3}}\;
\epij^{s_2\ast}(k_2)\;\epij^{s_3\ast}(k_3)\;
\l(\vkb\cdot\vkc\r)\nn\\
& &\times\, \l[\alpha_{k1}^{\ast}\,\cM_{2}(k_{1},k_{2},k_{3})
-\beta_{k1}^{\ast}\,\cM_{2}(-k_{1},k_{2},k_{3})\r],\\ 
\mathcal{G}^{3-}_{R\gamma\gamma}(k_1,k_2,k_3)  
&=&-\frac{i\,A_{-}}{24\,H_0\,\Mpl^4\,\sqrt{2\,k_1^3\,k_2^3,k_3^3}}\;
\epij^{s_2\ast}(k_2)\;\epij^{s_3\ast}(k_3)\nn\\
& &\times\,\biggl\{\frac{ \l(\vka\cdot\vkb\r)\,k_{3}^{2}}{k_{1}^{2}}\;
\l[\alpha_{k1}^{\ast}\,\cM_{3}(k_{1},k_{3},k_{2})
-\beta_{k1}^{\ast}\,\cM_{3}(-k_{1},k_{3},k_{2})\r]\nn\\
& & +\,\frac{\l(\vka\cdot\vkc\r)\,k_{2}^{2}}{k_1^2}\,
\l[\alpha_{k1}^{\ast}\, \cM_{3}(k_{1},k_{2},k_{3})
-\beta_{k1}^{\ast}\, \cM_{3}(-k_{1},k_{2},k_{3})\r]\biggr\}.
\end{eqnarray}
\end{subequations}
The forms of the expressions $\cM_i(k_1,k_2,k_3)$ with $i=1,2,3$ are given
in the next sub-section.


\subsection{Evaluation of integrals}

The quantity $\cI_{\sr\sr\g}^{11}(k_1,k_2,k_3)$ is described by the integral
\begin{equation}
\cI_{\sr\sr\g}^{11}(k_1,k_2,k_3) 
= \int_{-k_0^{-1}}^0 \f{\d\eta}{\eta^2}\, \l(1-i\,k_1\,\eta\r)\;
\l(1-i\,k_2\,\eta\r)\, \l(1-i\,k_3\,\eta\r)\,{\rm e}^{i\,\kT\,\eta},
\end{equation}
which can be easily evaluated to be
\begin{eqnarray}
\cI_{\sr\sr\g}^{11}(k_1,k_2,k_3)
&=& \lim_{\ee\to 0}\l(-\f{{\rm e}^{i\,\kT\,\ee}}{\ee}\r)\nn\\ 
& &-\, \biggl(k_0 + \f{i\,(k_1k_2 + k_1\,k_3 + k_2\,k_3)}{\kT}
-\f{k_1\,k_2\,k_3}{\kT\, k_0} 
+i\,\f{k_1\,k_2\,k_3}{\kT^2}\biggr)\;{\rm e}^{-i\,\kT/k_0}\nn\\
& & +\,\f{i\,\l(k_1\,k_2+k_1\,k_3+k_2\,k_3\r)}{\kT} 
+ \f{i\,k_1\,k_2\,k_3}{\kT^2}.\qquad
\end{eqnarray}
We find that the rest of the functions $\cI_{\sr\sr\g}^{1i}(k_1,k_2,k_3)$
with $i=2,3,4$ can be expressed in terms of $\cI_{\sr\sr\g}^{11}(k_1,k_2,k_3)$ 
as follows:
$\cI_{\sr\sr\g}^{12}(k_1,k_2,k_3)
=\cI_{\sr\sr\g}^{11}(k_1,-k_2,k_3)$, $\cI_{\sr\sr\g}^{13}(k_1,k_2,k_3) 
= \cI_{\sr\sr\g}^{11}(-k_1,k_2,k_3)$ and $\cI_{\sr\sr\g}^{14}(k_1,k_2,k_3)
=\cI_{\sr\sr\g}^{11}(-k_1,-k_2,k_3)$.

\par

The quantity $I_{\sr\sr\g}^{21}(k_1,k_2,k_3)$ is described by the integral
\begin{eqnarray}
\cI_{\sr\sr\g}^{21}(k_1,k_2,k_3) 
&=& \f{A_-^2}{18\,H_0^4\, \Mpl^2}\,
\int_{-k_0^{-1}}^{0}\f{\d\eta}{\eta^2}\,\l(1+\rho^3\eta^3\r)^{2}
\l(1-i\,k_3\,\eta\r)\nn\\
& &\times\, \l[\f{\epsilon_{2-}}{2\,\eta}\,\l(1-i\,k_1\,\eta\r) 
+ k_1^2\,\eta\r]\,
\l[\f{\epsilon_{2-}}{2\,\eta}\,\l(1-i\,k_2\,\eta\r) +k_2^2\,\eta\r],
\end{eqnarray}
with $\epsilon_{2-}$ being given by Eq.~(\ref{eq:e2m}).
We find that this quantity can be written as
\begin{eqnarray}
\cI_{\sr\sr\g}^{21}(k_1,k_2,k_3) 
&=& \f{A_-^2}{18\, H_0^4\,\Mpl^2}\, \biggl[\cA_1(k_1,k_2,k_3) 
+ \cA_2(k_1,k_2,k_3) \nonumber \\
& & +\, \cA_3(k_1,k_2,k_3) + \cA_4(k_1,k_2,k_3)\biggr],
\end{eqnarray}
where 
\begin{eqnarray}
\cA_1(k_1,k_2,k_3) 
&=& 9\,\rho^6\, 
\int_{-k_0^{-1}}^{0}\d\eta\; \eta^2\, \l(1-i\,k_1\,\eta\r)\,
\l(1-i\,k_2\,\eta\r)\, \l(1-i\,k_3\,\eta\r)\, {\rm e}^{i\,\kT\,\eta}\nn\\
&=& 9\,\rho^6\; 
\biggl\{\f{8\,i}{\kT^3} + \f{24\,i\,\l(k_1\,k_2+k_1\,k_3+k_2\,k_3\r)}{\kT^5}
+ \f{120\,i\,k_1\,k_2\,k_3}{\kT^6}\nn\\
& & +\,\biggl[\f{4\,i}{\kT\, k_0^2} 
- \f{8\,i}{\kT^3} -\f{1}{k_0^3} + \f{8}{\kT^2 k_0}\nn\\
& & -\,\l(k_1\,k_2+k_1\,k_3+k_2\,k_3\r)\, 
\l( \f{i}{\kT k_0^4} + \f{4}{\kT^2\, k_0^3} 
- \f{12\,i}{\kT^3\, k_0^2} -\f{24}{\kT^4 k_0} + \f{24\,i}{\kT^5}\r)\nn\\
& & +\, k_1\,k_2\,k_3\, 
\l(\f{1}{\kT k_0^5} -\f{5\,i}{\kT^2 k_0^4} - \f{20}{\kT^3\, k_0^3} 
+ \f{60\,i}{\kT^4 k_0^2} + \f{120}{\kT^5 k_0} 
- \f{120\,i}{\kT^6}\r) \biggr]\nn\\ 
& &\times\,{\rm e}^{-i\,\kT/k_0}\biggr\},
\end{eqnarray}
\begin{eqnarray}
\cA_2(k_1,k_2,k_3) 
&=& -3\,\rho^3\,k_2^2\, \int_{-k_0^{-1}}^{0} \d\eta\;\eta\,
\l(1+\rho^3\,\eta^3\r)\,\l(1-i\,k_1\,\eta\r)\,
\l(1-i\,k_3\,\eta\r)\; {\rm e}^{i\,\kT\,\eta}\nn\\
&=& -3\,\rho^3\,k_1^2\, 
\biggl(\f{1}{\kT^2} + \f{2\,(k_1+k_3)}{\kT^3} + \f{6\,k_1\,k_3}{\kT^4}
-\f{24\,i\,\rho^3}{\kT^5} - \f{120\,i\,\rho^3\, (k_1+k_3)}{\kT^6}\nn\\ 
& &-\, \f{720\,i\,\rho^3\,k_1\,k_3}{\kT^7}
+ \biggl\{-\f{i}{\kT\, k_0} -\f{1}{\kT^2} -i\,\l(k_1+k_3\r)\, 
\l(\f{i}{\kT\, k_0^2} + \f{2}{\kT^2\, k_0}-\f{2\,i}{\kT^3}\r)\nn\\
& & -k_1\,k_3\, \l(-\f{i}{\kT\, k_0^3}-\f{3}{\kT^2\,k_0^2} 
+ \f{6\,i}{\kT^3\,k_0}+ \f{6}{\kT^4}\r)\nn\\ 
& & +\, \rho^3\, \biggl[\f{i}{\kT\, k_0^4} + \f{4}{\kT^2\, k_0^3}
- \f{12\,i}{\kT^3\, k_0^2} -\f{24}{\kT^4\,k_0} + \f{24\,i}{\kT^5}\nn\\ 
& &-i\,\l(k_1 + k_3\r)\,\l(-\f{i}{\kT\, k_0^5} -\f{5}{\kT^2\, k_0^4}
+ \f{20\,i}{\kT^3\, k_0^3} + \f{60}{\kt^4\,\ko^2} 
-\f{120\,i}{\kt^5\,\ko} -\f{120}{\kt^6} \r)\nn\\ 
& &-k_1\,k_3\,\l(\f{i}{\kt\,\ko^6}+ \f{6}{\kt^2\,\ko^5}
- \f{30\,i}{\kt^3\,\ko^4} -\f{120}{\kt^4\, \ko^3}  
+ \f{360\,i}{\kt^5\,\ko^2} +\f{720}{\kt^6\,\ko}-\f{720\,i}{\kt^7}\r)\biggr]
\biggr\}\nn\\
& &\times\,{\rm e}^{-i\,\kt/\ko}\biggr),
\end{eqnarray}
\begin{eqnarray}
\cA_3(k_1,k_2,k_3) 
&=& -3\,\rho^3\,k_1^2\, 
\int_{-k_0^{-1}}^{0} \d\eta\, \eta\,
\l(1+\rho^3\,\eta^3\r)\, \l(1-i\,k_3\,\eta\r)\,
\l(1-i\,k_2\,\eta\r)\, {\rm e}^{i\,\kT\,\eta}\nn\\
&=& \cA_2(k_2,k_1,k_3),
\end{eqnarray}
\begin{eqnarray}
\cA_4(k_1,k_2,k_3) 
&=& k_1^2\, k_2^2\,
\int_{-k_0^{-1}}^{0}\d\eta\, \l(1+\rho^3\,\eta^3\r)^2\,
\l(1-i\,k_3\,\eta\r)\, {\rm e}^{i\,\kT\,\eta}\nn\\
&=& k_1^2\,k_2^2\,\biggl\{-\f{i}{\kt} -\f{12\,\rho^3}{\kt^4}
+\f{720\,i\,\rho^6}{\kt^7} - \f{i\,k_3}{\kt^2} - \f{48\,k_3\,\rho^3}{\kt^5}
+\f{5040\,i\,k_3\,\rho^6}{\kt^8}\nn\\
& &+\,\biggl[\f{i}{\kt} -\f{2\,i\,\rr3}{\kt\,\ko^3} -\f{6\,\rr3}{\kt^2\,\ko^2}
+ \f{12\,i\,\rr3}{\kt^3\,\ko} + \f{12\,\rr3}{\kt^4} + \f{i\,\rr6}{\kt\,\ko^6}
+\f{6\,\rr6}{\kt^2\,\ko^5} -\f{30\,i\,\rho^6}{\kt^3\,\ko^4}\nn\\ 
& &-\f{120\,\rr6}{\kt^4\,\ko^3} +\f{360\,i\,\rr6}{\kt^5\,\ko^2}
+ \f{720\,\rr6}{\kt^6\,\ko} - \f{720\,i\,\rr6}{\kt^7}
+ i\,k_3\,\l(\f{i}{\kt\,\ko} +\f{1}{\kt^2}\r)\nn\\ 
& &-\,2\,i\,k_3\,\rr3\, \l(\f{i}{\kt\,\ko^4}
+ \f{4}{\kt^2\,\ko^3} - \f{12\,i}{\kt^3\,\ko^2} 
-\f{24}{\kt^4\,\ko} + \f{24\,i}{\kt^5}\r)\nn\\ 
& &-\,i\,k_3\,\rr6\, \biggl(-\f{i}{\kt\,\ko^7} - \f{7}{\kt^2\,\ko^6} 
+ \f{42\,i}{\kt^3\,\ko^5} + \f{210}{\kt^4\,\ko^4} 
-\f{840\,i}{\kt^5\,\ko^3} - \f{2520}{\kt^6\,\ko^2}\nn\\ 
& &+\,\f{5040\,i}{\kt^7\,\ko} 
+\f{5040}{\kt^8}\biggr)\biggr]\, {\rm e}^{-i\,\kt/\ko}\biggr\}.
\end{eqnarray}
Moreover, it can be shown that $\cI_{\sr\sr\g}^{22}(k_1,k_2,k_3)
=\cI_{\sr\sr\g}^{21}(k_1,-k_2,k_3)$, $\cI_{\sr\sr\g}^{23}(k_1,k_2,k_3) 
= \cI_{\sr\sr\g}^{21}(-k_1,k_2,k_3)$ and $\cI_{\sr\sr\g}^{24}(k_1,k_2,k_3)
=\cI_{\sr\sr\g}^{11}(-k_1,-k_2,k_3)$.

\par

The quantity $\cI_{\sr\sr\g}^{31}(k_1,k_2,k_3)$ is described by the integral
\begin{equation}
\cI_{\sr\sr\g}^{31}(k_1,k_2,k_3) 
= \int_{-k_0^{-1}}^0\d\eta\,\f{\epsilon_{1-}}{\eta}\,
\l[\f{\epsilon_{2-}}{2\,\eta}\,(1-i\,k_1\,\eta)\,(1-i\,k_2\,\eta) 
+k_2^2\,\eta\,(1-i\,k_1\,\eta)\r]\,{\rm e}^{i\,\kt\,\eta},
\end{equation}
where $\epsilon_{1-}$ is the slow roll parameter after the transition, which
is given by Eq.~(\ref{eq:e1m}).
The above integral can be evaluated to yield
\begin{eqnarray}
\cI_{\sr\sr\g}^{31}(k_1,k_2,k_3) 
&=& \f{A_{-}^2}{18\, H_0^4\, \Mpl^2}\,
\biggl(-3\,\rr3\,\biggl\{\f{1}{\kt^2} + \f{2\,\l(k_1 + k_2\r)}{\kt^3}
+\f{6\,k_1\,k_2}{\ktr4} - \f{24\,i\,\rr3}{\kt^5}\nn\\
& &-\,\f{120\,i\,\rr3\,(k_1+k_2)}{\ktr6}-\f{720\,i\, k_1\, k_2\, \rr3}{\ktr7}\nn\\
& &+\, \biggl[-\f{i}{\kt\ko} -\f{1}{\ktr2} -i\,\l(k_1+k_2\r)\,
\biggl(\f{i}{\kt\kor2} + \f{2}{\ktr2\ko} - \f{i2}{\ktr3}\biggr)\nn\\ 
& &-\,k_1\,k_2\, \biggl(-\f{i}{\kt\,\kor3}-\f{3}{\ktr2\,\kor2} 
+\f{6\,i}{\ktr3\,\ko} +\f{6}{\ktr4}\biggr)\nn\\
& &+\, \rr3\,\l(\f{i}{\kt\,\kor4} + \f{4}{\ktr2\,\kor3}
- \f{12\,i}{\ktr3\,\kor2} - \f{24}{\ktr4\,\ko} + \f{24\,i}{\ktr5}\r)\nn\\
& &-\,i\,\l(k_1 + k_2\r)\,\rr3\, \l(-\f{i}{\kt\,\kor5} - \f{5}{\ktr2\,\kor4} 
+ \f{20\,i}{\ktr3\,\kor3}+ \f{60}{\ktr4\,\kor2} - \f{120\,i}{\ktr5\,\ko}
- \f{120}{\ktr6}\r)\nn\\
& &-\rr3\,k_1\,k_2\, \l[\f{i}{\kt\,\kor6} +\f{6}{\ktr2\,\kor5}
- \f{30\,i}{\ktr3\kor4} - \f{120}{\ktr4\,\kor3} 
+ \f{360\,i}{\ktr5\,\kor2} + \f{720}{\ktr6\,\ko} 
- \f{720\,i}{\ktr7}\r)\biggr]\nn\\ 
& &\times\,{\rm e}^{-i\,\kt/\ko}\biggr\}\nn\\
& & +\,k_2^2\; \biggl\{-\f{i}{\kt} - \f{12\,\rr3}{\ktr4} 
+ \f{720\,i\,\rr6}{\ktr7} - \f{i\,k_1}{\ktr2} -\f{48\,k_1\,\rr3}{\ktr5} 
+ \f{5040\,i\,k_1\,\rr6}{\ktr8}\nn\\
& &+\, \biggl[\f{i}{\kt} + 2\,\rr3\,
\l(-\f{i}{\kt\kor3} - \f{3}{\ktr2\,\kor2} + \f{6\,i}{\ktr3\,\ko} 
+ \f{6}{\ktr4}\r)\nn\\ 
& &+\, \rr6\, \biggl(\f{i}{\kt\,\kor6} + \f{6}{\ktr2\,\kor5} 
- \f{30\,i}{\ktr3\,\kor4} - \f{120}{\ktr4\,\kor3} 
+ \f{360\,i}{\ktr5\,\kor2} + \f{720}{\ktr6\,\ko} - \f{720\,i}{\ktr7}\biggr)\nn\\
& & +\, i\,k_1\,\l(\f{i}{\kt\,\ko} + \f{1}{\ktr2}\r)\nn\\  
& &-\,2\,i\,k_1\,\rr3\, \l(\f{i}{\kt\,\kor4}+ \f{4}{\ktr2\,\kor3}
-\f{12\,i}{\ktr3\,\kor2} - \f{24}{\ktr4\,\ko}  + \f{24\,i}{\ktr5}\r)\nn\\ 
& &-\,i\,k_1\,\rr6\,\biggl(-\f{i}{\kt\, \kor7} - \f{7}{\ktr2\,\kor6} 
+ \f{42\,i}{\ktr3\,\kor5} + \f{210}{\ktr4\,\kor4}  
- \f{840\,i}{\ktr5\,\kor3} - \f{2520}{\ktr6\,\kor2}\nn\\
& &+\, \f{5040\,i}{\ktr7\,\ko} + \f{5040}{\ktr8}\biggr)\biggr]\,
{\rm e}^{-i\,\kt/\ko}\biggr\}\biggr).
\end{eqnarray}
We find that the rest of the quantities can be written in terms of 
$\cI_{\sr\sr\g}^{31}(k_1,k_2,k_3)$ as follows: 
$\cI_{\sr\sr\g}^{32}(k_1,k_2,k_3)=\cI_{\sr\sr\g}^{31}(k_1,-k_2,k_3)$, 
$\cI_{\sr\sr\g}^{33}(k_1,k_2,k_3)=\cI_{\sr\sr\g}^{31}(-k_1,k_2,k_3)$, 
$\cI_{\sr\sr\g}^{34}(k_1,k_2,k_3)=\cI_{\sr\sr\g}^{31}(-k_1,-k_2,k_3)$, 
$\cJ_{\sr\sr\g}^{31}(k_1,k_2,k_3)=\cI_{\sr\sr\g}^{31}(k_2,k_1,k_3)$,
$\cJ_{\sr\sr\g}^{32}(k_1,k_2,k_3)=\cJ_{\sr\sr\g}^{31}(k_1,-k_2,k_3)$, 
$J_{\sr\sr\g}^{33}(k_1,k_2,k_3)=\cJ_{\sr\sr\g}^{31}(-k_1,k_2,k_3)$ and 
$\cJ_{\sr\sr\g}^{34}(k_1,k_2,k_3)=\cJ_{\sr\sr\g}^{31}(-k_1,-k_2,k_3)$.

\par

Lastly, the quantities $\cM_i(k_{1},k_{2},k_{3})$, with $i=1,2,3,$ are
given by
\begin{subequations}
\begin{eqnarray}
\cM_{1}(k_{1},k_{2},k_{3}) 
&=& \frac{1}{\kt}+\frac{k_{1}}{\kt^{2}}-\frac{6\,i\,\rho^{3}}{\kt^{4}}
-\frac{24\,i\,k_1\,\rho^{3}}{\kt^{5}}
- \biggl[\frac{1}{\kt}+k_{1}\,\l(\frac{i}{\kt\,k_{0}}
+\frac{1}{\kt^{2}}\r)\nn\\ 
& & +\,\rho^{3}\, \l(-\frac{1}{\kt\,k_{0}^{3}} 
+ \frac{3\,i}{\kt^{2}\,k_{0}^{2}} + \frac{6}{\kt^{3}\,k_{0}}
-\frac{6\,i}{\kt^{4}}\r)\nn\\
& & +\, \rho^{3}\,k_{1}\, \l(-\frac{i}{\kt
k_{0}^{4}}-\frac{4}{\kt^{2}\,k_{0}^{3}}
+\frac{12\,i}{\kt^{3}\,k_{0}^{2}} 
+\frac{24}{\kt^{4}\,k_{0}}-\frac{24\,i}{\kt^{5}}\r)\biggr]\;
{\rm e}^{-i\,\kT/k_0},\\
\cM_{2}(k_{1},k_{2},k_{3}) 
&=& \lim_{\ee\to 0}\, \l(-\f{i\,e^{i\,\kt\,\ee}}{\ee}\r) 
- \frac{k_1\,k_2+k_1\,k_3+k_2\,k_3}{\kt}
-\frac{k_1\,k_2\,k_3}{\kt^{2}} +\frac{3\,i\,\rho^{3}}{\kt^{2}}\nn\\
& & +\,\frac{6\,i\,\rho^{3}\,\l(k_1\,k_2+k_1\,k_3+k_2\,k_3\r)}{\kt^{4}} 
+\frac{24\,i\,\rho^{3}\,k_1\,k_2\,k_3}{\kt^{5}}\nn\\ 
& & -\,\biggl[i\,k_{0}-\frac{\l(k_1\,k_2+k_1\, k_3+k_2\, k_3\r)}{\kt}\nn\\
& &-\frac{i\,k_1\, k_2\, k_3}{\kt\, k_{0}}-\frac{k_1\,k_2\,k_3}{\kt^{2}}
-\frac{i\,\rho^{3}}{k_{0}^{2}}-\frac{3\,\rho^{3}}{\kt\,k_{0}}
+\frac{3\,i\,\rho^{3}}{\kt^{2}}\nn\\ 
& &-i\,\rho^{3}\,\l(k_1\, k_2+k_2\, k_3+k_2\, k_3\r)\, 
\l(\frac{i}{\kt\, k_0^3} +\frac{3}{\kt^{2}\,k_0^2}
-\frac{6\,i}{\kt^{3}\,k_0} -\frac{6}{\kt^{4}}\r)\\
& &-\,\rho^{3}\,k_1\, k_2\, k_3\,
\l(-\frac{i}{\kt\, k_0^4}
-\frac{4}{\kt^{2}\,k_{0}^{3}} + \frac{12\, i}{\kt^{3}\,k_0^{2}}
+\frac{24}{\kt^{4}\,k_{0}}-\frac{24\,i}{\kt^{5}}\r)\biggr]\,
{\rm e}^{-i\,\kT/k_0},\nn\\
\cM_{3}(k_{1},k_{2},k_{3}) 
&=& -\,\frac{3\,i\,\rho^{3}}{\kt^{2}} 
- \f{6\,i\,\rho^{3}\,\l(k_{1}+k_{3}\r)}{\kt^{3}}
-\f{18\,i\,\rho^{3}\,k_{1}\,k_{3})}{\kt^{4}}\nn\\ 
& &+\,k_{1}^{2}\, \l(\frac{1}{\kt} +\frac{k_{3}}{\kt^{2}}
-\frac{6\,i\,\rho^{3}}{\kt^{4}}
 -\frac{24\,i\,\rho^{3}\,k_{3}}{\kt^{5}}\r)\nn\\
& & +\,\biggl[-\frac{3\,\rho^{3}}{\kt\, k_{0}}
+\frac{3\,i\,\rho^3}{\kt^{2}} + 3\,\rho^{3}\,\l(k_{1}+k_{3}\r)\, 
\l(-\frac{i}{\kt\, k_{0}^{2}} -\frac{2}{\kt^{2}\,k_{0}}
+\frac{2\,i}{\kt^{3}}\r)\nonumber\\
&& -\,3\,i\,\rho^{3}\,k_{1}\,k_{3}\,
\l(\frac{i}{\kt\, k_{0}^{3}} +\frac{3}{\kt^{2}\,k_{0}^{2}}
-\frac{6\,i}{\kt^{3}\,k_{0}}-\frac{6}{\kt^{4}}\r)\nn\\
& & -\frac{k_{1}^{2}}{\kt}
-k_{1}^{2}\,k_{3}\,\l(\frac{i}{\kt\, k_{0}}+\frac{1}{\kt^{2}}\r)\nn\\
& & -\,i\,\rho^{3}\,k_{1}^{2}\, \l(\frac{i}{\kt k_{0}^{3}} +\frac{3}{\kt^{2}
k_{0}^{2}}-\frac{i6}{\kt^{3}k_{0}} -\frac{6}{\kt^{4}}\r)\\
& &-\,\rho^{3}\,k_{1}^{2}\,k_{3}\, \l(-\frac{i}{\kt\, k_{0}^{4}} 
-\frac{4}{\kt^{2}\,k_{0}^{3}} +\frac{12\,i}{\kt^{3}\,k_{0}^{2}}
+\frac{24}{\kt^{4}\,k_{0}} -\frac{24\,i}{\kt^{5}}\r)\biggr]\,
{\rm e}^{-i\,\kT/k_0}.\nn
\end{eqnarray}
\end{subequations}

\end{document}